\documentclass[12pt]{ociamthesis}  

\usepackage{amsmath}
\usepackage{amssymb}

\usepackage{dsfont} 
\usepackage{stmaryrd} 
\usepackage{tikz} 

\usepackage[shortlabels]{enumitem}

\usepackage{physics}
\usepackage{tensor}

\usepackage{multirow}
\usepackage{diagbox}

\usepackage{setspace}

\counterwithout{footnote}{chapter} 

\usepackage{hyperref}
\usepackage[noabbrev,capitalise]{cleveref}

\usepackage{xcolor}
     \definecolor{MagentaXD}{RGB}{204, 48, 152}
    \definecolor{MagentaXDdetail}{RGB}{150, 79, 126}
    \definecolor{BlueMGS}{RGB}{0, 0, 255}
    \definecolor{BlueMGSdetail}{RGB}{0, 50, 200}
    \definecolor{detail}{RGB}{110,110,110}

\newif\ifcomments
     \commentstrue
\newif\ifdetails
     \detailstrue

\newcommand*\Z{\mathbb{Z}}
\newcommand*\R{\mathbb{R}}
\newcommand*\C{\mathbb{C}}
\newcommand*\Sc{\mathbb{S}}

\newcolumntype{C}[1]{>{\centering\let\newline\\\arraybackslash\hspace{0pt}}m{#1}}

\newcommand{\SVseven}{\text{SV}^{\text{G}_2}}
\newcommand{\SVeight}{\text{SV}^\text{Spin(7)}}

\newcommand{\normord}[1]{:\mathrel{#1}:} 

\usepackage[numbers,compress]{natbib}
\renewenvironment{thebibliography}[1]{%
\begin{oldthebibliography}{#1}%
\small%
\raggedright%
\setlength{\itemsep}{5pt plus 0.2ex minus 0.05ex}%
}%
{%
\end{oldthebibliography}%
}

\title{The Geometry\\[1ex]  
and Superconformal Algebras\\[1ex]     
of String Compactifications\\[1ex]
with a $\boldsymbol{G}$-structure}   

\author{Mateo Galdeano Solans}             
\college{Wolfson College}  

\degree{Doctor of Philosophy}     
\degreedate{Trinity 2022}         

\begin{document}


\setcounter{secnumdepth}{3}
\setcounter{tocdepth}{3}

\maketitle                  

\pagenumbering{gobble}
\begin{dedication}

\flushright \textit{To my parents and my brother,}

\textit{con todo mi cari\~no.}

\end{dedication}        
\begin{abstract}

In this thesis we study string compactifications on manifolds equipped with a $G$-structure, placing a special emphasis on the interplay between geometry and physics. We follow two complementary approaches.

In the first part of the thesis we adopt a sigma model perspective and focus on the worldsheet superconformal field theory. We consider compactifications on 7-dimensional Extra Twisted Connected Sum (ETCS) G$_2$ manifolds as well as 8-dimensional Generalized Connected Sum (GCS) Spin(7) manifolds. These are special holonomy manifolds obtained by gluing together two open manifolds along isomorphic asymptotic ends in a particular fashion. We find that this geometric construction is reproduced in the worldsheet algebra via a diamond of algebra inclusions. A study of the automorphisms of these algebras leads us to conjecture new mirror maps for GCS manifolds. Finally, we explore---making use of the worldsheet algebras---whether these constructions produce manifolds of generic special holonomy.

In the second part of the thesis we change gears and consider string compactifications from a supergravity point of view. In particular, we focus on compactifications of the heterotic string down to three spacetime dimensions preserving minimal supersymmetry $\mathcal{N}=1$. These are described by the heterotic G$_2$ system, which is the 7-dimensional analogue of the Hull--Strominger system. Finding solutions to this system involves the study of 7-dimensional manifolds with integrable G$_2$-structures and the construction of G$_2$-instantons on bundles over them. In addition, the anomaly cancellation condition mixes the different degrees of freedom of the theory in a highly non-trivial way. As a result, explicit background solutions are difficult to obtain. We construct new families of solutions to the system on homogeneous 3-Sasakian manifolds---that means either the 7-sphere or the Aloff-Wallach space $N_{1,1}\,$---equipped with squashed metrics. Our solutions present constant dilaton and AdS$_3$ spacetime. Some of these families can be regarded as finite deformations from a given solution, providing an explicit description of a particular direction in the moduli space.

\end{abstract}          
\begin{originality}

The content of this thesis is based on joint work with different collaborators. In particular:

\bigskip

Chapter \ref{chap:scas} of the thesis is based on \cite{Fiset:2021ruv}, written in collaboration with Marc-Antoine Fiset.
    
\begin{enumerate}[label={[\arabic*]}, leftmargin=*]
\item M.-A. Fiset and M. Galdeano, (2021) \textit{Superconformal algebras for generalized Spin(7) and G$_2$ connected sums}, \href{https://doi.org/10.1007/JHEP10(2021)137}{\emph{JHEP} {\bfseries 10}
  (2021) 137}, [\href{https://arxiv.org/abs/2104.05716}{{\ttfamily
  2104.05716}}].
\end{enumerate}
  
\bigskip
  
Chapter \ref{chap:heterotic} of the thesis is based on \cite{delaOssa:2021qlt}, written in collaboration with Xenia de la Ossa.
    
\begin{enumerate}[label={[\arabic*]}, leftmargin=*]
  \setcounter{enumi}{1}
\item X. de la Ossa and M. Galdeano, (2021) \textit{Families of solutions of the heterotic G$_2$ system}, [\href{https://arxiv.org/abs/2111.13221}{{\ttfamily
  2111.13221}}].
\end{enumerate}

\end{originality}
\begin{acknowledgementslong}

First of all, I would like to thank my supervisor Xenia de la Ossa for introducing me to the world of research. She has been an extremely dedicated supervisor and I am deeply grateful for her constant support both in academic and non-academic matters. Her profound understanding of physics and mathematics will always be an inspiration to me.

\smallskip

My collaborator Marc-Antoine Fiset also deserves recognition. He is a fantastic researcher, he has taught me all I know about superconformal algebras, and even more importantly his advice has helped me to successfully navigate the DPhil and beyond. I would also like to praise my fellow DPhil student Enrico Marchetto: his passion and tenacity are contagious, it's a pleasure working with him, and the biggest flaw of this thesis is not including a chapter on our joint research.

\smallskip

During the course of the DPhil I have had the pleasure to discuss about research with many different people. Some of the names that come to mind are Chris Beem, Andreas Braun, Philip Candelas, Magdalena Larfors, Matt Magill, Suvajit Majumder, Challenger Mishra, Gon\c{c}alo Oliveira, Daniel Platt, Yuuji Tanaka and Atul Sharma. I would particularly like to thank Mario Garcia-Fernandez, both for the conversations and for his hospitality when I have visited him in Madrid.

\smallskip

I have spent countless hours with my officemates working in the Mathematical Institute. In my first year, Andrea, Diego and Mark gave me a warm welcome to the department. In my second year, Joseph and Sujay joined and the S1.02 office officially became the ``fun office''. Without them, there would have not been a meme wall or a Britney Spears calendar. And I would have not travelled to work every day with a smile on my face.

\smallskip

There are other fellow DPhil students who have been part of my life in the department and I would like to mention some of them: Andrea, Beppe, Carmen, Dewi, Horia, Lea, Maria, Max and Mohamed. Pyry deserves a special mention for introducing me to Finnish saunas. So does Palash (a.k.a. Splash), for being a great guy.

\smallskip

I have been incredibly lucky to share these four years with some friends who have been like a family to me. Filippo has made me laugh more than anybody else in Oxford. Giulia has shared with me two of my biggest passions: gossip and board games. Jean always transmits a wonderful, cheerful energy. Juan showed me the joy of playing music together (with our band \emph{Betti and the Numbers}), and I showed him the joy of watching the best football team in the world (\emph{SD Huesca}, of course). Marieke has a brilliant sense of humor and a heart of gold. And finally, Pietro has been not only a true friend, but also a source of inspiration for my own life. Thank you to all of you.

\smallskip

Life outside the department has been memorable and there are several people I would like to acknowledge. First, I would like to thank the ``Puntentados'' for the joyful board game sessions: David, Guillermo, Laura, Izar and Sam. I am also grateful to Paula for her support during my first year in Oxford as a master's student. I would specially like to thank \'Alex for all the walks in Port Meadow, all the video game sessions, and for coming to visit me in Toronto.

\smallskip

Thanks to the ``Guayabitas'', Alex and Andrew, for their support and all the times we have shared. They have done a great job at improving my English pronunciation.

\smallskip

During my stay in Oxford I have had the chance to play lots of ultimate. Thanks to the Wolfson college team and specially to OW! for all the great times. I have to highlight my friend Kay as well as \emph{der Sekret\"ar} Jonas, who is my only friend from Oxford who has visited me in Madrid. Ach so!

\smallskip

I thank Eli for meeting with me every month, for introducing me to so many nice people, and for reminding me that good friends remain in your life forever. I would also like to thank all my ``old'' friends, both in Madrid and spread around the world, particularly those who have taken the time to pay me a visit in Oxford.

\smallskip

A very, very special thank you goes to Kayla for all her love and kindness. Her encouragement and support during the process of writing this thesis gave me strength to succeed in this last push. Thank you!

\smallskip

I would also like to thank all the people I have met during my academic travels as a DPhil student: Jersey, Toronto, Hamburg, Rio de Janeiro, Madrid and Warsaw. I thank my students in Oxford as well: all of them brilliant and full of enthusiasm, it's been a pleasure teaching them.

\bigskip

Finally, I want to thank my family for their continuous support and love. I thank my parents, Virginia and Ram\'on, and my brother Ramiro. I thank my family in Huesca and other places of Spain, and I want to specially thank the ``Primotis'' who visited me in Oxford: Ana, Lu and Gabriela. I thank my ``fairy godmother'' Bel\'en for using her powers to make me smarter, and I also want to thank my grandmothers Sara and Carmen. Carmen said I would not finish the thesis on time with all the travelling I was doing. Luckily for me, she was wrong this time.

\bigskip

I gratefully acknowledge the support I have received during the completion of my thesis. During my first year I held a fellowship from ``la Caixa” Foundation (ID 100010434) with fellowship code LCF/BQ/EU17/11590062. The rest of my DPhil has been supported by a scholarship from the Mathematical Institute, University of Oxford. I would also like to thank the Fields Institute, Toronto for support during my long term visit to the institution.

\end{acknowledgementslong}  
\begin{dedication}

``Sit down, and I will tell you a tale\\
like none that you have ever heard.''

\flushright \textit{Prince of Persia: The Sands of Time.}

\end{dedication}

\begin{romanpages}          
\tableofcontents            
\end{romanpages}            

\pagenumbering{arabic}

\chapter{Introduction}

\label{chap:introduction}

\doublespacing

String compactifications were originally introduced to reconcile our experience of a four-dimensional world with the ten-dimensional nature of supersymmetric string theory \cite{Candelas:1985en}. Almost 40 years later their scope has developed well beyond the original phenomenological motivation, becoming one of the most important techniques to produce and understand quantum field theories in different dimensions.

One of the most powerful features of string compactifications is the intimate relationship between the geometry of the compactifying manifold $M$ and the physical properties of the theory obtained after compactification. This constitutes a bidirectional interplay: a good understanding of the geometry of $M$ provides new information about the physics of the compactification and vice-versa.

A well-known instance of this phenomenon is the correspondence between the number of supersymmetries that are preserved after compactification and the number of nowhere-vanishing spinors on $M$ that are covariantly constant under a---generically torsionful---connection $\nabla$. This statement is a direct consequence of the gravitino Killing spinor equation and it is important to note that the torsion of the connection $\nabla$ is determined by the background physical fluxes \cite{Strominger:1986uh, Hull:1986kz, Gauntlett:2001ur, Friedrich:2001nh, LopesCardoso:2002vpf, Gauntlett:2003cy, Grana:2005jc}.

This is a paradigmatic example of the deep interactions between the geometry and physics of string compactifications. The immediate upshot is that we must restrict our attention to spin manifolds to obtain supersymmetric backgrounds.

The existence of nowhere-vanishing spinors on $M$ can be equivalently formulated in terms of the existence of a particular $G$\emph{-structure} on $M$, see for example \cite{Joyce:2007}. This can be understood as follows: well-defined, nowhere-vanishing differential forms on $M$ can be constructed as bilinears of the spinors, and it turns out these forms precisely characterize the underlying $G$-structure.

The spinors and the forms that define the $G$-structure are both covariantly constant with respect to a family of connections with holonomy $G$. We say these connections are \emph{compatible} with the $G$-structure and, as we pointed out earlier, they play a major role in string compactifications: when background fluxes are turned on the resulting connection must be compatible with the $G$-structure to preserve supersymmetry. For the reader unfamiliar with the language of $G$-structures, a review is provided in \cref{chap:Gstructures}.

A useful observation is that $G$-structures constitute a generalization of the concept of $G$-holonomy. Indeed, in the case when the compatible connection has no torsion it reduces to the Levi-Civita connection and the manifold has special holonomy. This is the adequate setting for compactifications with vanishing flux.

It is natural then to explore further connections and implications between $G$-structures and supersymmetric vacuum solutions. The purpose of this thesis is to address this question from two complementary points of view. In the first part of the thesis we adopt a worldsheet perspective and rephrase the geometric structure of certain Spin(7) and G$_2$ manifolds in terms of superconformal algebras. In the second part of the thesis we employ G$_2$-structure manifolds to produce new background solutions of the heterotic string.

Let us be more precise about the compactifications we are interested in studying. In this thesis we consider supersymmetric backgrounds---of type II as well as heterotic string theories\footnote{Despite their importance, in this thesis we will not study compactifications of M-theory \cite{Witten:1995ex, Horava:1995qa} or F-theory \cite{Vafa:1996xn}.}---given by a warped product
\begin{equation}
\label{eq:compactificationansatz}
\mathcal{M}_d\times M \, ,
\end{equation}
where $\mathcal{M}_d$ is either Minkowski $\left(\mathbb{M}_d\right)$ or Anti-de Sitter $\left(\text{AdS}_d\right)$ space in $d$ dimensions, and $M$ is a $(10-d)$-dimensional compact manifold. We focus on two cases: $d=3$, for which $M$ is a 7-dimensional manifold endowed with a G$_2$-structure; and $d=2$, for which $M$ is an 8-dimensional manifold endowed with a Spin(7)-structure.

\bigskip

It is useful to regard string compactifications as non-linear sigma models.\footnote{See for example \cite{Hori:2003ic, Melnikov:2019tpl, Fiset:2019ecu} for more detailed presentations on sigma models.} Strings propagating in a given background can be parametrized via maps from the string worldsheet $\Sigma$ into the \emph{target space} consisting of the geometry \eqref{eq:compactificationansatz}
\begin{equation}
\label{eq:sigmamodel}
X:\,\Sigma\longrightarrow\mathcal{M}_d\times M \, .
\end{equation}
A two-dimensional action governs the motion of the string. For illustration purposes, consider the case where the string enjoys $\mathcal{N}=(1,1)$ supersymmetry. This is the sigma model that correctly describes type II string compactifications and its action can be locally written as
\begin{equation}
\label{eq:bosonicaction}
S=\int\dd z^+\dd z^- \dd\theta^+\dd\theta^-\left(g_{ij}+B_{ij}\right)D_+X^iD_-X^j \, ,
\end{equation}
where $\left(z^\pm,\theta^\pm\right)$ are the worldsheet lightcone superspace coordinates, $g$ and $B$ are the metric and $B$-field on the target space, $D_\pm$ are the superspace derivatives and $X^i$ denotes the sigma model superfields.

We focus on the case where $M$ has a $G$-structure and consider one of the nowhere-vanishing $p$-forms $\phi$ associated to that $G$-structure. Locally, we can write the form as\footnote{In this thesis we will always abbreviate $\dd x^{i_{1}}\wedge\cdots\wedge\dd x^{i_{p}}=\dd x^{i_{1}\cdots i_{p}}$.} $\phi=\frac{1}{p!}\phi_{i_{1}\cdots i_{p}}\dd x^{i_{1}\cdots i_{p}}$ and define the following infinitesimal transformation of the superfields
\begin{equation}
\delta^\phi X^i=\frac{\epsilon\left(z^\pm,\theta^\pm\right)}{\left(p-1\right)!}\,\phi\indices{^i_{i_2 \cdots i_p}}\, D_+X^{i_2}\cdots D_+X^{i_p}\, ,
\end{equation}
where $\epsilon\left(z^\pm,\theta^\pm\right)$ has even Grassmann parity if $p$ is odd and vice-versa. It can be shown \cite{Howe:1991vs, Howe:1991im, Howe:1991ic} that this constitutes a (non-linear) symmetry of the action \eqref{eq:bosonicaction} provided that $D_-\, \epsilon=0$ and $\nabla^+\phi=0$, where $\nabla^+$ is the connection with torsion equal to the physical flux $H=\dd B$. We call this a \emph{Howe--Papadopoulos symmetry}.\footnote{There exists an analogous symmetry in the other chiral sector upon exchanging the roles of $+$~and~$-$. Furthermore, in the case of $(1,0)$ sigma models---relevant for heterotic string compactifications---the symmetry can also be extended to include a gauge bundle \cite{delaOssa:2018azc}.}

Note the condition $\nabla^+\phi=0$ is satisfied when the connection $\nabla^+$ is compatible with the $G$-structure. This imposes a relation between the $G$-structure and the flux, and can be interpreted as the worldsheet realization of the gravitino Killing spinor equation.

This is a fundamental example of the philosophy of this thesis: \emph{there exists a correspondence between $G$-structures on the compactifying manifold and certain aspects of the physical theory after compactification}. We have described how the $G$-structure gives rise---upon imposing a physical condition on the geometry---to brand new symmetries on the worldsheet. We now explain how further physical structures emerge from the $G$-structure.

The action \eqref{eq:bosonicaction} is invariant under superconformal transformations. The presence of a $G$-structure in the target space enhances the original superconformal symmetry with the addition of Howe--Papadopoulos symmetries. The chiral algebra of commutators of all these symmetries can be computed \cite{Howe:1991im, Stojevic:2006pq} and it has been shown that in the $G$-holonomy case it closes, providing a classical non-linear algebra.

Upon quantization of the theory, a natural question is whether these symmetries develop an anomaly. It was shown recently in the context of $(1,0)$ sigma models that Howe--Papadopoulos symmetries are indeed anomalous. Nevertheless, it was also shown that these anomalies can be cancelled when $\alpha'$ corrections are taken into account \cite{delaOssa:2018azc}. Thus, when the compact manifold has $G$-holonomy we expect the worldsheet quantum theory to display an enhancement of the superconformal symmetry, meaning that the superconformal algebra can be extended beyond the super-Virasoro algebra by introducing additional operators. These operators are quantum-corrected versions of the currents associated to the Howe--Papadopoulos symmetries.

Such extended, non-linear, chiral, superconformal algebras---which we refer to as \emph{W-algebras}---were first discovered in the Calabi-Yau case by Odake \cite{Odake:1988bh}. The cases of holonomy G$_2$ and Spin(7) were first discussed by Shatashvili and Vafa \cite{Shatashvili:1994zw}, see also \cite{Figueroa-OFarrill:1990tqt, Figueroa-OFarrill:1990mzn, Blumenhagen:1991nm, Figueroa-OFarrill:1996tnk}. The more general situation where the target manifold has a $G$-structure with torsion has been recently tackled for the G$_2$ case by Fiset and Gaberdiel in \cite{Fiset:2021azq}, where a new family of extended algebras was found. This suggests that geometric structures beyond $G$-holonomy might be captured by these W-algebras.

This is indeed the case: in \cite{Fiset:2018huv} Fiset studies compactifications on Twisted Connected Sum (TCS) G$_2$ manifolds from the worldsheet. These manifolds are obtained by gluing together two open Calabi-Yau manifolds and two circles along isomorphic asymptotic ends, producing a compact manifold with G$_2$ holonomy \cite{Kovalev:2003, Corti:2012, Corti:2012kd}. As shown in \cite{Fiset:2018huv}, the piecewise nature of the construction can be interpreted as algebra inclusions in the realm of W-algebras. Similarly, the gluing morphism corresponds to an automorphism of W-algebras. These results are extended in \cref{chap:scas} to manifolds of holonomies G$_2$ and Spin(7) constructed via generalizations of the TCS procedure \cite{Crowley:2015ctv, Nordstrom:2018cli, Goette:2020, Braun:2018joh}. Additionally, we investigate in detail the rich interplay between geometry and superconformal algebras.

\bigskip

As we have explained, the existence of a $G$-structure in the compact manifold is essential for obtaining a supersymmetric string background solution. Nevertheless, this condition by itself is not sufficient to guarantee the existence of a solution to the equations of motion. It is then useful to adopt a complementary approach to the sigma model techniques previously discussed: we revisit the 10-dimensional point of view to understand which additional requirements are imposed by these equations.

Every string theory can be described as some type of supergravity in the low-energy limit. This is the perspective we adopt in what follows. Supergravity theories in ten dimensions have an explicit action that we can study. As an example, we present below the action of the common NS-NS sector of type II and heterotic theories \cite{Blumenhagen:2013fgp}
\begin{equation}
\label{eq:actioncommonsector}
S=\frac{1}{2 \ \tilde{\kappa}^2_{10}}\int\dd^{10}x \ \sqrt{-g} \ e^{-2\Phi}\left(R+4\abs{\dd\Phi}^2-\frac{1}{2}\abs{H}^2\right).
\end{equation}
Here $\tilde{\kappa}_{10}$ is the 10-dimensional gravitational constant, $R$ is the Ricci scalar, $\Phi$ is the dilaton and $H$ is the three-form flux (which depends on the B-field). The massless field content varies from one string theory to another and constitutes the set of fields that we need to determine in the background. All supergravity theories come equipped with a supergravity multiplet---the particular amount of supersymmetry depends on the theory under consideration---that always includes a graviton, two-form field and dilaton as well as their fermionic superpartners. Note how the bosonic fields indeed appear in the action \eqref{eq:actioncommonsector}.

In type II theories, the supergravity multiplet includes additional form fields and the action \eqref{eq:actioncommonsector} is supplemented with further terms featuring the corresponding field strengths. These additional fluxes are not present in heterotic and type I theories. Nevertheless, heterotic and type I theories have massless gauge fields and gauginos due to the presence of an extra vector multiplet. Even though the field content of the theories differ, the process to obtain a vacuum solution following the ansatz \eqref{eq:compactificationansatz} is the same in all cases: we begin by setting all the background fermionic degrees of freedom to zero and the bosonic fields are then constrained as we now explain.

Supersymmetry is preserved in the vacuum if the solution is invariant under the supersymmetry transformations. Therefore, we must study the conditions under which the supersymmetry variation of the background fields vanishes. As background fermions are set to zero in our ansatz, the variation of the background bosons vanishes automatically and does not impose any constraints. The equations governing the vanishing of the gravitino and dilatino variations receive the name of \emph{Killing spinor equations}. As we have already explained, the gravitino Killing spinor equation determines the $G$-structure on the compact manifold $M$ in terms of the fluxes. The dilatino equation typically imposes further restrictions to the $G$-structure.

In addition, the background fields must satisfy the equations of motion obtained from the supergravity action. However, it is usually the case that not all the equations of motion need to be explicitly checked. First of all, the physical fluxes are constrained to satisfy the so-called \emph{Bianchi identities}, which ensure the fluxes correctly describe the field strength of the form fields. If the fluxes satisfy their equations of motion together with these Bianchi identities and the Killing spinor equations, under mild assumptions the remaining equations of motion---such as the Einstein equations---are automatically satisfied \cite{Gauntlett:2002fz, Gauntlett:2002sc, Lust:2004ig, Ivanov:2009rh}. 

A field configuration satisfying all these conditions constitutes a supersymmetric vacuum solution of the corresponding supergravity theory. As we have motivated, the construction of these solutions is non-trivial and it can become rather involved when background fluxes are turned on. In the following we discuss in more detail heterotic string compactifications, illustrating some of their most compelling features and pointing out several captivating questions that have yet to be addressed.

An important caveat that should be stressed about the string backgrounds we have introduced is that they are only valid in the low-energy supergravity limit of string theory. Hence, the solutions are not going to be reliable at energy scales where higher order stringy corrections become relevant. Grasping a good understanding of these corrections---both at a perturbative and non-perturbative level---is an extremely challenging task.

We explore these stringy effects in a perturbative fashion: the low-energy effective field theory coming from the fully-fledged string theory can be obtained as an expansion in terms of the string parameter $\alpha'$. The supergravity action can then be corrected order by order in the perturbative parameter $\alpha'$, see for example \cite{Blumenhagen:2013fgp} for a more detailed explanation on how this is achieved.

In the heterotic case it has been argued \cite{delaOssa:2014msa} that corrections beyond first order can be understood simply in terms of field redefinitions. This suggests that an in-depth analysis of heterotic supergravity with just the first order $\alpha'$ corrections might suffice to capture the full set of stringy effects. Following this motivation, we focus on heterotic supergravity with $\alpha'$ corrections at first order. The bosonic part of the first order correction to the action is given by \cite{Bergshoeff:1988nn, Bergshoeff:1989de, Hull:1987pc}
\begin{equation}
S_{\text{correction}}=-\frac{1}{2 \ \tilde{\kappa}^2_{10}}\int\dd^{10}x \ \sqrt{-g} \ e^{-2\Phi} \ \frac{\alpha'}{4} \left(\tr\abs{F_A}^2-\tr\abs{R_\Theta}^2\right).
\end{equation}
Two new bosonic fields appear explicitly in the action: a vector bundle connection $A$ and a tangent bundle connection $\Theta$, each of them belonging to an $\mathcal{N}=1$ vector supermultiplet. These new fields give rise to extra conditions when looking for background solutions with $\alpha'$ corrections.

For instance, the presence of a gaugino results in an additional Killing spinor equation that constrains $A$ to be an instanton connection \cite{Candelas:1985en, Strominger:1986uh, Hull:1986kz, Gauntlett:2003cy, Ivanov:2003nd, ReyesCarrion:1998si}. Similarly, when $\Theta$ is an instanton the equations of motion of the action are automatically satisfied \cite{Ivanov:2009rh}. Furthermore, the NS flux receives an $\alpha'$ correction involving the gauge connections
\begin{equation}
\label{eq:anomalycancellationintro}
    H=\dd B+\frac{\alpha'}{4}\left(\mathcal{CS}(A)-\mathcal{CS}(\Theta)\right) \, ,
\end{equation}
where $\mathcal{CS}$ denotes the Chern-Simons three-forms. As a result, the Bianchi identity needs to be modified and the physical flux is in general not closed anymore \cite{Candelas:1985en, Strominger:1986uh, Hull:1986kz}
\begin{equation}
\label{eq:heteroticbianchiintro}
\dd H=\frac{\alpha'}{4}\big(\tr (F_A\wedge F_A) -\tr (R_\Theta\wedge R_\Theta)\big) \, .
\end{equation}
The heterotic Bianchi identity relates the flux---which is itself related to the $G$-structure of the compactifying manifold---and the gauge connections in a complicated way. It constitutes a highly non-trivial constraint not just for the geometry but also for the gauge theory of the solution, and it will play a major role in the solutions we are going to describe in this thesis.

Heterotic flux compactifications down to four dimensions are described by the Hull--Strominger system \cite{Strominger:1986uh, Hull:1986kz}. The compact manifold must have an SU(3)-structure with torsion,\footnote{When the flux $H$ is turned off, the traces in \eqref{eq:heteroticbianchiintro} must cancel and we are in the case of Calabi--Yau compactifications with the standard embedding \cite{Candelas:1985en}.} which posed a technical challenge when the system was originally discovered. This difficulty was overcome around 15 years later, resulting in a thorough study of the system and the construction of several explicit solutions. For a non-exhaustive list of references see \cite{Strominger:1990et, Dasgupta:1999ss, Ivanov:2000fg, LopesCardoso:2002vpf, Becker:2002sx, Goldstein:2002pg, Becker:2003yv, Becker:2003gq, LopesCardoso:2003dvb, LopesCardoso:2003sp, Becker:2003sh, Gukov:2003cy, Gurrieri:2004dt, Micu:2004tz, Li:2004hx, deCarlos:2005kh, Becker:2005nb, Fu:2006vj, Becker:2006et, Adams:2006kb, Benmachiche:2008ma, Fernandez:2008wa, Fu:2009zee, McOrist:2010jw, Andreas:2010qh, Martelli:2010jx, Andreas:2010cv, Klaput:2012vv, Cicoli:2013rwa, Garcia-Fernandez:2013gja, Anderson:2014xha, delaOssa:2014cia, Fernandez:2014kwa, Fei:2014aca, Garcia-Fernandez:2015hja, Fei:2015kua, Fei:2015yaa, Otal:2016bgn, Garcia-Fernandez:2016ofz, Fei:2017ctw, Ivanov:2017gpd, Larfors:2018nce, Garcia-Fernandez:2018qcl, Fino:2019mvp, Pujia:2021wsi, Tsuyuki:2021xqu} as well as the detailed reviews \cite{Blumenhagen:2006ci, Garcia-Fernandez:2016azr}.

Heterotic flux compactifications on manifolds that are not 6-dimensional have been considerably less studied---see for example \cite{Gunaydin:1995ku, Acharya:1996tw, Louis:2001uy, Ivanov:2001ma, Gauntlett:2002sc, Gauntlett:2003cy, Ivanov:2003nd, Gran:2005wf, Fernandez:2008wla, Fernandez:2008pf, Kunitomo:2009mx, Lukas:2010mf, Nolle:2010nn, Harland:2011zs, Gray:2012md, Fernandez:2014pfa, Clarke:2020erl, Lotay:2021eog, delaOssa:2021qlt}---and deserve further investigation: they have the potential of shedding light on features of heterotic theories that would otherwise go unnoticed.

Compactifications down to three dimensions constitute an archetypical example of this, as they are the only heterotic compactifications that allow for Anti-de Sitter (AdS) solutions \cite{Beck:2015gqa}.

This possibility was overlooked for some years in the literature and it has exciting consequences. For instance, one could hope to develop some heterotic version of holography, and some work has been done in this direction \cite{Dabholkar:2007gp, Lapan:2007jx, Hohenegger:2009zz, DominisPrester:2009vvp, Gemmer:2012pp}.

The conditions to obtain a heterotic compactification down to three dimensions preserving minimal supersymmetry are encoded in the \emph{heterotic G$_2$ system} \cite{Gauntlett:2002sc, Friedrich:2001yp, Gauntlett:2003cy, Ivanov:2003nd}, see also \cite{delaOssa:2017pqy}. The 7-dimensional compact manifold must have an \emph{integrable}\footnote{Since the only physical flux of heterotic theories is the NS flux, the $G$-structure in the compact manifold must admit a compatible connection with totally antisymmetric torsion \cite{Friedrich:2001nh}. A G$_2$-structure only satisfies this in the integrable case, see \cref{chap:Gstructures}.} G$_2$-structure, and the connections $A$ and $\Theta$ must both be G$_2$-instantons. Therefore, these compactifications are a natural playground to explore the geometry and gauge theory of G$_2$ manifolds, which have received much attention in the mathematics community recently.\footnote{\label{foot:g2references}For a non-exhaustive list of references see \cite{Joyce:1996a, Joyce:1996b, Cabrera:1996, Donaldson:1996kp, Friedrich:1997, Fernandez:1998, Friedrich:2001nh, Friedrich:2001yp, Cleyton:2001yf, Lee:2002fa, Bilal:2003bf, Cleyton:2003, Kovalev:2003, Bryant:2005mz, Karigiannis:2005, Chiossi:2004ci, Verbitsky:2011, Friedrich:2007, Karigiannis:2007, Cleyton:2007, Nordstrom:2007, Donaldson:2009yq, Kovalev:2010, Grigorian:2009ge, Conti:2011, Earp:2011dh, Grigorian:2011ap, Walpuski:2011xla, Corti:2012, Corti:2012kd, Crowley:2012, Fino:2013, SaEarp:2013sgz, Lotay:2015, Lotay:2017, Walpuski:2015wdf, Crowley:2015ctv, Menet:2015yjx, Calvo-Andrade:2016fti, Joyce:2016fij, Haydys:2017gag, Joyce:2017, Lotay:2018gxe, Nordstrom:2018cli, Karigiannis:2020, Driscoll:2019gad, Goette:2020, DelZotto:2021ydd}.}

There are very few explicit solutions of the heterotic G$_2$ system available in the literature. Most of them possess a 3-dimensional Minkowski spacetime \cite{Fernandez:2008wla, Nolle:2010nn, Fernandez:2014pfa, Clarke:2020erl}---although note that some of these solutions have been recently shown to preserve more than one supersymmetry \cite{delaOssa:2021cgd}---whereas AdS solutions have only been obtained recently \cite{Lotay:2021eog}. Thus, the construction of further explicit solutions constitutes an interesting research direction.

There exists yet another motivation for the search of heterotic G$_2$ vacua. The infinitesimal moduli space of the heterotic G$_2$ system has been the subject of recent study\footnote{Moduli spaces of string theory backgrounds have been studied since the early days of string compactifications, see for example \cite{DHoker:1988pdl, Seiberg:1988pf, Strominger:1990pd, Candelas:1990rm, Candelas:1990pi, Aspinwall:1993nu, Giveon:1994fu, Seiberg:1994bz, Aspinwall:1994rg, Polchinski:1995sm, Katz:1997eq, Gaiotto:2009we, Anderson:2014xha, delaOssa:2014cia, Candelas:2016usb}.} \cite{Clarke:2016qtg, delaOssa:2016ivz, delaOssa:2017pqy, delaOssa:2017gjq, Fiset:2017auc, Clarke:2020erl}. This moduli space combines geometric moduli---for example, the metric moduli of the compactifying manifold---and moduli with an origin in physics---such as the B-field moduli. This results in yet another rich interplay between different aspects of the string compactification.

A notable consequence is that this infinitesimal moduli space is finite-dimensional.  This is a remarkable property: the infinitesimal moduli space of integrable G$_2$-structures on a manifold is generically infinite-dimensional,\footnote{In fact, it is finite-dimensional only when the manifold presents G$_2$ holonomy \cite{Joyce:1996a}.} yet the heterotic G$_2$ system is so stringent that most of these infinitesimal deformations are not allowed. This is due to the heterotic Bianchi identity \eqref{eq:heteroticbianchiintro}, which intertwines the geometric moduli with the moduli of the G$_2$-instanton connections $A$ and $\Theta$ in a non-trivial way.

The understanding of the moduli space beyond the infinitesimal case remains elusive, one of the reasons being the very limited number of explicit solutions available in the literature. With this problem in mind, we construct in \cref{chap:heterotic} the first AdS heterotic G$_2$ solutions that can be regarded as finite deformations of a given solution, providing a first glimpse into the (finite) moduli space.

\bigskip

The two perspectives on string compactifications we have described complement each other. In the worldsheet approach, we have access to the toolkit of superconformal field theories and we find a correspondence between the geometry of the compact manifold and superconformal worldsheet algebras. We can then use these algebras to make very general statements about these backgrounds. On the other hand, in the supergravity approach we have a detailed description of the conditions that a string compactification must satisfy. Using tools from geometry we can then construct completely explicit background solutions.

In this thesis, we follow both approaches to explore compactifications on manifolds with a G$_2$-structure, although we will also study some manifolds with Spin(7)-holonomy. The thesis is organized as follows.

We begin in \cref{chap:Gstructures} introducing $G$-structures, the key geometric concept that will permeate the whole thesis. The purpose is twofold: on the one hand we hope to provide a clear and concise exposition for the reader unfamiliar with the topic and on the other hand we will set the notation for the remainder of the thesis.

In \cref{chap:scas} we present the results of \cite{Fiset:2021ruv}, which is joint work with Marc-Antoine Fiset. In this chapter we adopt a sigma model perspective and analyze the worldsheet superconformal algebras corresponding to Extra Twisted Connected Sum (ETCS) G$_2$ manifolds and Generalized Connected Sum (GCS) Spin(7) manifolds. We find that the geometric construction is directly reflected in the algebras in the guise of a diamond of inclusions. We study automorphisms of these algebras and conjecture new mirror maps for GCS manifolds. Finally, we discuss the genericity of the geometric construction using information from the worldsheet.

In \cref{chap:heterotic} we turn our attention to the heterotic G$_2$ system and follow \cite{delaOssa:2021qlt}, which is joint work with Xenia de la Ossa. We construct families of solutions to the system on homogeneous 3-Sasakian manifolds with squashed metrics. We work with integrable G$_2$-structures on these manifolds and we construct different G$_2$-instanton connections over them. All our solutions present non-zero flux, constant dilaton and AdS$_3$ spacetime. Some of our families can be regarded as finite deformations from a given solution, thus describing an unobstructed direction in the moduli space of the heterotic G$_2$ system.

We finish with \cref{chap:conclusions}, which is considerably shorter. We provide a summary of the results of the thesis and supply outlook on future research directions.

Each of the main chapters in this thesis includes a short introduction containing relevant background. In addition, we incorporate a conclusion as well as some useful appendices at the end of those chapters. Therefore, Chapters~\ref{chap:scas} and \ref{chap:heterotic} can essentially be read independently.

\chapter{Review of $G$-structures}

\label{chap:Gstructures}


In this first chapter we briefly recall some basics of $G$-structures that will be ubiquitous in this thesis. Consider a string compactification on a compact manifold $M$. As we explained in \cref{chap:introduction}, a $G$-structure on $M$ is intimately related with the physics of the background solution. Thus, the existence of $G$-structures on $M$ will be the most fundamental source of geometric information for us. Every single manifold we encounter in this thesis will be endowed with a particular $G$-structure.

Following \cite{Joyce:2007}, a $G$\emph{-structure} on $M$ is a reduction of the frame bundle\footnote{The \emph{frame bundle} of a manifold $M$ is the principal bundle associated to the tangent bundle. The fibre over a point is given by the set of all ordered basis elements of the tangent space at that point. The general linear group acts on the fibre by changing the basis and provides the principal bundle structure.} of $M$ to a $G$-principal subbundle. Nevertheless, this definition turns out not to be practical for our purposes in this thesis. Still following \cite{Joyce:2007}, a $G$-structure is alternatively characterized by the existence of certain well-defined, nowhere-vanishing invariant forms on the manifold, that we call $G$-forms. Thus, we define each $G$-structure that will appear in this thesis using the language of $G$-forms.


Before doing so, let us stress that the concept of $G$-structure is a natural generalization of the concept of $G$-holonomy, which might be more familiar to the reader. In the language of principal bundles this can be seen as follows: when a connection $\nabla$ on the frame bundle---generically with torsion---reduces to a connection on the $G$-principal subbundle we say that $\nabla$ is \emph{compatible} with the $G$-structure. In this case, the holonomy of $\nabla$ is contained in $G$.\footnote{The existence of a connection on $M$ with holonomy contained in $G$ is actually equivalent to the existence of a $G$-structure on $M$. Therefore, given a $G$-structure there always exists at least one compatible connection.} If the Levi-Civita connection turns out to be compatible with the $G$-structure, we say that the $G$-structure is \emph{torsion-free} and we find that $M$ has holonomy contained in $G$. If the holonomy of $M$ is precisely $G$, we say that $M$ has $G$\emph{-holonomy}.

In terms of $G$-forms, the holonomy of $M$ is contained in $G$ when the $G$-forms are covariantly constant with respect to the Levi-Civita connection. It can be shown this condition is equivalent to the vanishing of the exterior derivatives of the $G$-forms. For the study of worldsheet superconformal algebras in \cref{chap:scas} we will restrict ourselves to the case of $G$-holonomy. On the other hand, in \cref{chap:heterotic} we need to work with more general G$_2$-structures in order to find solutions preserving minimal supersymmetry.

The existence of a $G$-structure has a direct impact on the bundles over the manifold $M$. A major consequence is that tensors on a $G$-structure manifold can be decomposed into irreducible representations of the group $G$. The $G$-forms characterize these representations since they can be used to construct projector operators onto the irreducible subspaces.

Perhaps the most important case is the decomposition of the exterior derivatives of the $G$-forms into $G$ representations. These derivatives can be rewritten in terms of irreducible differential forms that we call \emph{torsion classes}. These torsion classes can be used to classify $G$-structures, for example the vanishing of all torsion classes corresponds to a torsion-free $G$-structure.

With the purpose of illustrating these statements and fixing notation, we proceed to define each of the $G$-structures appearing in this thesis. In each case we include for completeness a description of the torsion classes as well as a comment on connections with totally antisymmetric torsion, which are relevant for string compactifications with NS flux.

\section{G$_2$-structures}
\label{sec:G2Structures}

Let $Y$ be a 7-dimensional manifold and let $\lbrace e^1,\dots,e^7\rbrace$ denote an orthonormal coframe on $Y$. We say a three-form $\varphi$ on $Y$ is \emph{positive} if it can be locally written as
\begin{equation}
\label{eq:varphi0}
\varphi=e^{123}+e^{145}+e^{167}+e^{246}-e^{257}-e^{347}-e^{356} \, .
\end{equation}
A G$_2$\emph{-structure}\footnote{For more detailed accounts on G$_2$-structures, see \cite{Joyce:2000, Bryant:2005mz, Grigorian:2011ap, Karigiannis:2020}.} on $Y$ is equivalent to the existence of a (nowhere-vanishing) positive three-form $\varphi$ on $Y$. We call this the \emph{associative} three-form.\footnote{The subgroup of $\text{SO}(7)$ preserving the associative three-form is precisely G$_2\,$.}

The associative three-form defines a metric and an orientation on $Y$ that we can use to construct the dual four-form $\psi=*\varphi$ on $Y$. This is called the \emph{coassociative} four-form and it locally takes the form
\begin{equation}
\psi=e^{1357}+e^{2345}+e^{2367}+e^{4567}-e^{1247}-e^{1256}-e^{1346} \, .
\end{equation}
Thus, the G$_2$-forms are the associative and coassociative form. As we mentioned earlier, tensors on $Y$ decompose into irreducible G$_2$ representations. For differential forms, the decomposition is given by
\begin{equation}
\label{eq:splittingofformseq}
\Lambda^0=\Lambda^0_1 \, ,\qquad
\Lambda^1=\Lambda^1_7 \, ,\qquad
\Lambda^2=\Lambda^2_7\oplus\Lambda^2_{14} \, ,\qquad
\Lambda^3=\Lambda^3_1\oplus\Lambda^3_7\oplus\Lambda^3_{27} \, ,
\end{equation}
where $\Lambda^k$ denotes the space of $k$-forms on $Y$ and $\Lambda^k_p$ denotes the subspace of $\Lambda^k$ consisting of $k$-forms transforming in the $p$-dimensional irreducible representation of G$_2\,$. The decomposition for higher degrees follows from Hodge duality. Each $\Lambda^k_p$ can be described in terms of the G$_2$-forms, for example $\Lambda^2_{14} \, $---which corresponds to the two-forms contained in the Lie algebra of G${}_2\,$---is given by
\begin{equation}
\label{eq:splitofg2liealgebra}
\Lambda^2_{14}=\lbrace\beta\in\Lambda^2 : \beta\lrcorner \, \varphi=0\rbrace=\lbrace\beta\in\Lambda^2 : \beta\wedge\psi=0\rbrace \, .
\end{equation}
The exterior derivatives of the associative and coassociative forms can be decomposed into G$_2$ representations as follows \cite{Gray:1980}
\begin{align}
\label{eq:torsionclassequation}
\begin{split}
\dd\varphi &= \tau_0 \, \psi+3\,\tau_1\wedge\varphi+*\tau_3 \, ,\\
\dd\psi &= 4\,\tau_1\wedge\psi+*\tau_2 \, .
\end{split}
\end{align}
Here the \emph{torsion classes} $\tau_k$ are $k$-forms with $\tau_3\in\Lambda^3_{27}$ and $\tau_2\in\Lambda^2_{14} \, $. Note the vanishing of all torsion classes indeed corresponds to a torsion-free G$_2$-structure
\begin{equation}
    \dd\varphi=0 \, , \qquad \dd\psi=0 \, .
\end{equation}
In \cref{chap:heterotic} we will be interested in G$_2$-structures such that $\tau_2=0 \, $, which we call \emph{integrable} following Fern\'andez--Ugarte \cite{Fernandez:1998}. Given an integrable G$_2$-structure there exists a unique metric connection compatible with the G$_2$-structure with totally antisymmetric torsion \cite{Friedrich:2001nh} given by the three-form
\begin{equation}
\label{eq:torsiong2}
T(\varphi)=\frac{1}{6}\,\tau_0 \, \varphi-\tau_1\lrcorner \, \psi-\tau_3 \, .
\end{equation}
We call this the \emph{torsion} of the G$_2$-structure.

\section{Spin(7)-structures}
\label{sec:Spin7Structures}

Let $Z$ be an 8-dimensional manifold and let $\lbrace e^1,\dots,e^8\rbrace$ denote an orthonormal coframe on $Z$. We say a four-form $\Psi$ on $Z$ is \emph{admissible} if it can be locally written as
\begin{align}
\label{eq:spin7fourformintro}
    \Psi=\; &e^{1234} +e^{1256} +e^{1278} +e^{1357} -e^{1368} -e^{1458} -e^{1467} \nonumber \\
    & -e^{2358} -e^{2367} -e^{2457} +e^{2468} +e^{3456} +e^{3478} +e^{5678} \, .
\end{align}
A Spin(7)\emph{-structure}\footnote{For more detailed accounts on Spin(7)-structures, see \cite{Ivanov:2001ma, Karigiannis:2005}.} on $Z$ is equivalent to the existence of a (nowhere-vanishing) admissible four-form $\Psi$ on $Z$. We call this the \emph{Cayley} four-form.\footnote{The subgroup of $\text{SO}(8)$ preserving the Cayley four-form is precisely Spin(7).}

The Cayley four-form is the only Spin(7)-form. It determines a metric and an orientation on $Z$ and we note the Cayley four-form is \emph{self-dual} with respect to them, $*\Psi=\Psi$. The decomposition of differential forms into irreducible Spin(7) representations is given by
\begin{align}
\Lambda^0&=\Lambda^0_1 \, ,\qquad
\Lambda^1=\Lambda^1_8 \, ,\qquad
\Lambda^2=\Lambda^2_7\oplus\Lambda^2_{21} \, ,\qquad
\Lambda^3=\Lambda^3_8\oplus\Lambda^3_{48} \, , \nonumber\\
\label{eq:splittingofspin7forms}
\Lambda^4&=\Lambda^4_1\oplus\Lambda^4_7\oplus\Lambda^4_{27}\oplus\Lambda^4_{35} \, ,
\end{align}
where $\Lambda^k$ denotes the space of $k$-forms on $Z$ and $\Lambda^k_p$ denotes the subspace of $\Lambda^k$ consisting of $k$-forms transforming in the $p$-dimensional irreducible representation of Spin(7). The decomposition for higher degrees follows from Hodge duality. Each $\Lambda^k_p$ can be described in terms of the Cayley four-form, for example $\Lambda^2_{21} \, $---which corresponds to the two-forms contained in the Lie algebra of Spin(7)---is given by
\begin{equation}
\label{eq:splitofspin7liealgebra}
\Lambda^2_{21}=\lbrace\beta\in\Lambda^2 : *\left(\Psi\wedge\beta\right)=-\beta \rbrace \, .
\end{equation}

The decomposition in Spin(7) representations of the exterior derivative of the Cayley four-form is \cite{Fernandez:1986}
\begin{equation}
\dd\Psi=\sigma_1\wedge\Psi+*\sigma_3
\end{equation}
Here the \emph{torsion classes} $\sigma_k$ are $k$-forms with $\sigma_3\in\Lambda^3_{48}\, $. Note the vanishing of all torsion classes indeed corresponds to a torsion-free Spin(7)-structure
\begin{equation}
    \dd\Psi=0 \, .
\end{equation}
Given a Spin(7)-structure there exists a unique metric connection compatible with the Spin(7)-structure with totally antisymmetric torsion \cite{Ivanov:2001ma} given by the three-form
\begin{equation}
\label{eq:torsionspin7}
T(\varphi)=-\,\frac{1}{6}\,\sigma_1\lrcorner \, \Psi-\sigma_3 \, .
\end{equation}
We call this the \emph{torsion} of the Spin(7)-structure.

\section{SU($n$)-structures}
\label{sec:SUnstructures}

Let $X$ be a $2n$-dimensional manifold and let $\lbrace e^1,\dots,e^{2n}\rbrace$ denote an orthonormal coframe on $X$. An SU($n$)\emph{-structure}\footnote{For more detailed accounts on SU($n$)-structures, see \cite{LopesCardoso:2002vpf, Cabrera:2005, Becker:2015wga, Prins:2016zhg, Larfors:2018nce}.} on $X$ is equivalent to the existence of a nowhere-vanishing, real two-form $\omega$ and a nowhere-vanishing, complex $n$-form $\Omega$ satisfying the following relations
\begin{equation}
\label{eq:relationsSUNstructures}
\omega\wedge\Omega=0, \qquad \frac{1}{n!}\,\omega^n=\frac{(-1)^{n(n-1)/2}\, i^n}{2^n}\,\Omega\wedge\bar{\Omega}=\dd\text{vol}_{X}.
\end{equation}
We call $\omega$ the \emph{Hermitian} form and $\Omega$ the \emph{holomorphic volume form}.\footnote{We will sometimes add a subindex $\omega_n$, $\Omega_n$ to indicate the dimension of $X$ and the dimension of the SU($n$)-structure.} These SU($n$)-forms are globally well-defined on the manifold and have particular local expressions. In this thesis we encounter SU(2), SU(3) and SU(4) structures, we include the local expressions of the forms for completeness. For SU(2) we have
\begin{equation}
\label{eq:su2localforms}
\omega_2= e^{12}+e^{34}, \qquad
\Omega_2= e^{13}- e^{24} + i\left( e^{14}+ e^{23} \right).
\end{equation}
For SU(3) we find\footnote{The local expression of the SU($n$)-forms can be obtained from the local expression of SU($n-1$)-forms as follows
\begin{equation}
\label{eq:constructsunfromsun-1}
\omega_n=\omega_{n-1}+e^{2n-1}\wedge e^{2n}, \qquad \Omega_n=(e^{2n}-i\, e^{2n-1})\wedge\Omega_{n-1}.
\end{equation}
It can be checked these forms do indeed satisfy \eqref{eq:relationsSUNstructures} and represent locally an SU($n$)-structure.}
\begin{align}
\label{eq:su3localomega}
\omega_3 &= e^{12}+e^{34}+e^{56},\\
\Omega_3 &= e^{136}+ e^{145} +e^{235}- e^{246}+ i\left( -e^{135}+ e^{146} +e^{236}- e^{245} \right).
\end{align}
Finally, for SU(4) we obtain
\begin{align}
\label{eq:su4localomega}
\omega_4 &= e^{12}+e^{34}+e^{56}+e^{78},\\
\Omega_4 &= e^{1357}- e^{1368} -e^{1458}- e^{1467}- e^{2358}- e^{2367} +e^{2457}+ e^{2468} \nonumber \\
&+ i\big( e^{1358}+ e^{1367} +e^{1457}- e^{1468} + e^{2357}- e^{2368} +e^{2458}- e^{2467} \big).
\end{align}
The holomorphic volume form determines an almost complex structure $J$ on $X$. For example, in the SU(3) case the explicit expression is given by \cite{Hitchin:2000jd}
\begin{equation}
J\indices{^i_j}=\frac{I\indices{^i_j}}{\sqrt{-\frac{1}{6}\tr(I^2)}}\, , \qquad I\indices{^i_j}=\Re(\Omega)_{jkl}\Re(\Omega)_{mnr}\varepsilon^{iklmnr}\, ,
\end{equation}
where $\varepsilon$ is the Levi-Civita tensor. The form $\Omega$ is of type $(n,0)$ with respect to $J$, whereas $\omega$ is of type $(1,1)$. The Hermitian form together with the almost complex structure define a metric on $X$ by the formula $g(v,w)=\omega(v,J(w))$ where $v,w$ are vector fields on $X$.

Differential forms decompose into irreducible SU($n$) representations. As expected, the decomposition depends on the value of $n$. We present below the SU(3) decomposition for illustration purposes
\begin{equation}
\Lambda^0=\Lambda^0_1 \, ,\qquad
\Lambda^1=\Lambda^1_{3\,\oplus\,\bar{3}} \, ,\qquad
\Lambda^2=\Lambda^2_1\oplus\Lambda^2_{3\,\oplus\,\bar{3}}\oplus\Lambda^2_8 \, ,\qquad
\Lambda^3=\Lambda^3_{1\,\oplus\, 1}\oplus\Lambda^3_{3\,\oplus\,\bar{3}}\oplus\Lambda^3_{6\,\oplus\,\bar{6}} \, .
\end{equation}
Here $\Lambda^k$ denotes the space of $k$-forms on $X$ and $\Lambda^k_p$ denotes the subspace of $\Lambda^k$ consisting of $k$-forms transforming in the $p$-dimensional irreducible representation of SU(3). The decomposition for higher degrees follows from Hodge duality. Each $\Lambda^k_p$ can be described in terms of the SU(3)-forms, for example $\Lambda^2_{8} \, $---which corresponds to the two-forms contained in the Lie algebra of SU(3)---is given by
\begin{equation}
\Lambda^2_{8}=\lbrace\beta\in\Lambda^{1,1} : \omega\lrcorner \, \beta=0 \rbrace \, 
\end{equation}
where $\Lambda^{p,q}$ denotes the space of $(p,q)$-forms under the almost complex structure $J$.

The exterior derivatives of the SU($n$)-forms decompose into representations as well. In the SU(3) case these are given by \cite{Chiossi:2002, Larfors:2018nce}
\begin{align}
\begin{split}
\label{eq:torsionclassSUn}
\dd\omega &=\frac{3\, i}{4}\left(\bar{W}_1\,\Omega+W_1\,\bar{\Omega}\right)+W_4\wedge\omega+W_3\, , \\
\dd\Omega &=W_1\,\omega\wedge\omega+W_2\wedge\omega+W_5\wedge\Omega\, .
\end{split}
\end{align}
Here the \emph{torsion classes} $W_k$ are as follows: $W_1$ is a complex function, $W_2$ is a complex two-form in the $8\oplus 8$ representation, $W_3$ is a real three-form such that $W_3\in \Lambda^3_{6\,\oplus\,\bar{6}}\,$, $W_4$ is a real one-form, and $W_5$ is a real one-form as well. For SU($n$) structures with $n>3$ the decomposition in torsion classes is analogous \cite{Gray:1980, Cabrera:2005} although the degree and representations of some torsion classes change. For $n=2$ the torsion classes $W_1$, $W_3$ and $W_5$ vanish automatically \cite{Gray:1980, Cabrera:2005}. For any $n$, the vanishing of all torsion classes corresponds to a torsion-free SU($n$)-structure\footnote{If the holonomy of $X$ is precisely SU($n$), we say that $X$ is a \emph{Calabi--Yau} manifold.}
\begin{equation}
    \dd\omega=0 \, , \qquad \dd\Omega=0\, .
\end{equation}
A manifold with an SU(3)-structure admits a unique metric connection compatible with that SU(3)-structure and with totally antisymmetric torsion if and only if the Nijenhuis tensor $N_J$ is totally antisymmetric \cite{Friedrich:2001nh}. This condition is equivalent to the vanishing of the torsion class $W_2$ \cite{Gray:1980}, and in this case the torsion of the connection is given by
\begin{equation}
T(\omega,\Omega)=-N_J-J(\dd\omega)\, .
\end{equation}
We call this the \emph{torsion} of the SU(3)-structure.

\section{Hermitian and hyper-Hermitian structures}
\label{sec:HyperkahlerStructures}

Let $X$ be a $2n$-dimensional manifold with an almost complex structure $J$. We say a two-form $\omega$ on $X$ is \emph{positive} if $\omega(v,J(v))>0$ for all nonzero vector fields $v$ on $X$. We say a two-form $\omega$ on $X$ is \emph{compatible} with $J$ if $\omega(J(v),J(w))=\omega(v,w)$ for all vector fields $v,w$ on $X$.

A U($n$)-\emph{structure}\footnote{More details can be found in \cite{Gray:1980, Falcitelli:1994, Alexandrov:2004cp, Schoemann:2007}.} on $X$, also known as a \emph{Hermitian} structure, is equivalent to the existence of a nowhere-vanishing, real, positive two-form $\omega$ which is compatible with $J$ and has type $(1,1)$. This form is called the \emph{Hermitian form} and it defines a metric on $X$ by $g(v,w)=\omega(v,J(w))$ where $v,w$ are vector fields on $X$.


In this thesis, all manifolds we encounter which are equipped with a U($n$)-structure will actually present an SU($n$)-structure. Hence, we will not discuss these structures in detail and we just point out the main similarities and differences between the U($n$) and SU($n$) cases. First of all, the U($n$) Hermitian form has the same local expression as in \eqref{eq:su2localforms}, \eqref{eq:su3localomega} or \eqref{eq:su4localomega}.

As for the torsion classes, they are analogous to the SU($n$) case we discussed in \cref{sec:SUnstructures} with the exception that the class $W_5$ is absent \cite{Gray:1980, Chiossi:2002, Cabrera:2005}. The exterior derivative of the Hermitian form is essentially as in \eqref{eq:torsionclassSUn}. Note the torsion class $W_2$ does not appear in this derivative, this means that $\dd\omega$ does not encode all torsion classes of the U($n$)-structure---one needs to compute the Nijenhuis tensor to obtain $W_2$ \cite{Cabrera:2005}. Similarly to the SU($n$) case, when $W_2$ vanishes the U($n$)-structure admits a compatible connection with totally antisymmetric torsion \cite{Gray:1980, Schoemann:2007}, the \emph{Bismut} connection \cite{Bismut:1989}. A manifold with U($n$)-holonomy is a \emph{K\"ahler} manifold.

\smallskip

Let $X$ be a 4$n$-dimensional manifold. An \emph{almost hyper-complex structure} on $X$ is given by a triple of almost complex structures\footnote{These almost complex structures can be combined to give rise to a two-dimensional sphere worth of almost complex structures.} $J^1$, $J^2$, $J^3$ satisfying the quaternionic identities
\begin{equation}
\label{eq:hypercomplexrelations}
(J^1)^2=(J^2)^2=(J^3)^2=-\text{Id}\,, \qquad J^1 J^2=-J^2 J^1=J^3.
\end{equation}
An Sp($n$)-\emph{structure}\footnote{For more detailed accounts on Sp($n$)-structures, see \cite{Boyer:1988, Howe:1996kj, Grantcharov:1999kv, Barberis:2008}. However, note that many of these references only discuss some particular types of Sp($n$)-structures.} on $X$, also known as a \emph{hyper-Hermitian} structure, is equivalent to a triple of Hermitian structures---that is, a triple of Hermitian forms $\omega^1$, $\omega^2$, $\omega^3\,$---each of them associated to one of the almost complex structures, and giving rise to the same Riemannian metric
\begin{equation}
g(v,w)=\omega^1(v,J^1(w))=\omega^2(v,J^2(w))=\omega^3(v,J^3(w))\, ,
\end{equation}
where $v,w$ are vector fields on $X$. For $n=1$, let $\lbrace e^1,\dots,e^{4}\rbrace$ denote an orthonormal coframe on $X$. The Hermitian forms can then be locally written as
\begin{equation}
\label{eq:sp1localforms}
\omega^1= e^{12}+e^{34}, \qquad
\omega^2= e^{13}- e^{24}, \qquad \omega^3=e^{14}+ e^{23}.
\end{equation}
The local expression of the forms for $n>1$ can be obtained generalizing \eqref{eq:sp1localforms}. Nevertheless, the case $n=1$ will be the only one relevant for this thesis.

The torsion classes of a hyper-Hermitian manifold can be understood in terms of the Hermitian forms. First note that $\text{Sp}(n)\subset\text{SU}(2n)$ so a hyper-Hermitian manifold actually has a triple of SU($2n$)-structures.\footnote{Analogously to the case of almost complex structures, there is a two-dimensional sphere worth of SU($2n$)-structures.} Indeed, we can define a holomorphic volume form as \cite{Cabrera:2005}
\begin{equation}
\label{eq:hyperkahlerholvolform}
\Omega^1=\frac{(-1)^{n(n-1)/2}}{n!}\left( \omega^2+i\,\omega^3 \right)^n \, ,
\end{equation}
and $\left(\omega^1,\Omega^1\right)$ defines an SU($2n$)-structure on $X$. The remaining two structures are obtained analogously. Since the SU($2n$)-forms are defined by the Hermitian forms, it is enough to compute the exterior derivatives of the Hermitian forms to obtain the SU($2n$) torsion classes \eqref{eq:torsionclassSUn}. When all the torsion classes vanish, the Sp($n$)-structure is torsion-free.\footnote{If the holonomy of $X$ is precisely Sp($n$), we say that $X$ is a \emph{hyper-K\"ahler} manifold \cite{Cabrera:2005}.}

The existence of a connection compatible with the Sp($n$)-structure---also known as a Hermitian connection---with totally antisymmetric torsion also depends directly on the Hermitian forms. The connection exists if and only if the following condition is satisfied \cite{Gauduchon:1997, Grantcharov:1999kv}
\begin{equation}
J^1\,\dd\omega^1=J^2\,\dd\omega^2=J^3\,\dd\omega^3 \, .
\end{equation}
In this case, the torsion of the connection is given by $T=J^1\,\dd\omega^1$. These manifolds are known as \emph{hyper-K\"ahler with torsion} or simply \emph{HKT} \cite{Howe:1996kj}.

The case $n=1$ is especially relevant since we have the accidental isomorphism $\text{Sp}(1)\cong\text{SU}(2)$. This means that an Sp(1)-structure on a 4-dimensional manifold is equivalent to an SU(2)-structure. We will consider by convention the following SU(2)-structure constructed from the Sp(1)-structure
\begin{equation}
\label{eq:equivalencesp1su2}
\omega=\omega^1\, , \qquad \Omega=\omega^2+i\,\omega^3\, .
\end{equation}
The hyper-complex relations relations \eqref{eq:hypercomplexrelations} imply these useful identities
\begin{equation}
\label{eq:hyperkahlerrelations}
(\omega^1)^2=(\omega^2)^2=(\omega^3)^2\neq 0 \, , \qquad \omega^1\wedge\omega^2=\omega^2\wedge\omega^3=\omega^3\wedge\omega^1=0 \, .
\end{equation}
The Sp(1)-structure only has two torsion classes---$W_2$ and $W_4$ in the SU(2) language. A manifold with Sp(1)-holonomy is a \emph{K3 surface}.

\section{Relations between $G$-structures}
\label{sec:RelationsStructures}

After introducing the different $G$-structures that will appear in this thesis, we finish the chapter with a collection of relations between them.

Let $M$ be an $n$-dimensional manifold with a $G$-structure and let $K$ be a subgroup of SO($n$) such that $G\subset K\subset \text{SO}(n)$. Since the frame bundle of $M$ admits a reduction to a $G$-bundle it must also admit a reduction to a $K$-bundle. Thus, $M$ has a $K$-structure and we should be able to describe it in terms of the $G$-structure. We will do so---locally---following the language of differential forms.

We have already presented an example of this in the previous section: a $4n$-dimensional manifold $X$ with an Sp($n$)-structure always has an SU($2n$)-structure. The holomorphic volume form is constructed using the Hermitian forms of the hyper-Hermitian structure \eqref{eq:hyperkahlerholvolform}.

Another simple example is the case of manifolds with an SU($n$)-structure, which automatically come equipped with a U($n$)-structure. The Hermitian form of the SU($n$)-structure defines the corresponding U($n$)-structure.

A case that will be relevant for us is provided by the inclusion $\text{SU}(4)\subset\text{Spin}(7)\,$. This means an 8-dimensional manifold with an SU($4$)-structure can be given a Spin(7)-structure as follows
\begin{equation}
\label{eq:inclusionsu4spin7}
\Psi=\Re(\Omega)+\frac{1}{2}\,\omega\wedge\omega\, .
\end{equation}
Following \cite{Joyce:2007}, an important class of examples are obtained by considering the direct product of a certain $G$-structure manifold with a number of copies of the real line $\R$ and/or the circle $\Sc^1$. The product manifold inherits the $G$-structure and it might be possible to combine the $G$-forms with the coordinate one-forms of the $\R$ and $\Sc^1$ factors to construct a new structure on the product manifold.

As a first example, consider a G${}_2$-structure manifold $Y$ times a circle $\Sc^1$---or a line $\R$---and denote this additional coordinate by $\theta$. We have the chain of inclusions $\text{G}_2\subset\text{Spin}(7)\subset\text{SO}(8)$ and as a result the G$_2$-forms and $\dd \theta$ can be combined to define a Spin(7)-structure
\begin{equation}
\label{eq:inclusiong2spin7}
\Psi=\dd\theta\wedge\varphi+\psi\,.
\end{equation}
Consider now the case of an SU($n-1$)-structure manifold times one of the following: $\R^2$, $\R\times\Sc^1$ or $\Sc^1\times\Sc^1$, and denote these additional coordinates by $(t,\theta)$. Since we have the chain of inclusions $\text{SU}(n-1)\subset\text{SU}(n)\subset\text{SO}(2n)$ we can produce an SU($n$)-structure in the product manifold in the same fashion as \eqref{eq:constructsunfromsun-1}
\begin{equation}
\label{eq:inclusionsun-1sun}
\omega_n=\omega_{n-1}+\dd t\wedge\dd \theta\, , \qquad \Omega_n=(\dd \theta-i\, \dd t)\wedge\Omega_{n-1}\, .
\end{equation}
Finally, one can analyze the product of an SU(3)-structure manifold $X$ with a line $\R$ or a circle $\Sc^1$. We denote this additional coordinate by $t$. We have $\text{SU}(3)\subset\text{G}_2\subset\text{SO}(7)$ and we can construct a G${}_2$-structure in the product manifold by
\begin{equation}
\label{eq:inclusionsu3g2}
\varphi=\Re(\Omega)-\dd t\wedge\omega\, , \qquad \psi=\frac{1}{2}\,\omega\wedge\omega+\dd t\wedge\Im(\Omega)\, .
\end{equation}
It is straightforward to check that the $G$-forms given by the local formulas \eqref{eq:hyperkahlerholvolform}, \eqref{eq:inclusionsu4spin7}, \eqref{eq:inclusiong2spin7}, \eqref{eq:inclusionsun-1sun} and \eqref{eq:inclusionsu3g2} satisfy the requirements to define the corresponding $G$-structures.

\chapter{Superconformal algebras for generalizations of connected sums}
\label{chap:scas}

\section{Introduction}

This chapter focuses on worldsheet aspects of string compactifications on manifolds constructed via generalizations of the Twisted Connected Sum (TCS) approach \cite{Kovalev:2003, Corti:2012, Corti:2012kd}. The relevant compactification ansatz is slightly less general than the one we presented in \eqref{eq:compactificationansatz}: we consider type II backgrounds of the form
\begin{equation}
\mathbb{M}_d \times M \, ,
\end{equation}
where $\mathbb{M}_d$ denotes $d$-dimensional Minkowski space and $M$ is a $\left(10-d\right)$-dimensional compact manifold.\footnote{We will write $M_{10-d}$ when we want to stress the dimension of the compact manifold.} We will only consider manifolds with special holonomy so the intrinsic torsion of all the $G$-structures we describe in this chapter will be set to zero.

Worldsheet considerations of such backgrounds were initiated with the work of Shatashvili and Vafa \cite{Shatashvili:1994zw}, see also \cite{Figueroa-OFarrill:1990tqt, Figueroa-OFarrill:1990mzn, Blumenhagen:1991nm, Figueroa-OFarrill:1996tnk}, who identified the chiral symmetry W-algebras characteristic of $M$ having holonomy either G$_2$ or Spin(7) but being otherwise generic. These algebras are two specific extensions of the chiral $\mathcal{N}=1$ superconformal symmetry of the sigma model with target space $M$, {and we shall denote them} by $\SVseven$ and $\SVeight$. Subsequent research focused on various aspects of theories with these symmetries \cite{Partouche:2000uq, Gepner:2001px, Eguchi:2001xa, Sugiyama:2001qh, Noyvert:2002mc, Sugiyama:2002ag, Eguchi:2003yy, deBoer:2005pt, deBoer:2006bp, Sriharsha:2006zc, Benjamin:2014kna, Cheng:2015fha}, particularly orbifolds of free CFTs by a finite group \cite{Shatashvili:1994zw, Acharya:1996fx, Acharya:1997rh, Gaberdiel:2004vx, Chuang:2004th}, and also realizing $M_7=\big(\Sc^1\times (\text{Gepner model})\big)/\mathbb{Z}_2$ \cite{Roiban:2001cp, Eguchi:2001ip, Blumenhagen:2001jb, Roiban:2002iv} as well as $M_8=\big(\text{Gepner model}\big)/\mathbb{Z}_2$ \cite{Blumenhagen:2001qx}. Both of these reflect geometric constructions pioneered by Joyce \cite{Joyce:1996a, Joyce:1996b, MR1383960, Joyce:2007}.

Important steps have been taken in recent years to understand string theory implications of compactifications on TCS manifolds and their generalizations.\footnote{A non-exhaustive sample of these interesting works is  \cite{Halverson:2014tya, Halverson:2015vta, Braun:2016igl, Braun:2017ryx, Guio:2017zfn, Braun:2017uku, Braun:2017csz, Braun:2018fdp, Braun:2018joh, Acharya:2018nbo, Braun:2018vhk, Braun:2019lnn, Barbosa:2019bgh, Braun:2019wnj, Cvetic:2020piw, Xu:2020nlh, Hubner:2020yde}.} It is natural to then also consider the CFT emerging when $M$ is a connected sum; this was initiated in \cite{Braun:2017ryx, Braun:2017csz, Fiset:2018huv, Braun:2019lnn}. A particularly intriguing application of these ventures was to test the G$_2$ and Spin(7) analogues of mirror symmetry conjectured in \cite{Shatashvili:1994zw, Papadopoulos:1995da}.

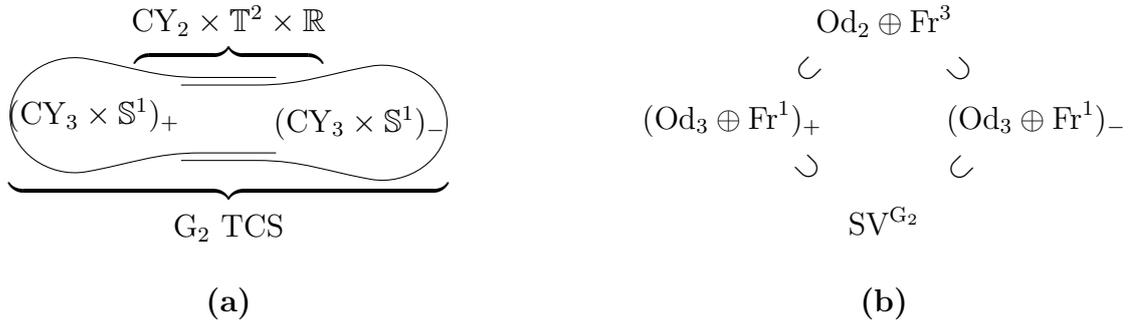
\begin{figure}[h]
\begin{center}
\begin{tikzpicture}
\draw [black]
(2,0.5) to (1,0.5)
to[out=180, in=-10] (-0.5,0.75)
to[out=170, in=90] (-1.5,0)
to[out=-90, in=190] (-0.5,-0.75)
to[out=10, in=180] (1,-0.5) to (2,-0.5);
\draw [black]
(0.75,0.4) to (1.75,0.4)
to[out=0, in=190] (3.25,0.65)
to[out=10, in=90] (4.25,-0.1)
to[out=-90, in=-10] (3.25,-0.85)
to[out=170, in=0] (1.75,-0.6) to (0.75,-0.6);
\node at (1.375,0.8) {$\overbrace{\qquad\qquad\qquad}$};
\node at (1.375,-1.0) {$\underbrace{\qquad\qquad\qquad\qquad\qquad\qquad\qquad}$};
\node at (1.375,-2.5) {\textbf{(a)}};
\node at (10,-2.5) {\textbf{(b)}};
\node at (9,0.6) {\rotatebox{45}{$\subset$}};
\node at (11,0.6) {\rotatebox{135}{$\subset$}};
\node at (9,-0.7) {\rotatebox{135}{$\subset$}};
\node at (11,-0.7) {\rotatebox{45}{$\subset$}};

\node at (1.375,1.25) {$\text{CY}_2\times \mathbb{T}^2\times \mathbb{R}$};
\node at (-0.4,0) {$(\text{CY}_3 \times \Sc^1)_+$};
\node at (3.1,-0.1) {$(\text{CY}_3 \times \Sc^1)_-$};
\node at (1.375,-1.5) {G$_2$ TCS};

\node at (10,1.25) {$\text{Od}_2\oplus \text{Fr}^3$};
\node at (8,-0.05) {$(\text{Od}_3\oplus \text{Fr}^1)_+$};
\node at (12,-0.05) {$(\text{Od}_3\oplus \text{Fr}^1)_-$};
\node at (10,-1.4) {$\SVseven$};
\end{tikzpicture}
\caption{\textbf{(a)} Sketch of a compact 7-dimensional G$_2$-holonomy manifold obtained as Twisted Connected Sum (TCS). \textbf{(b)} Diamond of algebra inclusions corresponding to a sigma model whose target space is a TCS, to be described in \cref{sec:Idea}.}
\label{fig:TCS}
\end{center}
\end{figure}

In this chapter we extend upon \cite{Fiset:2018huv}, where general symmetry aspects of the sigma model in a generic TCS  manifold were first investigated. Very superficially a G$_2$ TCS has the structure sketched in \cref{fig:TCS}(a): two open manifolds $M_\pm = (\text{CY}_3 \times \Sc^1)_\pm$ glued together by an appropriate isomorphism along a ``neck'' region where they have asymptotically the form of a cylinder with cross-section $\text{CY}_2\times \mathbb{T}^2$. (By $\text{CY}_n$ we mean Calabi--Yau manifold of complex dimension $n$).

These local geometric models were argued to translate to specific W-algebras on the worldsheet, in which $\SVseven$ was found to sit as a subalgebra. The transition map between the two halves of the geometry was said to correspond to automorphisms of the various chiral algebras. While this is essentially correct we find it useful to revisit that interpretation in \cref{sec:Idea}, stressing the uniqueness of the $\SVseven$ subalgebra arising in these considerations. We explain that the worldsheet theory displays, at least in the appropriate regime, a network of algebra inclusions in the shape of a diamond, see \cref{fig:TCS}(b). The upper tip of the diamond is the symmetry of the sigma model excitations localized on the neck of the TCS, whereas the left and right lateral tips reflect the symmetry of the two open subsets $(\text{CY}_3\times\Sc^1)_\pm$. In the intersection of the latter two subalgebras, a unique $\SVseven$ sits, reflecting symmetries enjoyed by the sigma model whose target space is the whole compact $M_7$.

We also generalize this set-up to the case where $M_7$ is an Extra Twisted Connected Sum (ETCS) \cite{Crowley:2015ctv, Nordstrom:2018cli, Goette:2020}. This is a worthy addition given that certain topological types of G$_2$-manifolds are known to admit the ETCS construction, but not the TCS construction \cite{Nordstrom:2018cli}. We briefly describe the ETCS construction in \cref{sec:G2}, concentrating on differences with the ordinary TCS case which are perceptible in CFT language. We then find that the diamond of algebra inclusions is unaffected by these changes, thus asserting its general validity.

Our other focus is on Spin(7)-holonomy manifolds $M_8$. In that case a conjectural construction called ``Generalized Connected Sum'' (GCS) was proposed by Braun and Sch\"afer-Nameki, mimicking the TCS construction \cite{Braun:2018joh}. We review it in \cref{sec:Spin7Geom}; it is essentially like in \cref{fig:TCS}(a), except that the local models are different, see \cref{fig:GCS}(a). While various arguments were provided in \cite{Braun:2018joh}, no mathematical proof currently shows that this systematically yields Spin(7)-holonomy manifolds. We provide here what can be regarded as proof from worldsheet string theory. We describe the diamond of chiral algebra inclusions it yields in \cref{sec:Spin7Alg}. Similarly to the G$_2$ case we find that a unique $\SVeight$ features in an appropriate intersection of subalgebras, at the bottom of the diamond. In the rest of \cref{sec:Spin7} we examine various algebra automorphisms and interpret them with a view towards Spin(7) mirror symmetry.
Studying four examples of Joyce orbifolds which admit a GCS description leads us to propose three new mirror maps, as well as recovering some aspects of \cite{Braun:2019lnn} where similar questions were tackled.

Finally in \cref{sec:Num} we study numerically whether the Shatashvili-Vafa algebra at the bottom tip of all the diamonds we have mentioned so far, which sits in the intersection between the algebras on the lateral tips, is in fact equal to that intersection. This indeed seems to be the case, at least up to level 5 in the vacuum module in the TCS case; and up to level 6 in the GCS case. If true, this would mean that TCS G$_2$-manifolds are representative of generic G$_2$-manifolds, and that GCS Spin(7)-manifolds are representative of generic Spin(7)-manifolds, at least as far as chiral symmetries of strings into them are concerned.

\subsection{The general idea} \label{sec:Idea}

In this section we set up notations and the dictionary between target space and worldsheet symmetries which will be useful throughout the rest of the chapter.

The general philosophy is in fact not specific to two dimensions; consider a quantum field theory whose fields are valued, by definition, in configuration space $M$. Then the ``simpler'' $M$ is, the more symmetric the field theory will be. Various invariances can be counted in the set of ``symmetries'', but for us these are the chiral algebra, say on the holomorphic side, of an $\mathcal{N}=(1,1)$ superconformal sigma model with Riemannian target space $M$. Let us denote them by
\begin{equation}
\mathcal{A}(M) \, .
\end{equation}
Similarly ``simplicity'' can take various meanings:
for us it refers to a reduction of the $G$-structure of $M$ from $G=\text{O}(d)$ to a subgroup.\footnote{O($d$) reflects the Riemannian metric $g$ on $M$.} Equivalently,\footnote{See for instance \cite{Joyce:2007}.} there is a tangent bundle connection on $M$ with holonomy strictly contained in $\text{O}(d)$. As we explained in \cref{chap:introduction}, covariantly constant differential $p$-forms under that connection then lead to worldsheet symmetries \cite{Howe:1991ic}.\footnote{This principle was recently shown \cite{delaOssa:2018azc} to hold even in the most general $\mathcal{N}=(1,0)$ non-linear sigma model, and at one loop in $\sqrt{\alpha'}$ perturbation theory.} 
The worldsheet currents come in supersymmetric pairs, and have spin $\frac{p}{2}$ and $\frac{p+1}{2}$. The basic holomorphic $\mathcal{N}=1$ superconformal symmetry then gets replaced by some other $\mathcal{N}=1$ superconformal W-algebra.

The simplest example is when $M = \mathbb{R}\times \mathcal{X}$ is a cylinder over an arbitrary manifold $\mathcal{X}$.
The one-form $\dd t$ along $\mathbb{R}$ is constant and thus leads to worldsheet currents
\begin{equation}
\psi_t \qquad \text{and}\qquad
j_t = i\partial_t \, ,
\end{equation}
which are a free Majorana--Weyl fermion and a $\widehat{\mathfrak{u}}(1)$ current. Together they generate the chiral algebra that we shall denote $\text{Fr}^1$; in general
\begin{equation}
\text{Fr}^n = \underbrace{\big(\text{free fermion} \oplus \widehat{\mathfrak{u}}(1)\big)\oplus\ldots}_{n \text{ times}} \, .
\end{equation}
The superconformal symmetry is given by the standard expressions\footnote{We reuse these notations throughout the chapter. If $t$ parametrizes $\mathbb{R}$ or $\Sc^1$ in target space, $G_t$ is the chiral supersymmetry current {and} $T_t$ is the holomorphic part of the stress-tensor. Colons represent normal ordering and $\partial$ is the derivative with respect to the holomorphic worldsheet coordinate.}
\begin{equation} \label{eq:virasorostandard}
G_t = \normord{j_t\psi_t} \, , \qquad
T_t = \frac{1}{2}\normord{\Big(\partial\psi_t\psi_t + j_t j_t\Big)} .
\end{equation}

More generally this correspondence yields a uniform and satisfying interpretation of the chiral algebras we will need in this chapter. They are listed in \cref{tab:dictionary}, where we also provide the relevant holonomy groups, covariantly constant tensors, and our notations for the corresponding worldsheet currents. For instance while Calabi--Yau $n$-folds are widely known to lead to $\mathcal{N}=2$ superconformal symmetry on the worldsheet, their true W-algebra is in fact larger. There is one algebra for each $n\in\mathbb{N}$ and we denote them by $\text{Od}_n$ after Odake \cite{Odake:1988bh}. The $\mathcal{N}=2$ generators are related to the Hermitian form associated to $G=\text{U}(n)$; the extra generators are due to the holomorphic volume form which is specific to $G=\text{SU}(n)$. 

\bigskip

\begin{table}[h]
{\footnotesize
\renewcommand\arraystretch{1.3}
\centering
\begin{tabular}{|c|c|c|l|l|c|}
\hline
Dim.\ &
$G$ &
Target space &
Cov.\ const.\ tensors &
Generators of $\mathcal{A}$ {(with weights)} &
Algebra $\mathcal{A}$\\
 &
 &
 &
 &
\& SUSY partners {(with weights)} &
\& c.\ charge
\\\hline\hline
$1$ &
$\mathds{1}$ &
$\mathbb{R}$ or S$^1$ &
$\dd t$ &
$\psi_t$~($\tfrac{1}{2}$) &
$\text{Fr}^1$\\
 &
 &
 &
 &
$j_t=i\partial t$~($1$) &
$c=3/2$
\\\hline
$d$ &
O($d$) &
Riemannian &
$g$ (metric) &
$G$~($\tfrac{3}{2}$) &
$\mathcal{N}=1$\\
 &
 &
 &
 &
$T$~($2$) &
$c=3d/2$
\\\hline
$2n$ &
U($n$) &
K\"ahler &
$g$, $\omega$ &
$G_n$~($\tfrac{3}{2}$), $J^3_n$~($1$) &
$\mathcal{N}=2$\\
 &
 &
 &
(Hermitian form)
 &
$T_n$~($2$), $G^3_n$~($\tfrac{3}{2}$) &
$c=3n$
\\\hline
$2n$ &
SU($n$) &
Calabi--Yau &
$g$, $\omega$, $\Omega$ &
$G_n$~($\tfrac{3}{2}$), $J^3_n$~($1$), $A_n+iB_n$~($\tfrac{n}{2}$) &
$\text{Od}_n$\\
 &
 &
 &
(holom.\ vol. form) &
$T_n$~($2$), $G^3_n$~($\tfrac{3}{2}$), $C_n+iD_n$~($\tfrac{n+1}{2}$) &
$c=3n$
\\\hline
$7$ &
G$_2$ &
G$_2$ &
$g$, $\varphi$, $\psi$ &
$G_7$~($\tfrac{3}{2}$), $P$~($\tfrac{3}{2}$), $X_7$~($2$) &
$\SVseven$\\
 &
 &
holonomy &
(3-form and 4-form) &
$T_7$~($2$), $K$~($2$), $M_7$~($\tfrac{5}{2}$) &
$c=21/2$
\\\hline
$8$ &
Spin(7) &
Spin(7) &
$g$, $\Psi$ (4-form) &
$G_8$~($\tfrac{3}{2}$), $X_8$~($2$) &
$\SVeight$\\
 &
 &
holonomy &
 &
$T_8$~($2$), $M_8$~($\tfrac{5}{2}$) &
$c=12$
\\\hline
\end{tabular}
\caption{Notations and correspondences between covariantly constant tensors on the sigma model target space with holonomy G and (supersymmetric pairs of) generators of the worldsheet chiral symmetry W-algebra $\mathcal{A}$. The conformal weight of the generators is indicated in parentheses.
}
\label{tab:dictionary}
}
\end{table}

We will only need Od$_n$ for $n=4$, $n=3$ and $n=2$; the latter being actually isomorphic to the small $\mathcal{N}=4$ superconformal algebra at $c=6$. Our conventions for their OPEs can be found in Appendix~\ref{app:odakeOPEs}. We have also collected the OPEs of the $\SVseven$ algebra in Appendix~\ref{app:SVG2OPEs} and the ones of the $\SVeight$ algebra in Appendix~\ref{app:SVSpin7OPEs}.

The algebras Od$_3$, Od$_4$ and $\SVseven$ are only associative modulo certain singular fields \cite{Odake:1988bh, Figueroa-OFarrill:1996tnk}, respectively
\begin{equation}
N^1_n = \partial A_n - \normord{J^3_n B_n} \, , \qquad
N^2_n = \partial B_n + \normord{J^3_n A_n} \, ,
\end{equation}
for $n=3$ and $n=4$, and
\begin{equation}
N_7 = 4\normord{G_7X_7} -\,2\normord{P_7K_7}-\,4\,\partial M_7-\partial^2 G_7
\end{equation}
for $\SVseven$. More details can be found in \cite{Fiset:2018huv}.

\bigskip

We now give the reasoning behind the diamonds of algebra inclusions discussed in this chapter. This is a motivation rather than a derivation, and serves to intuitively appreciate the origin of our main algebraic results presented later.

Notice that our considerations are so far independent of global features of $M$. Let $U$ be an open subset in the manifold of interest $M$. Generally we expect
\begin{equation} \label{eq:SMinSU}
\mathcal{A}(M) ~\subset~ \mathcal{A}(U) \, .
\end{equation}
Indeed $U$ will generically be simpler, more symmetric, than $M$ so $\mathcal{A}(U)$ should be larger---for example a $\text{U}(n)$-structure may be definable locally but not globally. Moreover it should be possible to realize $\mathcal{A}(M)$, reflecting the global structure, in terms of degrees of freedom of the theory into $U$. There may however be some flexibility in the way the global structure is reflected in the local theory, leading to some freedom in the embedding \eqref{eq:SMinSU}. For example if $M$ has $G=\text{O}(2n)$ and $U$ has $G=\text{U}(n)$, then there is a full $\Sc^1$ orbit worth of $\mathcal{A}(M)=(\mathcal{N}=1) ~\subset~ (\mathcal{N}=2)= \, \mathcal{A}(U)$, which is actually reflecting the R-symmetry.\footnote{In this example, we are implicitly assuming that $M$ and $U$ are of the kind necessary for a quantum CFT to exist in the first place (e.g.\ Ricci-flatness to leading order, etc.).}

Now let $V\subset M$ be another open subset overlapping with $U$. The same logic yields
\begin{align}
\mathcal{A}(U) ~\subset~ \mathcal{A}(U\cap V) \qquad \text{and} \qquad
\mathcal{A}(V) ~\subset~ \mathcal{A}(U\cap V) \, .
\end{align}
As one transitions from $U$ to $V$, it can be expected that the freedom in the embedding \eqref{eq:SMinSU} needs to be restricted. Indeed symmetries emerging locally start being lost if one wants to cover a larger part of $M$. With the restriction, one achieves that $\mathcal{A}(M)$ also sits in $\mathcal{A}(V)$, and thus we have produced the diamond picture, \cref{fig:Diamond}. There may remain freedom in the realization of $\mathcal{A}(M)$, but as one covers more and more patches of $M$, one should be left with a single embedding, since $\mathcal{A}(M)$ are by assumption the symmetries of the theory with configuration space \emph{all} of $M$.

\begin{figure}[h]
\begin{center}
\begin{tikzpicture}
\node at (9,0.6) {\rotatebox{45}{$\subset$}};
\node at (11,0.6) {\rotatebox{135}{$\subset$}};
\node at (9,-0.7) {\rotatebox{135}{$\subset$}};
\node at (11,-0.7) {\rotatebox{45}{$\subset$}};
\node at (10,1.25) {$\mathcal{A}(U\cap V)$};
\node at (8,-0.05) {$\mathcal{A}(U)$};
\node at (12,-0.05) {$\mathcal{A}(V)$};
\node at (10,-1.4) {$\mathcal{A}(M)$};
\end{tikzpicture}
\caption{Diamond of algebra inclusions.}
\label{fig:Diamond}
\end{center}
\end{figure}

The connected sum constructions all have in common to be described by only two open patches, so a single diamond reflects the whole geometry $M=M_+\cup M_-$. As a consequence, $\mathcal{A}(M)$ is given precisely by $\mathcal{A}(M_+) \cap \mathcal{A}(M_-)$, where the intersection is in $\mathcal{A}(M_+ \cap M_-)$.\footnote{$M_+ \cap M_-$ is precisely what we have been calling the ``neck'' of the connected sum.} We find moreover in the examples below that geometric transition functions from $M_+$ to $M_-$ translate to some algebra automorphism of $\mathcal{A}(M_+\cap M_-)$ preserving the $\mathcal{A}(M)$ subalgebra.

The results of \cite{Fiset:2018huv} imply indeed that the $\SVseven$ algebra expected for G$_2$-manifolds is present in two distinct $(\text{Od}_3\oplus\text{Fr}^1)_\pm$ subalgebras of $(\text{Od}_2\oplus\text{Fr}^3)$, see \cref{fig:TCS}, and thus in their intersection.
The reverse inclusion, unaddressed in \cite{Fiset:2018huv}, is perhaps even more interesting, because it informs on worldsheet symmetries of TCS G$_2$-manifolds, which could conceivably be larger than those of a generic G$_2$-manifold. We investigate this in \cref{sec:Num}.

\section{G$_2$ Extra Twisted Connected Sums} \label{sec:G2}

In this section we extend the results of \cite{Fiset:2018huv} to the Extra Twisted Connected Sums (ETCS) of \cite{Crowley:2015ctv, Nordstrom:2018cli, Goette:2020}. Let us first briefly summarize the geometry.

\subsection{ETCS geometry} \label{sec:G2geom}

The situation for ETCS is very similar to the TCS case shown in \cref{fig:TCS}(a), since in both cases we glue two Asymptotically Cylindrical Calabi--Yau 3-folds times a circle along a common asymptotic neck region of the form $\text{CY}_2\times \mathbb{T}^2\times \mathbb{R}$. An \emph{Asymptotically Cylindrical (ACyl) Calabi--Yau $n$-fold} ($n$ complex dimensions) has a compact region whose complement is diffeomorphic to a cylinder, here with cross-section a closed Calabi--Yau $(n-1)$-fold times a circle. We represent this asymptotic behaviour by an arrow
\begin{equation}
\text{ACyl CY}_n \longrightarrow \text{CY}_{n-1}\times \R^+\times\Sc^1  \, .
\end{equation}
In addition, the metric $g_n$ and the Hermitian and holomorphic volume forms $\omega_n$ and $\Omega_n$ of $\text{CY}_n$ asymptote to those of the cylinder, see \eqref{eq:inclusionsun-1sun}. If we parametrize $\R^+$ by $t$ and $\Sc^1$ by $\theta$, we can write the asymptotic relations between them as follows:
\begin{align}
g_n\longrightarrow g_{n, \infty} &=
g_{n-1} + \dd t^2 + \dd \theta^2  \, , \\
\label{eq:cyintermsofk31}
\omega_n\longrightarrow\omega_{n, \infty} &=\omega_{n-1}+\dd t\wedge\dd\theta \, ,\\
\label{eq:cyintermsofk32}
\Omega_n\longrightarrow\Omega_{n, \infty} &=(\dd\theta-i\dd t)\wedge\Omega_{n-1} \, ,
\end{align}
where the subscript $\infty$ refers to the forms in the limit $t\rightarrow\infty$.

As a first step to generalize the TCS construction we assume there exist cyclic groups $\Gamma_\pm=\mathbb{Z}/k_\pm\mathbb{Z}$
acting diagonally
on the two sides to be glued,  $\text{ACyl~CY}_{3,\pm}\times \Sc^1_\pm$, in such a way that the Calabi--Yau structure is preserved and that the action is free on the $\Sc^1_\pm$. We call the latter $\Sc^1_\pm$ \emph{external circles} and we parametrize them with coordinates $\xi_\pm$. Furthermore, in the neck region where the ACyl~CY$_{3,\pm}$ asymptote to  CY$_{2,\pm}\times\mathbb{R}_\pm\times\Sc^1_\pm$,
we demand that the groups $\Gamma_\pm$ act trivially on $\text{CY}_{2,\pm}\times\mathbb{R}_\pm$ and freely on $\Sc^1_\pm$. We call the latter $\Sc^1_\pm$ \emph{internal circles} and parametrize them with coordinates $\theta_\pm$. Note that, in the asymptotic region, $\Gamma_\pm$ is only acting non-trivially on the torus formed by the internal and external circles.
The quotient $\mathbb{T}^2_\pm/\Gamma_\pm$ is still a torus, however, the group action modifies the original torus lattice, effectively ``twisting'' its structure and changing the length of its sides.

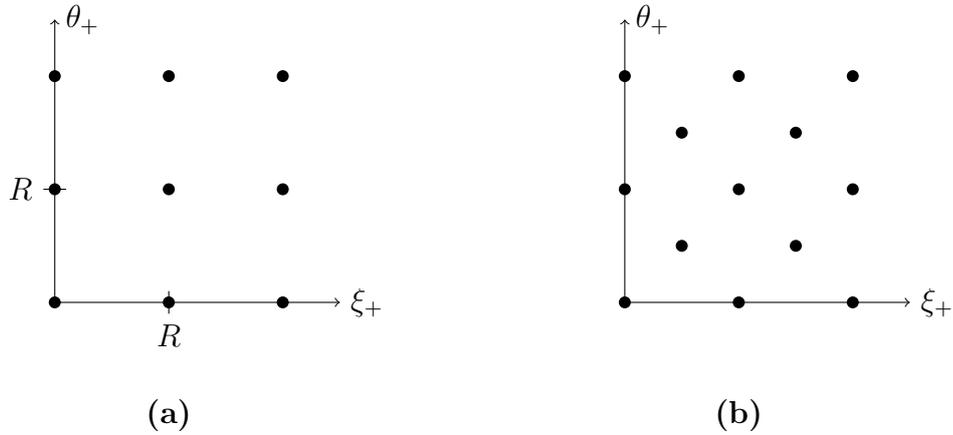
\begin{figure}[h]
\begin{center}
\begin{tikzpicture}[scale=1.5]
\draw[->] (0,0)--(2.5,0) node[right] {$\xi_+$};
\draw[->] (0,0)--(0,2.5) node[right] {$\theta_+$};
\draw[-] (1,0.1)--(1,-0.1) node[below] {$R$};
\draw[-] (0.1,1)--(-0.1,1) node[left] {$R$};
\draw[->] (5,0)--(7.5,0) node[right] {$\xi_+$};
\draw[->] (5,0)--(5,2.5) node[right] {$\theta_+$};
\foreach \y in {0,...,2}{
    \foreach \x in {0,...,2}
    {
    \fill (\x,\y) circle (1.5pt);
    }
    \foreach \x in {5,...,7}
    {
    \fill (\x,\y) circle (1.5pt);
    }
    }
    
\foreach \y in {0.5,1.5}{
    \foreach \x in {5.5,6.5}
    {
    \fill (\x,\y) circle (1.5pt);
    }}

\node at (1,-1) {\textbf{(a)}};
\node at (6,-1) {\textbf{(b)}};
\end{tikzpicture}
\caption{\textbf{(a)} Torus lattice of $\mathbb{T}^2_+$. \textbf{(b)} Torus lattice of $\mathbb{T}^2_+/\Gamma_+$.
}
\label{fig:lattices}
\end{center}
\end{figure}

Let us illustrate these features with an example. The simplest ETCS in \cite{Crowley:2015ctv} involves no quotient on one side of the construction, $\Gamma_-=\lbrace 1 \rbrace$, and a $\mathbb{Z}_2$ quotient on the other side, $\Gamma_+=\lbrace 1,\tau \rbrace$. Note that $\tau$ acts on the CY$_{3,+}$ as an involution which in the asymptotic end performs a rotation of the internal circle by an angle of $\pi$ leaving the rest fixed. The action of $\tau$ on the external circle is also a rotation of angle $\pi$. Take the radius of the internal and external circles
to be of the same length $R$, so that the torus $\mathbb{T}^2_+$ is 
obtained from the square lattice in \cref{fig:lattices}(a).
The identification under the action of $\tau$
appears in the lattice as a new set of points, see \cref{fig:lattices}(b). The lattice of
$\mathbb{T}^2_+/\Gamma_+$
still represents a square torus, however the lattice is tilted with respect to the original one and the radius of the circles is now $R/\sqrt{2}$.

The tangent vectors $(\partial_{\theta\pm},\partial_{\xi\pm})$ define an orthonormal frame even after the quotient so we will use them to describe the gluing.
The tori are glued by an orientation-reversing isometry; we call such a map a \emph{torus matching} following \cite{Nordstrom:2018cli}. This was achieved in the TCS case by identifying the internal circle on one side with the external circle on the other side. ETCS require in general a different alignment of the internal and external circles. We assume there exists a torus matching $\mathfrak{t}$ between the tori $\mathbb{T}^2_\pm/\Gamma_\pm$ such that the orthogonal frames are related by
\begin{align}
\label{eq:partialxi} \partial_{\xi-}=\cos\vartheta\,\partial_{\xi+} +\sin\vartheta\,\partial_{\theta+} \, ,\\
\label{eq:partialtheta} \partial_{\theta-}=\sin\vartheta\,\partial_{\xi+} -\cos\vartheta\,\partial_{\theta+} \, ,
\end{align}
for some $\vartheta\in(0,\pi)$ called the \emph{gluing angle}. This is determined from the tori $\mathbb{T}^2_\pm/\Gamma_\pm$: they are described by the same lattice up to a rotation which essentially
{fixes} the gluing angle. The systematic process to extract this information is described in \cite{Goette:2020}, but for our purposes it is enough to declare that the lattice of $\mathbb{T}^2_+/\Gamma_+$ is kept fixed whereas the lattice of $\mathbb{T}^2_-/\Gamma_-$ is rotated so that they can be glued together. Then, we can express $(\partial_{\theta-},\partial_{\xi-})$ in terms of $(\partial_{\theta+},\partial_{\xi+})$.
This will later be the key to describe the diamond of algebras for the ETCS case. Note also that the usual TCS corresponds to a gluing angle of $\vartheta=\pi/2$.

\begin{figure}[h]
\begin{center}
\begin{tikzpicture}[scale=1.5]
\foreach \y in {0,...,2}{
    \foreach \x in {0,...,2}
    {
    \fill (\x,\y) circle (1.5pt);
    }    }
    
\foreach \y in {0.5,1.5}{
    \foreach \x in {0.5,1.5}
    {
    \fill (\x,\y) circle (1.5pt);
    }}
    
\draw[->,thick,blue](1,1) -- (1,1.66);
\draw[->,thick,red](1,1) -- (1.66,1);
\draw[->,thick,red](1,1) -- (1.5,1.5);
\draw[->,thick,blue](1,1) -- (1.5,0.5);

\draw[->](1.33,1) arc (0:45:0.33);

\node at (0.75,1.75) {$\partial_{\theta+}$};
\node at (1.5,1.75) {$\partial_{\xi-}$};
\node at (1.75,0.75) {$\partial_{\xi+}$};
\node at (1.5,0.25) {$\partial_{\theta-}$};

\node at (1.5,1.25) {$\vartheta$};
\end{tikzpicture}
\caption{Torus matching between $\mathbb{T}^2_+/\Gamma_+$ and $\mathbb{T}^2_-/\Gamma_-$ with $\vartheta=\pi/4$.}
\label{fig:matching}
\end{center}
\end{figure}
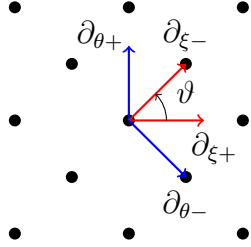

Returning to our example, the torus $\mathbb{T}^2_+/\Gamma_+$ is described by the square lattice in \cref{fig:lattices}(b) which corresponds to circles of radii $R/\sqrt{2}$. Since the quotient for $\mathbb{T}^2_-/\Gamma_-$ is trivial in our example, we take the radius of the internal and external circles
on this side
to be $R/\sqrt{2}$ (therefore the lattice of $\mathbb{T}^2_-/\Gamma_-$ is \cref{fig:lattices}(a) with lengths reduced by a factor of $\sqrt{2}$). Then the lattices of $\mathbb{T}^2_+/\Gamma_+$ and $\mathbb{T}^2_-/\Gamma_-$ coincide up to a rotation. We can find a torus matching with $\vartheta=\pi/4$ between the lattices, as illustrated in \cref{fig:matching}.

The K3 surfaces CY$_{2,\pm}$ in the neck region possess hyper-K\"ahler structures---see \cref{sec:HyperkahlerStructures}---given by a triple of closed two-forms $\omega^1_\pm$, $\omega^2_\pm$ and $\omega^3_\pm$. As we pointed out in \eqref{eq:hyperkahlerrelations}, the Hermitian forms satisfy
\begin{equation}
(\omega^1_\pm)^2=(\omega^2_\pm)^2=(\omega^3_\pm)^2\neq 0 \, , \qquad \omega^1_\pm\wedge\omega^2_\pm=\omega^2_\pm\wedge\omega^3_\pm=\omega^3_\pm\wedge\omega^1_\pm=0 \, .
\end{equation}
We can then describe SU(2)-structures on the K3 surfaces constructing the Hermitian forms $\omega_\pm$ and holomorphic volume forms $\Omega_\pm$ as we explained in \eqref{eq:equivalencesp1su2}
\begin{equation}
\omega_\pm=\omega^1_\pm \, , \qquad \Omega_\pm=\omega^2_\pm+i\, \omega^3_\pm \, .
\end{equation}
The nontrivial angle in the torus matching implies that a similar rotation must be applied to the isometry used to glue CY$_{2,+}$ with CY$_{2,-}$ so as to achieve a global G$_2$-structure. We must use a \emph{hyper-K\"ahler matching with angle} $\vartheta$, which we denote by $\mathfrak{r}$ and which satisfies
\begin{equation} \label{eq:hypermatching} \mathfrak{r}^*\omega^3_-=-\,\omega^3_+ \, ,
\qquad
\mathfrak{r}^*(\omega^1_-+i\, \omega^2_-)=e^{i\vartheta}(\omega^1_+-i \omega^2_+) \, .
\end{equation}
The TCS hyper-K\"ahler matching is recovered by $\vartheta=\pi/2$, see e.g. \cite[Sect.~1.1]{Fiset:2018huv}.

These isometries are used to define a gluing of the two sides of the ETCS construction $M_\pm=(\text{ACyl~CY}_{3,\pm}\times \Sc^1_\pm)/\Gamma_\pm$ along the neck region\footnote{Strictly speaking, the ACyl~CY$_{3,\pm}$ must be first truncated at finite distance $t_0$ in the asymptotic direction before being glued. This truncation gives rise to a non-vanishing G$_2$ torsion, going to zero as $t_0\rightarrow\infty$.}
\begin{equation}\label{eq:gluingmapETCS}
F=(-\text{Id}_\mathbb{R})\times\mathfrak{t}\times\mathfrak{r}: \, (\mathbb{R}\times\mathbb{T}^2_+\times\text{CY}_{2,+})/\Gamma_+ \longrightarrow \, (\mathbb{R}\times\mathbb{T}^2_-\times\text{CY}_{2,-})/\Gamma_- \, .
\end{equation}
The resulting manifold after the gluing has a globally-defined G$_2$-structure which is described in the neck region by the same associative three-form as for the TCS case
\begin{equation} \label{eq:neckg2structure}
\varphi_\pm=\dd\xi_\pm\wedge\omega^1_\pm +\dd\theta_\pm\wedge\omega^2_\pm +\dd t_\pm\wedge\omega^3_\pm+\dd t_\pm\wedge\dd\theta_\pm\wedge\dd\xi_\pm \, .
\end{equation}
The gluing identifies the G$_2$ forms on both sides: $F^*(\varphi_-)=\varphi_+$. This G$_2$-structure has torsion localized around the neck. However, techniques from analysis show that the G$_2$-structure can be slightly deformed to remove the torsion and obtain a torsion-free G$_2$-holonomy ETCS manifold.

\subsection{Chiral algebra viewpoint}

Let us now translate this construction to chiral algebras in the worldsheet.
First of all, note that the building blocks of an ETCS possess the same geometric properties as those of a TCS since the quotient does not spoil the Calabi--Yau structure. As a result we
will have
a diamond of inclusions
essentially identical
to the TCS one, see \cref{fig:TCS}(b), except for how the relevant subalgebras are concretely realized.
Since the neck region
has geometrically the form $\text{CY}_2\times \mathbb{T}^2\times \mathbb{R}$, the
algebra sitting on top of our diamond is $\text{Od}_2\oplus \text{Fr}^3$, consistently with the generalities laid out in \cref{sec:Idea}. That is, as in \cite[Sect.~2.2]{Fiset:2018huv}, there is a correspondence between generators of $\text{Od}_2\oplus \text{Fr}^3$ and invariant forms in the neck region, see again \cref{tab:dictionary} for our notations.
To be completely explicit in what follows, we can think of this algebra as being associated to the neck region of $M_+=(\text{ACyl~CY}_{3,+}\times \Sc^1_+)/\Gamma_+$, so for example $\dd \theta_+$ is associated with $\psi_\theta$ and $j_\theta$, and so on for the other covariantly constant tensors.

Now the realization of $\text{Od}_3\oplus \text{Fr}^1\subset\text{Od}_2\oplus \text{Fr}^3$ corresponding to the $M_+$ side is precisely the one found in \cite{Fiset:2018huv}.\footnote{In order to perform all the computations involving operator algebras in this work, we used the package \emph{OPEdefs} by Thielemans \cite{Thielemans:1994er}.} This is because the quotient by $\Gamma_+$ preserves the Calabi--Yau structure and the asymptotic description of the different forms. Explicitly the generators of $(\text{Od}_3\oplus \text{Fr}^1)_+$ are given by $\psi_\xi$ and
\begin{equation} \label{eq:Od3inOd2F2}
\begin{split}
G_3 &= G_2 + G_t + G_\theta \, , \\
J^3_3 &= J^3_2 \, + \normord{\psi_t\,\psi_\theta} \, , \\
A_3+iB_3 &= \, \normord{(\psi_\theta-i\psi_t)(A_2+iB_2)} \, ,
\end{split}
\end{equation}
along with their supersymmetric partners, see \cref{tab:dictionary}, which can all be reconstructed from \eqref{eq:Od3inOd2F2}.

Moreover the expression \eqref{eq:neckg2structure} of the G$_2$-structure in the neck region  is the same as in the TCS case, so the realisation of SV$^{\text{G}_2}\subset\text{Od}_2\oplus \text{Fr}^3$ \cite{Figueroa-OFarrill:1996tnk} used in \cite{Fiset:2018huv} also applies to this ETCS case.
Explicitly\footnote{Note that it is sufficient to specify the $\text{SV}^{G_2}$ generators $G_7$ and $P$, as the others can be deduced from them by taking operator product expansions.}
\begin{equation}
G_7 = G_3 + G_\xi \, , \qquad
P = A_3 \, + \normord{J^3_3\, \psi_\xi} \, ,
\end{equation}
so combining with \eqref{eq:Od3inOd2F2},
\begin{equation} \label{eq:SV7inOd2F3}
G_7 = G_2 + G_\theta + G_\xi + G_t \, , \qquad
P = \, \normord{\psi_\theta\, A_2} +\normord{\psi_t\, B_2} + \normord{J^3_2 \psi_\xi} + \normord{\psi_t\,\psi_\theta\,\psi_\xi} .
\end{equation}
The other subalgebra $(\text{Od}_3\oplus \text{Fr}^1)_- \subset\text{Od}_2\oplus \text{Fr}^3$ corresponding to the open manifold $M_-=(\text{ACyl~CY}_{3,-}\times \Sc^1_-)/\Gamma_-$ is obtained from \eqref{eq:Od3inOd2F2} by applying the following automorphism of $\text{Od}_2\oplus \text{Fr}^3$, which is inferred from the geometric gluing map \eqref{eq:gluingmapETCS}:
\begin{equation} \label{eq:GluingETCS}
\begin{split}
G_2 &\longmapsto G_2 \\
\begin{pmatrix}
J^3_2 \\ A_2
\end{pmatrix} &\longmapsto
\begin{pmatrix}
\cos\vartheta & \sin\vartheta\\
\sin\vartheta & -\cos\vartheta
\end{pmatrix}
\begin{pmatrix}
J^3_2 \\ A_2
\end{pmatrix}
\qquad\qquad
\begin{pmatrix}
\psi_\xi \\ \psi_\theta
\end{pmatrix} \longmapsto
\begin{pmatrix}
\cos\vartheta & \sin\vartheta\\
\sin\vartheta & -\cos\vartheta
\end{pmatrix}
\begin{pmatrix}
\psi_\xi \\ \psi_\theta
\end{pmatrix}
\\
B_2 &\longmapsto -B_2
\qquad\qquad\qquad
\qquad\qquad\qquad
\quad\ \,
\psi_t \longmapsto -\psi_t
\end{split}
\end{equation}
(with the same action on supersymmetric partners). Indeed applying this to the expressions \eqref{eq:Od3inOd2F2} generates $(\text{Od}_3 \oplus \text{Fr}^1)_-$ in terms of  $\cos\vartheta\,\psi_\xi + \sin\vartheta\,\psi_\theta$ and
\begin{equation} \label{eq:Od3inOd2F2Prime}
\begin{split}
G_3 &= G_2 + \sin^2\vartheta\, G_\xi +\cos^2\vartheta\, G_\theta -\sin\vartheta\cos\vartheta \,(\normord{j_\xi\,\psi_\theta}+\normord{j_\theta\,\psi_\xi}) + G_t \, ,
\\
J^3_3 &= \cos\vartheta \,(J^3_2\, +\normord{\psi_t\,\psi_\theta})+\sin\vartheta\,(A_2\,-\normord{\psi_t\,\psi_\xi}) \, ,
\\
A_3+iB_3 &= \;
\normord{(\sin\vartheta\,\psi_\xi-\cos\vartheta\,\psi_\theta+i\,\psi_t)(\sin\vartheta\, J^3_2-\cos\vartheta\, A_2 - iB_2)} .
\end{split}
\end{equation}
The ordinary TCS automorphism of \cite{Fiset:2018huv} is recovered from \eqref{eq:GluingETCS} by setting $\vartheta=\pi/2$.

It can be checked that, for any $\vartheta$, the map \eqref{eq:GluingETCS} leaves SV$^{\text{G}_2}$ in \eqref{eq:SV7inOd2F3} invariant. This shows that the diamond of inclusions, \cref{fig:TCS}, is indeed correct: SV$^{\text{G}_2}$ sits in the intersection of $(\text{Od}_3 \oplus \text{Fr}^1)_+$ with $(\text{Od}_3 \oplus \text{Fr}^1)_-$ inside $\text{Od}_2 \oplus \text{Fr}^3$ for any $\vartheta$.

\bigskip

We stress that these various statements rely on a careful treatment of null vectors. The OPEs of $\text{Od}_3 \oplus \text{Fr}^1$ are only satisfied by \eqref{eq:Od3inOd2F2} upon quotienting by \cite{Odake:1988bh}
\begin{equation}
N^1_3 = \partial A_3 \; -\normord{J^3_3\, B_3} \, , \qquad
N^2_3 = \partial B_3 \; +\normord{J^3_3\, A_3} \, ,
\end{equation}
where $J^3_3$, $A_3$, $B_3$ are given by \eqref{eq:Od3inOd2F2}. Similarly the image $(\text{Od}_3 \oplus \text{Fr}^1)_-$ \eqref{eq:Od3inOd2F2Prime} under the gluing automorphism \eqref{eq:GluingETCS} is only valid up to the image of $N^1_3$, $N^2_3$. When regarded as elements of $\text{Od}_2 \oplus \text{Fr}^3$, the fields $N^1_3$, $N^2_3$, and their images under \eqref{eq:GluingETCS}, all descend from the null vectors
\begin{equation}
N^1_2 = \partial A_2 \; -\normord{J^3_2\, B_2} \, , \qquad
N^2_2 = \partial B_2 \; +\normord{J^3_2\, A_2} .
\end{equation}
We always assume that null vectors of $\text{Od}_2\oplus \text{Fr}^3$ are quotiented out. This also ensures that the null vector $N$ modulo which the $\SVseven$ subalgebra is associative, see \cite{Figueroa-OFarrill:1996tnk}, is indeed zero.

It is worth stressing that the diamond diagram is well-defined for any value of the gluing angle $\vartheta$. ETCS have been constructed so far only for a discrete set of gluing angles. Our results show that, at least from the worldsheet algebra perspective, there is no reason to exclude ETCS with somewhat more general gluing angles.

\section{Spin(7) Generalized Connected Sums} \label{sec:Spin7}

\subsection{GCS geometry} \label{sec:Spin7Geom}

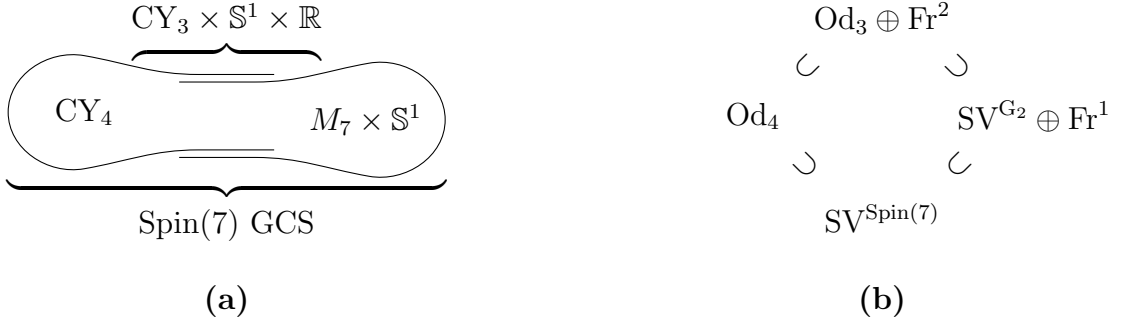
\begin{figure}[h] 
\begin{center}
\begin{tikzpicture}
\draw [black]
(2,0.5) to (1,0.5)
to[out=180, in=-10] (-0.5,0.75)
to[out=170, in=90] (-1.5,0)
to[out=-90, in=190] (-0.5,-0.75)
to[out=10, in=180] (1,-0.5) to (2,-0.5);
\draw [black]
(0.75,0.4) to (1.75,0.4)
to[out=0, in=190] (3.25,0.65)
to[out=10, in=90] (4.25,-0.1)
to[out=-90, in=-10] (3.25,-0.85)
to[out=170, in=0] (1.75,-0.6) to (0.75,-0.6);
\node at (1.375,0.8) {$\overbrace{\qquad\qquad\qquad}$};
\node at (1.375,-1.0) {$\underbrace{\qquad\qquad\qquad\qquad\qquad\qquad\qquad}$};
\node at (1.375,-2.5) {\textbf{(a)}};
\node at (10,-2.5) {\textbf{(b)}};
\node at (9,0.6) {\rotatebox{45}{$\subset$}};
\node at (11,0.6) {\rotatebox{135}{$\subset$}};
\node at (9,-0.7) {\rotatebox{135}{$\subset$}};
\node at (11,-0.7) {\rotatebox{45}{$\subset$}};

\node at (1.375,1.25) {$\text{CY}_3\times \Sc^1\times \mathbb{R}$};
\node at (-0.5,0) {$\text{CY}_4$};
\node at (3.2,-0.1) {$M_7 \times \Sc^1$};
\node at (1.375,-1.5) {Spin(7) GCS};

\node at (10,1.25) {$\text{Od}_3\oplus \text{Fr}^2$};
\node at (8.3,-0.05) {$\text{Od}_4$};
\node at (12,-0.05) {$\SVseven \oplus \text{Fr}^1$};
\node at (10,-1.4) {$\SVeight$};
\end{tikzpicture}
\caption{\textbf{(a)} Sketch of a compact 8-dimensional Spin(7)-holonomy manifold obtained as Generalized Connected Sum (GCS).{ $M_7$ is an ACyl G$_2$-manifold.} \textbf{(b)} Diamond of algebra inclusions corresponding to a sigma model whose target space is a GCS.}
\label{fig:GCS}
\end{center}
\end{figure}

Strong evidence for the existence of a generalization of G${}_2$ TCS for the case of Spin(7)-manifolds, called ``Generalized Connected Sum'' (GCS), was provided in \cite{Braun:2018joh}. From specific examples (Joyce orbifolds of $\mathbb{T}^8$), the authors show that in this case the two manifolds to be glued along the neck region must be different, see \cref{fig:GCS}(a).

One of the manifolds is an ACyl Calabi--Yau 4-fold $M_+$, see \cref{sec:G2geom}. We sometimes use the subindex $+$ to distinguish any object relative to it.

For the other half of the construction we need to introduce a similar notion. An \emph{ACyl G$_2$-manifold $M_7$}, see e.g.\ \cite{Nordstrom:2007, Kovalev:2010}, is a non-compact manifold of holonomy G${}_2$ with a compact region whose complement is diffeomorphic to a closed Calabi--Yau 3-fold times the positive real line. We represent this by
\begin{equation}
\text{ACyl }M_7 \longrightarrow \text{CY}_3 \times \R^+ \, .
\end{equation}
In addition, the associative three-form $\varphi$ and coassociative four-form $\psi$ of $M_7$ are asymptotic to the ones that can be constructed from the Calabi--Yau times the real line, see \eqref{eq:inclusionsu3g2}. Parametrizing $\mathbb{R}$ by $t$, we have
\begin{align}
\label{eq:g2intermsofcy1}
\varphi\longrightarrow \varphi_\infty &=\Re\Omega_3 -\dd t\wedge\omega_3 \, ,\\
\label{eq:g2intermsofcy2}
\psi\longrightarrow\psi_\infty &=\frac{1}{2}\,\omega_3\wedge\omega_3+\dd t\wedge\Im\Omega_3 \, ,
\end{align}
where again the subscript $\infty$ refers to the forms in the asymptotic limit and where $\Re$ and $\Im$ are the real and imaginary parts.

The manifolds we are going to glue together are, on the one hand, an ACyl CY$_4$ $M_+$ and, on the other hand, the product of a circle with an ACyl G$_2$ manifold $M_- = M_7 \times \Sc^1$. The manifolds $M_+$ and $M_-$ can be glued along their asymptotic ends, which are compatible. Introducing a boundary at $t=t_0+1$ (on both sides), we can define the gluing map along the interval $I=[t_0,t_0+1]$:
\begin{align}
\label{eq:gluingmapSpin7}
\begin{split}
F_{t_0}: 
\text{CY}_{3, +} \times \Sc^1_+ \times I_+
&\longrightarrow 
\text{CY}_{3, -} \times \Sc^1_- \times I_-
\\
\big(z,~\theta,~t\big) &\longmapsto \big(\phi(z),~-\theta,~2t_0+1-t\big) \, ,
\end{split}
\end{align}
where $\phi$ is a biholomorphic map between the Calabi--Yau 3-folds such that the holomorphic volume form changes global sign, that is, $F_{t_0}^*(\Omega_{3,-})=-\Omega_{3,+}$. {With our conventions we also need to reverse the sign in the identification of the circles in order to have the correct gluing.}

Next we have to specify a Spin(7)-structure on the whole manifold via the Cayley four-form $\Psi$. First of all, since $M_+$ has holonomy SU(4) and $\text{SU(4)}\subset\text{Spin(7)}$, we can define a torsion-free Spin(7)-structure on $M_+$, see \eqref{eq:inclusionsu4spin7}, via
\begin{equation}
\label{eq:psiintermsofcy4}
\Psi_+=\Re\Omega_{4,+}+\frac{1}{2}\,\omega_{4,+}\wedge\omega_{4,+} \, .
\end{equation}
In the asymptotic neck region $\omega_{4,+}$ and $\Omega_{4,+}$ decompose according to \eqref{eq:cyintermsofk31} and \eqref{eq:cyintermsofk32}. Thus, the Spin(7)-structure takes the form
\begin{equation}
\label{psi+}
\Psi_{\infty,+}=\dd\theta_+\wedge\Re\Omega_{3,+}+\dd t_+\wedge\Im\Omega_{3,+}+\frac{1}{2}\,\omega_{3,+}\wedge\omega_{3,+}+\dd t_+\wedge\dd\theta_+\wedge\omega_{3,+} \, .
\end{equation}
On the other hand, the manifold $M_-$ has holonomy G${}_2$ and ${\rm G}_2\subset\text{Spin(7)}$, so that we can construct a torsion-free Spin(7)-structure as follows, see \eqref{eq:inclusiong2spin7}
\begin{equation}
\label{eq:psiintermsofg2circle}
\Psi_-=\dd\theta_-\wedge\varphi_-+\psi_- \, .
\end{equation}
In the asymptotic region, the associative and coassociative forms can be decomposed as in \eqref{eq:g2intermsofcy1} and \eqref{eq:g2intermsofcy2}, so the Spin(7)-structure takes the form
\begin{equation}
\label{psi-}
\Psi_{\infty,-}=\dd\theta_-\wedge\Re\Omega_{3,-}+\dd t_-\wedge\Im\Omega_{3,-}+\frac{1}{2}\,\omega_{3,-}\wedge\omega_{3,-}+\dd t_-\wedge\dd\theta_-\wedge\omega_{3,-} \, .
\end{equation}
The diffeomorphism $F_{t_0}$ identifies the Spin(7)-structures $\Psi_-$ and $\Psi_+$ along the gluing, so that the resulting manifold has a global Spin(7)-structure. As in the G${}_2$ TCS case, the structure is torsion-free except around the gluing region (because of the truncation at finite $t_0$). It is believed that---analogously to the G$_2$ case---for large enough $t_0$, a small deformation of the structure can be found such that the resulting manifold has not only a torsion-free Spin(7)-structure but also Spin(7)-holonomy.

\subsection{Chiral algebra viewpoint} \label{sec:Spin7Alg}

We now describe how the GCS construction is reflected in the worldsheet chiral algebras, see \cref{fig:GCS}(b). The neck region where the open patches of a GCS overlap is given by $\text{CY}_3\times\mathbb{R}\times\Sc^1$ so the top algebra is given by $\text{Od}_3\oplus \text{Fr}^2$. Now, one of the open patches has a $\text{CY}_4$ structure so we expect an $\text{Od}_4$ subalgebra in one of the sides of the diamond, whereas the other open patch has the structure of a G$_2$-manifold times a circle so we expect a $\SVseven\oplus\text{Fr}^1$ on that side. Since the manifold has holonomy Spin(7) after the gluing, we expect a $\SVeight$ algebra at the bottom of the diagram.

Let us now provide these inclusions explicitly. First of all, we describe $\SVseven\oplus\text{Fr}^1\subset\text{Od}_3\oplus \text{Fr}^2$. The $\text{Fr}^1$ part corresponding geometrically to the circle is trivially identified, so we only require a realization of $\SVseven\subset\text{Od}_3\oplus \text{Fr}^1$. This was already found in \cite{Figueroa-OFarrill:1996tnk} and in fact constitutes the bottom inclusion of the TCS diagram in \cref{fig:TCS}(b). To be completely explicit, the ACyl G$_2$-manifold definition \eqref{eq:g2intermsofcy1}--\eqref{eq:g2intermsofcy2} suggests the following ansatz, where the subindex $t$ stands for the $\text{Fr}^1$ corresponding to the real line $\mathbb{R}$:
\begin{equation}\label{eq:g2frsubod3fr2}
\begin{split}
&G_7=G_3 + G_t \, , \qquad
P= A_3 \, -\normord{\psi_t\, J^3_3} \, , \qquad
X_7=\frac{1}{2}\normord{J^3_3\,J^3_3}+\normord{\psi_t\, B_3}-\, \frac{1}{2}\normord{\partial\psi_t\,\psi_t} \, ,
\end{split}
\end{equation}
These operators and their descendants indeed satisfy the $\SVseven$ OPE relations up to null vectors of Od$_3$.

A realization of $\SVeight\subset\SVseven\oplus\text{Fr}^1$ was already found in \cite{Shatashvili:1994zw, Gepner:2001px}. Explicitly, an ansatz is given by \eqref{eq:psiintermsofg2circle} and denoting the fields from the $\text{Fr}^1$ associated to the circle with a subindex $\theta$ we find
\begin{equation}\label{eq:spin7subg2fr}
\begin{split}
&G_8=G_7+G_\theta \, , \qquad 
X_8=-\left(\normord{\psi_\theta\, P}+X_7\right)+\frac{1}{2}\normord{\partial\psi_\theta\,\psi_\theta}.
\end{split}
\end{equation}
These operators and their descendants satisfy the $\SVeight$ algebra OPEs up to null vectors of $\SVseven$.

We now turn our attention to the other side of the diamond: $\text{Od}_4\subset\text{Od}_3\oplus \text{Fr}^2$. In order to find a realization in $\text{Od}_3\oplus \text{Fr}^2$, geometry again provides inspiration. From the asymptotic formulae \eqref{eq:cyintermsofk31} and \eqref{eq:cyintermsofk32}, we write
\begin{equation}\label{eq:od4subod3fr2}
\begin{split}
G_4 &=G_3+G_t+G_\theta \, ,\\
J^3_4 &=J^3_3\,+\normord{\psi_t\,\psi_\theta} \, ,\\
A_4+iB_4 &=\,\normord{(\psi_\theta-i\,\psi_t)(A_3+iB_3)} .
\end{split}
\end{equation}
We find that these operators and their descendants indeed satisfy the Od${}_4$ OPEs up to null fields of $\text{Od}_3$.\footnote{Moreover the null fields $N^1_4$, $N^2_4$ modulo which Od${}_4$ is associative, see \cref{sec:Idea}, are indeed null since they can be rewritten in terms of null fields of Od${}_3$:
$
N^1_4=\,\normord{\psi_\theta\, N^1_3}+\normord{\psi_t\, N^2_3} $ and
$
N^2_4=\,\normord{\psi_\theta\, N^2_3}-\normord{\psi_t\, N^1_3}
$.
}

The last relation of the diagram that we must give is $\SVeight\subset\text{Od}_4$. The existence of this subalgebra was already shown in \cite{Figueroa-OFarrill:1996tnk}; it again follows from geometric considerations, in particular \eqref{eq:psiintermsofcy4}:\footnote{Note the overall minus sign in $X_8$ as compared to the geometric formula \eqref{eq:psiintermsofcy4} for {$\Psi$}. This is because of an unfortunate convention in the field theory side. See \cite[Sect.~2]{Fiset:2019ecu} for other instances of the same mismatch.}
\begin{equation}\label{eq:spin7subod4}
\begin{split}
&G_8=G_4 \, ,\qquad 
X_8=-\Big( A_4+\frac{1}{2}\normord{J^3_4J^3_4}\Big) \, .
\end{split}
\end{equation}
Again these operators and their descendants satisfy the $\SVeight$ OPE relations up to the Od${}_4$ null fields.

There is one last check we have to perform. We have described two embeddings of the $\SVeight$ algebra inside $\text{Od}_3\oplus \text{Fr}^2$, one of them via an $\text{Od}_4$ subalgebra and the other via a $\SVseven\oplus\text{Fr}^1$ subalgebra. For the diamond to hold we must ensure that these embeddings describe the same subalgebra. This is indeed the case: once the $\text{Od}_4$ and $\SVseven\oplus\text{Fr}^1$ generators are rewritten in terms of $\text{Od}_3\oplus \text{Fr}^2$ operators we find that the generators of both $\SVeight$ algebras are exactly the same.

\bigskip

Recall that in the (E)TCS case the gluing morphism \eqref{eq:GluingETCS} inferred from geometry mapped the two sides of the diamond to one another. In the GCS case however the sides of the diamond have non-isomorphic subalgebras so we find a different behaviour. The gluing map \eqref{eq:gluingmapSpin7} suggests the following automorphism of $\text{Od}_3\oplus \text{Fr}^2$:
\begin{equation}
\label{eq:spin7gluingmorphism}
\begin{split}
G_{3} &\longmapsto G_{3} \, ,
\qquad\qquad\qquad
\psi_\theta \longmapsto-\psi_\theta \, ,\\
J^3_{3} &\longmapsto J^3_{3} \, ,
\qquad\qquad\qquad~
\psi_t \longmapsto-\psi_t \, ,\\
A_{3}+iB_3 &\longmapsto -(A_{3}+iB_3) \, .\\
\end{split}
\end{equation}
This map induces automorphisms of the algebras at the sides of the diamond: the identity map on the Od$_4$ side, and the following automorphism on the $\SVseven\oplus\text{Fr}^1$ side:
\begin{equation}
G_7 \longmapsto G_7 \, ,
\qquad
P \longmapsto -P \, ,
\qquad
\psi_\theta \longmapsto - \psi_\theta \, .
\end{equation}
Just like in the (E)TCS case, the Shatashvili-Vafa algebra at the bottom of the diamond is left invariant by the gluing morphism \eqref{eq:spin7gluingmorphism}.

\bigskip

This realization of the diamond of subalgebras heavily relies on the geometry of the GCS construction. A natural question is what happens to the diamond in the case where different conventions can be chosen for the geometric structures. Different conventions change the explicit embeddings of the subalgebras inside $\text{Od}_3\oplus \text{Fr}^2$. For example, rotating by a phase $A_3+iB_3$ (akin to the holomorphic volume form) is an automorphism of $\text{Od}_3$ producing from \eqref{eq:g2frsubod3fr2} a U(1)-family of $\SVseven\oplus\text{Fr}^1$ subalgebras.
The $\SVeight$ embedding is also modified. Nevertheless it is reassuring that we always find a diamond structure with a single $\SVeight$ in the intersection of the algebras in the lateral tips. Once conventions are fixed we find a single diamond of subalgebras, as we expected.

\subsection{Automorphisms and mirror symmetry} \label{sec:Spin7Aut}

Mirror symmetry for exceptional holonomy manifolds was first suggested in \cite{Shatashvili:1994zw}, see also \cite{Papadopoulos:1995da}, and later examined through Joyce orbifold examples \cite{Acharya:1996fx, Acharya:1997rh, Gaberdiel:2004vx, Chuang:2004th} and examples of the form $\big(\Sc^1\times \text{CY}_3\big)/\mathbb{Z}_2$ \cite{Partouche:2000uq, Salur:2007ev}.  The case of TCS manifolds was fruitfully explored in \cite{Braun:2017ryx} and \cite{Braun:2017csz}, whereas GCS mirror symmetry was recently addressed in \cite{Braun:2019lnn}. Some of the TCS mirror maps were interpreted in terms of chiral algebras in \cite{Fiset:2018huv} and in this section we perform a similar study for GCS manifolds, using the diamond of chiral algebras. We also exploit that some Joyce orbifolds admit a GCS description in order to compare and propose new mirror constructions.

A mirror symmetry map may alter drastically the geometry of the target manifold, but the sigma model theory is preserved. In particular, the mirror map corresponds to an automorphism of the chiral algebra. Furthermore, for the case of GCS manifolds it is natural to look for mirrors which also possess a GCS structure, as was the case in \cite{Braun:2019lnn}. We have shown that this geometric structure is encoded in the diamond of chiral algebras, see \cref{fig:GCS}(b), so a mirror map respecting the GCS structure has to correspond to an automorphism of the top algebra $\text{Od}_3\oplus \text{Fr}^2$ preserving the diamond. In particular, it has to reduce to automorphisms of the algebras at the lateral tips and at the bottom of the diamond.

We {performed} a systematic search for these automorphisms. An important observation is that only two automorphisms of $\SVeight$ exist:\footnote{We demand that the Virasoro subalgebra generated by $T_8$ and $G_8$ should also be preserved.} the identity map and the parity map {$(-1)^F$ (where $F$ acts as $0$ on bosons and $1$ on fermions)}. It turns out we do not miss interesting information by restricting ourselves to automorphisms that reduce to the identity on $\SVeight$.\footnote{Any automorphism reducing to the identity on $\SVeight$ can be composed with $(-1)^F$ on $\text{Od}_3\oplus \text{Fr}^2$ to produce an automorphism reducing to $(-1)^F$ on $\SVeight$, and vice versa.} Moreover it is not too hard to see that the automorphisms must act diagonally on $\text{Od}_3\oplus \text{Fr}^1_t \oplus \text{Fr}^1_\theta$.\footnote{This follows from the condition that the Virasoro algebra of $\text{Od}_3\oplus \text{Fr}^2$ has to be preserved by the automorphism, and that the map should reduce to an automorphism of the lateral algebra $\SVseven\oplus\text{Fr}^1$.}

There are only four automorphisms of $\text{Od}_3\oplus\text{Fr}^2$ which satisfy these constraints. One of them is the identity, which we denote by $\textbf{A}_0$. The nontrivial automorphisms $\textbf{A}_1$, $\textbf{A}_2$, $\textbf{A}_3$ and their restriction to the subalgebras of the diamond are described in \cref{tab:automorphisms} below. {We remark that the $\mathbf{A}_i$ form the group $\mathbb{Z}_2^2$ under composition.} In the table, $\textbf{Ph}^\pi$ is defined as
\begin{equation}
\textbf{Ph}^\pi ~:~
A_n+iB_n \longmapsto -(A_n+iB_n) \, , \qquad
C_n+iD_n \longmapsto -(C_n+iD_n) \, ,
\end{equation}
(with the other generators invariant) and it is interpreted geometrically as a phase rotation by $\pi$ of the Calabi--Yau volume form. The other boldface maps will be explained shortly.

\renewcommand{\arraystretch}{1.2}
\begin{table}[h]
\begin{center}
\begin{tabular}{ |c|c|c|c|c| } 
\hline
 $\text{Automorphisms}$  & $\textbf{A}_0$ & $\textbf{A}_1$ & $\textbf{A}_2$ & $\textbf{A}_3$ \\
 \hline
 \hline
 $\text{Od}_3\oplus \text{Fr}^2$ & $\textbf{Id}$ & $\textbf{Ph}^\pi\circ\textbf{T}_t\circ\textbf{T}_\theta$ & $\textbf{M}\circ\textbf{T}_t$ & $\textbf{M}\circ\textbf{Ph}^\pi\circ\textbf{T}_\theta$ \\
 \hline
$\SVseven\oplus\text{Fr}^1$ & $\textbf{Id}$ & $\textbf{GK}\circ\textbf{T}_\theta$ & $\textbf{Id}$ & $\textbf{GK}\circ\textbf{T}_\theta$ \\ 
\hline
$\text{Od}_4$ & $\textbf{Id}$ & $\textbf{Id}$ & $\textbf{M}$ & $\textbf{M}$ \\ 
\hline
$\SVeight$ & $\textbf{Id}$ & $\textbf{Id}$ & $\textbf{Id}$ & $\textbf{Id}$ \\
\hline
\end{tabular}
\end{center}
\caption{Candidates for mirror automorphisms and their action on the algebras of the diamond.}
\label{tab:automorphisms}
\end{table}

Recall that T-duality along a direction $\Sc^1_\theta$ is accompanied on the worldsheet by the automorphism
\begin{equation}
\label{eq:tdualityautomorphism}
\textbf{T}_\theta ~:~
\psi_\theta \longmapsto -\psi_\theta \, , \qquad
j_\theta \longmapsto -j_\theta \, ,
\end{equation}
acting on, say, the left-moving $\text{Fr}^1_\theta$ (but not on the right-moving $\overline{\text{Fr}^1_\theta}$).
A direction $\R_t$ gives rise to a worldsheet algebra $\text{Fr}^1_t$, so by a map $\textbf{T}_t$ we mean the analogous of the worldsheet automorphism \eqref{eq:tdualityautomorphism} acting on the currents $(\psi_t,j_t)$.\footnote{When the target manifold is a Joyce orbifold the map $\textbf{T}_t$ arises from a T-duality along the $t$ direction on the underlying torus. In the general case, the global geometric interpretation of $\textbf{T}_t$ is not clear even though the automorphism is perfectly well-defined. This was to be expected because the chiral algebra only captures the local behaviour of the target manifold: for example a line and a circle both give rise to the same worldsheet currents and the chiral algebras can not be told apart.} Mirror symmetry for Joyce orbifolds is essentially a combination of $\textbf{T}$ maps, as we now recall.

For the G$_2$ Joyce orbifolds two different maps were described in \cite{Acharya:1997rh}: one is obtained by T-dualizing along {associative} $\mathbb{T}^3$ fibres and the other one by T-dualizing along {coassociative} $\mathbb{T}^4$ fibres. We denote them by $\mathcal{T}^3$ and $\mathcal{T}^4$ respectively. In \cite{Gaberdiel:2004vx} it was shown that $\mathcal{T}^4$ leads to the identity automorphism of $\SVseven$, and that $\mathcal{T}^3$ leads to the automorphism
\begin{equation}
\textbf{GK} ~:~
P \longmapsto -P \, , \qquad
K \longmapsto -K \, ,
\end{equation}
of $\SVseven$, which we call the Gaberdiel--Kaste mirror map. These considerations are G$_2$ analogues of the familiar SYZ conjecture, wherein T-duality along {a supersymmetric $\mathbb{T}^n$ fibration of a} $\text{CY}_n$ gives rise to mirror symmetry and the following worldsheet automorphism of Od$_n$:
\begin{equation}
\textbf{M} ~:~ 
\big(J^3_n,G^3_n,B_n,D_n\big)
\longmapsto
\big(-J^3_n,-G^3_n,-B_n,-D_n\big)
\, .
\end{equation}

For Joyce orbifolds of Spin(7) holonomy, {we give a detailed account in the next section.}

\subsubsection{Spin(7) Joyce orbifolds}

\cite{Acharya:1997rh} describes a single {type of} mirror map in that case, obtained by T-dualizing along {supersymmetric} $\mathbb{T}^4$ fibres. For {any} given Joyce orbifold there are 14 such fibrations, corresponding to toroidal fibres which are calibrated by the Cayley four-form \eqref{eq:spin7fourformintro}
\begin{align} \label{eq:spin7fourform}
    \Psi=\; &\dd x^{1234} +\dd x^{1256} +\dd x^{1278} +\dd x^{1357} -\dd x^{1368} -\dd x^{1458} -\dd x^{1467} \nonumber \\
    & -\dd x^{2358} -\dd x^{2367} -\dd x^{2457} +\dd x^{2468} +\dd x^{3456} +\dd x^{3478} +\dd x^{5678} \, ,
\end{align}
where the $x^i$ denote the coordinates of the tori.
The combinations of {four} T-dualities that give a mirror map can be read off the terms of the four-form:
\begin{align} \label{eq:listofTdualities}
    \lbrace &(1, 2, 3, 4), (1, 2, 5, 6), (1, 2, 7, 8), (1, 3, 5, 7), (1, 3, 6, 8), (1, 4, 5, 8), (1, 4, 6, 7), \nonumber \\
    &(2, 3, 5, 8), (2, 3, 6, 7), (2, 4, 5, 7), (2, 4, 6, 8), (3, 4, 5, 6), (3, 4, 7, 8), (5, 6, 7, 8) \rbrace .
\end{align}

As shown in \cite{Braun:2018joh}, there are some Joyce orbifolds which also admit a GCS description, so a natural step for us is to consider what automorphisms of the diamond are generated by the mirror maps \eqref{eq:listofTdualities} in these particular orbifolds (see \cite{Chuang:2004th, Braun:2019lnn} for partial results). We thus consider Spin(7) orbifolds of the form $\mathbb{T}^8/\mathbb{Z}_2^4$ \cite{MR1383960}. The coordinates $x^i$ on the torus range from 0 to 1 and the action of the generators {$\alpha, \beta, \gamma, \delta$} of the quotient group are described in \cref{tab:discreteaction}. We focus on four particular orbifolds, described in \cref{tab:exampleslist}.

{\renewcommand{\arraystretch}{1.2}
\begin{table}[h]
\begin{center}
\begin{tabular}{ |c||c|c|c|c|c|c|c|c| } 
\hline
   & $x^1$ & $x^2$ & $x^3$ & $x^4$ & $x^5$ & $x^6$ & $x^7$ & $x^8$ \\
 \hline
 \hline
 $\alpha$ & $-$ & $-$ & $-$ & $-$ & $+$ & $+$ & $+$ & $+$ \\
 \hline
$\beta$ & $+$ & $+$ & $+$ & $+$ & $-$ & $-$ & $-$ & $-$ \\ 
\hline
$\gamma$ & $c_1-$ & $c_2-$ & $+$ & $+$ & $c_5-$ & $c_6-$ & $+$ & $+$ \\ 
\hline
$\delta$ & $d_1-$ & $+$ & $d_3-$ & $+$ & $d_5-$ & $+$ & $d_7-$ & $+$ \\
\hline
\end{tabular}
\end{center}
\caption{Action of $\mathbb{Z}^4_2$ on $\mathbb{T}^8$. We need to specify the parameters $c_j$ and $d_k$, which are allowed to take the values 0 or $\frac{1}{2}$. This is done in \cref{tab:exampleslist} for four different orbifolds. The $\pm$ entries correspond to a global $\pm$ sign action whereas $\frac{1}{2}-$ entries correspond to $x^i\mapsto-x^i+\frac{1}{2}$.}
\label{tab:discreteaction}
\end{table}}

{\renewcommand{\arraystretch}{1.3}
\begin{table}[h]
\begin{center}
\begin{tabular}{ |c||c|c|c|c|c|c|c|c| } 
\hline
   & $c_1$ & $c_2$ & $c_5$ & $c_6$ & $d_1$ & $d_3$ & $d_5$ & $d_7$ \\
 \hline
 \hline
 $I$ & $\frac{1}{2}$ & $\frac{1}{2}$ & $\frac{1}{2}$ & $\frac{1}{2}$ & $0$ & $\frac{1}{2}$ & $\frac{1}{2}$ & $\frac{1}{2}$ \\
 \hline
$II$ & $\frac{1}{2}$ & $0$ & $\frac{1}{2}$ & $0$ & $0$ & $\frac{1}{2}$ & $\frac{1}{2}$ & $\frac{1}{2}$ \\ 
\hline
$III$ & $\frac{1}{2}$ & $\frac{1}{2}$ & $\frac{1}{2}$ & $0$ & $0$ & $\frac{1}{2}$ & $\frac{1}{2}$ & $0$ \\ 
\hline
$IV$ & $\frac{1}{2}$ & $0$ & $\frac{1}{2}$ & $0$ & $0$ & $\frac{1}{2}$ & $\frac{1}{2}$ & $0$ \\
\hline
\end{tabular}
\end{center}
\caption{Coefficients for different orbifold examples (see \cref{tab:discreteaction}).}
\label{tab:exampleslist}
\end{table}}

As explained in \cite{Braun:2018joh}, all these orbifolds admit a GCS realization pulling them apart along the coordinate $x^3=t$. In this case, the external circle corresponds to the coordinate $x^4=\theta$.\footnote{Alternative choices for the coordinates $(t,\theta)$ are possible depending on the orbifold of \cref{tab:exampleslist} and we will comment on them later.}

From \cref{tab:dictionary} we know that the chiral algebra corresponding to each coordinate $x^i$ of the orbifold is just a free algebra $(\psi_i,j_i)$. This can be used to provide a free field realization of the diamond of algebras associated to the GCS decomposition of the orbifolds. The top algebra $\text{Od}_3\oplus \text{Fr}^1_t\oplus \text{Fr}^1_\theta$ is given as follows. We have $\text{Fr}^1_t=\text{Fr}^1_3$ and $\text{Fr}^1_\theta=\text{Fr}^1_4$. The generators of $\text{Od}_3$ are given by
\begin{equation} \label{eq:Od3freefields}
\begin{split}
J^3_3 &= \, \normord{\psi_1\,\psi_2} +\normord{\psi_5\,\psi_6} +\normord{\psi_7\,\psi_8} \, , \\
A_3 &= \, \normord{\psi_1\,\psi_5\,\psi_8} +\normord{\psi_1\,\psi_6\,\psi_7} +\normord{\psi_2\,\psi_5\,\psi_7} -\normord{\psi_2\,\psi_6\,\psi_8} \, , \\
B_3 &= -\normord{\psi_1\,\psi_5\,\psi_7} +\normord{\psi_1\,\psi_6\,\psi_8} +\normord{\psi_2\,\psi_5\,\psi_8} +\normord{\psi_2\,\psi_6\,\psi_7} \, ,
\end{split}
\end{equation}
with the Virasoro generators $(T_3,G_3)$ given by the standard expressions \eqref{eq:virasorostandard} combining the coordinates 1, 2, 5, 6, 7 and 8. The realizations of the lateral tip algebras $\SVseven\oplus\text{Fr}^1$ and $\text{Od}_4$ are obtained directly from \eqref{eq:g2frsubod3fr2} and \eqref{eq:od4subod3fr2} respectively using the realization \eqref{eq:Od3freefields}. The bottom $\SVeight$ algebra is obtained either through \eqref{eq:spin7subg2fr} or \eqref{eq:spin7subod4} and it can be checked that the operator $-X_8$ matches the geometric expectation from \eqref{eq:spin7fourform}.

The composition of four T-duality automorphisms in any of the directions of \eqref{eq:listofTdualities} provides an automorphism that leaves invariant the diamond realization we have obtained. Each of the 14 different automorphisms therefore reduces to one of the possibilities we described in \cref{tab:automorphisms}.

\begin{itemize}
    \item When no T-dualities are applied to $t$ or $\theta$, the automorphism corresponds to the identity $\textbf{A}_0$. This is the case for $\lbrace (1, 2, 5, 6), (1, 2, 7, 8), (5, 6, 7, 8) \rbrace$.
    \item When T-dualities are applied to both $t$ and $\theta$, the automorphism corresponds to $\textbf{A}_1$. This is the case for $\lbrace (1, 2, 3, 4), (3, 4, 5, 6), (3, 4, 7, 8)\rbrace$.
    \item When T-duality is applied to $t$ but not to $\theta$, the automorphism corresponds to $\textbf{A}_2$. This is the case for $\lbrace (1, 3, 5, 7), (1, 3, 6, 8), (2, 3, 5, 8), (2, 3, 6, 7)\rbrace$.
    \item When T-duality is applied to $\theta$ but not to $t$, the automorphism corresponds to $\textbf{A}_3$. This is the case for $\lbrace (1, 4, 5, 8), (1, 4, 6, 7), (2, 4, 5, 7), (2, 4, 6, 8)\rbrace$.
\end{itemize}

\noindent This shows that all our candidates for mirror automorphisms explicitly appear in these four orbifold examples. It is moreover satisfying to examine the geometric implementation of these maps as in \cite{Braun:2017ryx, Braun:2017csz}, since they agree with taking localized mirrors of the components of the GCS decomposition as suggested by \cref{tab:automorphisms}. We do this presently.

\bigskip

Consider the four different $\textbf{A}_3$ mirror maps. In the ACyl G$_2$ end of the construction the external circle $\Sc^1_\theta$ is T-dualized and the remaining $\mathbb{T}^3$ fibre is in all cases calibrated by the associative three-form $\varphi.$\footnote{Let us illustrate this for the case where we apply T-dualities in the $(1, 4, 5, 8)$ directions. The coordinate $x^4$ corresponds to the external circle, so the $\mathbb{T}^3$ fibre in the ACyl G$_2$ end is given by the coordinates $x^1$, $x^5$ and $x^8$. It can be checked that the associative form $\varphi$ for this GCS realization has a $\dd x^{158}$ term. This means the restriction of $\varphi$ to the fibre is the volume form of the $\mathbb{T}^3$ and the fibre is calibrated by $\varphi$.} As explained earlier, this corresponds to a $\mathcal{T}^3$ mirror map in the G$_2$ manifold, which manifests in the algebra as a \textbf{GK} automorphism. In the neck region, the $\mathbb{T}^3$ fibre within the CY$_3$ is always calibrated by $\Re\Omega_3$ and it is therefore special Lagrangian. This means we are performing a mirror symmetry in the CY$_3$, which corresponds to an automorphism \textbf{M} in the chiral algebra. Finally, for the ACyl CY$_4$ end the whole $\mathbb{T}^4$ fibre is found to be calibrated by $\Re\Omega_4$, thus we have a mirror symmetry on the CY$_4$ and an \textbf{M} automorphism in the algebra.

Now let us explore the four automorphisms which reduce to $\textbf{A}_2$. Here in the ACyl G$_2$ end the $\mathbb{T}^4$ fibre is always calibrated by the coassociative four-form $\psi$. This corresponds to a $\mathcal{T}^4$ mirror map in the G$_2$ manifold, which reduces to the identity in the chiral algebra. The neck region and the ACyl CY$_4$ end are similar to the $\textbf{A}_3$ case: we have T-duality in the $t$ direction and a $\mathbb{T}^3$ fibre calibrated by $\Im\Omega_3$ in the CY$_3$, whereas the whole $\mathbb{T}^4$ fibre is calibrated by $\Re\Omega_4$ in the CY$_4$. This means we expect mirror symmetry on the Calabi--Yau manifolds, which produces \textbf{M} automorphisms.

A general proposal to construct GCS mirror manifolds was given in \cite{Braun:2019lnn}: the idea is to apply mirror maps to the open ends of the construction and glue the manifolds back together. The mirror of the ACyl G$_2$ is obtained via a $\mathcal{T}^3$ map, and for the orbifolds presented this is precisely what $\textbf{A}_3$ describes. Our discussion above suggests the existence of an alternative mirror construction, based on the $\textbf{A}_2$ automorphism, where the map employed to obtain the ACyl G$_2$ mirror is $\mathcal{T}^4$. We discuss this further in \cref{sec:NewMirrors}.

\bigskip

We now turn our attention to the three maps producing the $\textbf{A}_1$ automorphism. In the ACyl G$_2$ end we dualize the external circle and an associative $\mathbb{T}^3$ fibration, so we have a $\mathcal{T}^3$ mirror map and a \textbf{GK} automorphism. On the neck CY$_3$ we have a $\mathbb{T}^2$ fibration which is calibrated by the Hermitian form $\omega_3$. This means that this fibration is just a complex submanifold and is not supersymmetric, therefore these T-dualities do not correspond to a mirror symmetry on the CY$_3$ and the associated chiral algebra automorphism is just the identity. For the ACyl CY$_4$ end, the $\mathbb{T}^4$ fibration is calibrated by $\frac{1}{2}\,\omega_4\wedge\omega_4$ so again we find a complex submanifold and not a supersymmetric fibration, resulting in an identity automorphism.

Finally, let us study the three maps corresponding to $\textbf{A}_0$. The $\mathbb{T}^4$ fibre in the ACyl G$_2$ end turns out to be a coassociative fibration for the three maps. This means that even though we see an identity automorphism in the algebra, there is a non-trivial $\mathcal{T}^4$ mirror map acting on the G$_2$ manifold. The $\mathbb{T}^4$ fibre is calibrated by $\frac{1}{2}\,\omega_3\wedge\omega_3$ in the neck CY$_3$ and by $\frac{1}{2}\,\omega_4\wedge\omega_4$ in the ACyl CY$_4$, therefore it corresponds to complex submanifolds and the associated automorphisms are the identity in both cases.

Once again the geometric description is consistent with the automorphisms of \cref{tab:automorphisms}. Moreover, the interpretation of these mirror maps is clear in these examples: we construct the mirror orbifold by applying a mirror map to the ACyl G$_2$ end of the construction whereas no mirror map is applied to the ACyl CY$_4$. When the $\mathcal{T}^3$ mirror map is applied we obtain an $\textbf{A}_1$ automorphism in the chiral algebra, whereas when the $\mathcal{T}^4$ map is applied the automorphism $\textbf{A}_0$ is obtained. We return to these in \cref{sec:NewMirrors}.

\bigskip

Our choice of pulling the Joyce orbifolds along the $t=x^3$ direction crucially influenced the previous discussion, yet it is somewhat arbitrary. Some Joyce orbifolds admit more than one GCS decomposition. Let us briefly consider what changes if we stretch the orbifold along the $x^6$ direction. This can be done for the orbifold $I$, with the external circle given by the coordinate $x^8$.\footnote{Note that there are yet more possible GCS structures: orbifolds $I$ and $II$ can be pulled apart along the coordinate $x^7$, with the external circle in the coordinate $x^8$ and orbifolds $I$ and $III$ can be pulled along $x^2$ with a circle in $x^4$.} A realization of the diamond of algebras can be obtained and upon studying the action of T-dualities on it we find:

\begin{itemize}
    \item $\textbf{A}_0$ is obtained from $\lbrace (1, 2, 3, 4), (1, 3, 5, 7), (2, 4, 5, 7) \rbrace$.
    \item $\textbf{A}_1$ is obtained from $\lbrace (1, 3, 6, 8), (2, 4, 6, 8), (5, 6, 7, 8)\rbrace$.
    \item$\textbf{A}_2$ is obtained from $\lbrace (1, 2, 5, 6), (1, 4, 6, 7), (2, 3, 6, 7),  (3, 4, 5, 6)\rbrace$.
    \item$\textbf{A}_3$ is obtained from $\lbrace (1, 2, 7, 8), (1, 4, 5, 8), (2, 3, 5, 8), (3, 4, 7, 8)\rbrace$.
\end{itemize}

Note that most of the T-duality combinations are now assigned a different $\textbf{A}_i$. This illustrates that all automorphisms $\textbf{A}_i$ should equally be considered as mirror maps, since they may be exchanged into each other when more than one GCS decomposition is available.

To conclude, we mention for completeness the existence of combinations of T-dualities not included in \eqref{eq:listofTdualities} preserving the Cayley four-form in these examples. The only possibilities are the trivial map, associated to $\textbf{A}_0$, and performing T-dualities along the eight coordinates, associated to $\textbf{A}_1$. These together with the maps \eqref{eq:listofTdualities} form a $\mathbb{Z}_2^4$ group under composition.

\subsubsection{New mirror maps} \label{sec:NewMirrors}

It is satisfying that T-dualities in Joyce orbifolds lead to recognizable mirror maps being applied to components of the GCS decomposition. 
Note that the interpretation of the diamond automorphisms $\mathbf{A}_i$ remains valid for more general GCS manifolds which do not present a description as a Joyce orbifold.
Extrapolating, this suggests that the mirror maps exist even in the general GCS setting.

We explained above how the GCS mirror construction proposed in \cite{Braun:2019lnn} corresponds to an $\textbf{A}_3$ automorphism in the associated diamond. We propose new methods to obtain mirrors of GCS manifolds based on the $\textbf{A}_2$, $\textbf{A}_1$ and $\textbf{A}_0$ automorphisms.

We begin with the $\textbf{A}_2$ construction. Consider a GCS manifold with a 4-tori fibration which is supersymmetric all over the manifold and which does not involve the external circle in the ACyl G$_2$ end. We can separate the two ends of the construction and, by dualizing the supersymmetric fibre, take a mirror map of type $\mathcal{T}^4$ in the ACyl G$_2$ and a mirror map in the ACyl CY$_4$. A mirror GCS manifold is obtained by gluing the ends back together after the mirror maps are applied.

The construction of mirrors based on the $\textbf{A}_1$ and $\textbf{A}_0$ automorphisms is different. Consider in this case a GCS manifold with a 4-tori fibration which is supersymmetric for the ACyl G$_2$ manifold but not for the ACyl CY$_4$. We can then construct a mirror GCS manifold by separating the two ends of the construction, taking a mirror map in the ACyl G$_2$ and four T-dualities in the ACyl CY$_4$, and gluing the ends back together. When the fibration includes the external circle in the ACyl G$_2$ end, the mirror map is of type $\mathcal{T}^3$ and the construction corresponds to an automorphism $\textbf{A}_1$. When the external circle is not included, the mirror map is $\mathcal{T}^4$ and the construction corresponds to $\textbf{A}_0$.

Note that in the $\textbf{A}_2$ construction we apply mirror maps to both ends of the GCS manifold, whereas for $\textbf{A}_1$ and $\textbf{A}_0$ this only occurs for the ACyl G$_2$ manifold. A similar phenomenon was observed in \cite{Braun:2017csz} for the mirror constructions of TCS manifolds associated to $\mathcal{T}^4$ and $\mathcal{T}^3$.
It is natural to ask if we could construct mirror GCS manifolds by applying a mirror map only to the ACyl CY$_4$ end of the construction and not to the ACyl G$_2$. Such a map would have to be associated to an $\textbf{A}_2$ automorphism with a non-coassociative fibre in the ACyl G$_2$. We do not find such a map amongst the examples we studied.

There is another piece of evidence for our GCS mirror symmetry proposal. One of the main arguments provided in \cite{Braun:2019lnn} in support of their GCS mirror symmetry construction ($\mathbf{A}_3$ in our notation) was the invariance of the dimension of the sigma model moduli space under the mirror maps, at least under some simplifying assumptions,
\begin{equation} \label{eq:modulispacedim}
    b^2(M)+b^4_-(M)+1=b^2(M^\vee)+b^4_-(M^\vee)+1 \, ,
\end{equation}
where $b^i$ denotes the Betti number of dimension $i$, the subscript $-$ indicates a restriction to anti-self dual forms, $M$ denotes the GCS manifold and $M^\vee$, its mirror. The proof of \eqref{eq:modulispacedim} detailed in \cite[Sect.~4.2]{Braun:2019lnn} {works as follows: as a first step the Betti numbers appearing in the LHS of \eqref{eq:modulispacedim} are rewritten in terms of cohomology groups of the open ends and the neck region by a Mayer-Vietoris argument. One then studies the effect on these cohomologies of applying mirror maps on both open ends of $M$, eventually reaching the RHS of \eqref{eq:modulispacedim}, as we now describe.

Consider the ACyl CY$_4$ end. By gluing together two copies of this end one obtains a compact CY$_4$. The mirror map of the ACyl CY$_4$ extends to a mirror map of the compact CY$_4$ which changes the Hodge numbers in the usual way: $h^{i,j}(\text{CY}_4)=h^{4-i,4-j}(\text{CY}_4^\vee)$. These numbers are related to those of the ACyl CY$_4$ and the neck region by another Mayer-Vietoris sequence, and one deduces from here restrictions on how the cohomologies change under mirror symmetry. It turns out the LHS of \eqref{eq:modulispacedim} is invariant under these changes.

An analogous argument for the ACyl G$_2$ end using the fact that G$_2$ mirror symmetry preserves the combination $b^2+b^3$ shows that the LHS of \eqref{eq:modulispacedim} is also preserved by a mirror symmetry on the ACyl G$_2$ end, showing that the equality \eqref{eq:modulispacedim} holds. This proof works for both $\mathcal{T}^3$ and $\mathcal{T}^4$ mirror maps, so in particular it also holds for the $\mathbf{A}_2$ construction.}

{Moreover, the invariance of the LHS of \eqref{eq:modulispacedim} under mirror symmetry on the ACyl CY$_4$ end is independent from the ACyl G$_2$ end and vice versa. This means that if a mirror map is applied to just one of the open ends, \eqref{eq:modulispacedim} remains valid. This is the case of the proposed $\mathbf{A}_0$ and $\mathbf{A}_1$ constructions, where the mirror map is applied only to the ACyl G$_2$ manifold, leaving the cohomologies of the ACyl CY$_4$ intact. Therefore, the dimension of the moduli space is preserved by all our proposed mirror constructions. Note also that \eqref{eq:modulispacedim} would still hold in a construction where a mirror map is applied only to the ACyl CY$_4$ end.}

\section{Are Connected Sums Generic?} \label{sec:Num}

In this section we compare the algebra at the bottom of the diamond diagram---either \cref{fig:TCS}(b) for (E)TCS or \cref{fig:GCS}(b) for GCS---and the intersection of the algebras on the lateral tips. By construction the bottom algebra is contained in the intersection algebra reflecting the fact that the manifold has holonomy G$_2$ or Spin(7) respectively.

Suppose this inclusion was strict, say for the GCS diagram for definiteness. This would mean there are fields in the intersection algebra that do not appear in the bottom algebra, leading to additional chiral symmetries. As a result, we would conclude that GCS manifolds have a set of symmetries larger than a generic manifold of holonomy Spin(7), whose chiral algebra is simply $\SVeight$. We postulate however that the opposite is true, i.e.\ that the intersection algebra perfectly agrees with the bottom algebra of the diamond both for (E)TCS and GCS. This would mean that, at least from a chiral algebra viewpoint, these constructions are representative of generic manifolds with holonomy G$_2$ or Spin(7).

The vacuum module character of a chiral algebra is defined as
\begin{equation}
    \chi=\tr\left(q^{L_0-\frac{c}{24}}\right)\, ,
\end{equation}
where the trace is taken over the vacuum module. The number of independent fields at level $h$ of the chiral algebra can be read off from the expansion of this character in powers of $q=e^{2\pi i \tau}$, as the coefficient of $q^{h-c/24}$.

An analytic proof of our proposal would require a good grasp of the vacuum module characters not only of the $\SVseven$ and $\SVeight$ algebras but also of the intersections involved. Unfortunately the latter seem very challenging to obtain, and only the character of $\SVeight$ is known analytically \cite{Benjamin:2014kna}.

We therefore rely on numerical checks to support the proposal. For each chiral subalgebra appearing in the diamond, the corresponding character expansion is obtained by listing and counting the fields in the subalgebra level by level. For the intersection of the lateral tip subalgebras, we search at each level for all linear combinations of fields which are contained in both algebras. The number of linearly independent combinations at each level provides the coefficients in the character expansion of the intersection algebras.

The numerical manipulations become more involved as we look into higher levels due to the dramatic increase in the number of null fields. Indeed the top algebras of both diagrams have singular fields whose descendants are null and must be quotiented out at each subsequent levels. When listing the fields in a subalgebra or an intersection, one has to ensure that this quotient is taken into account.

Once this is achieved we obtain vacuum character expansions of the different algebras and these can be compared to test the proposal. We verify the agreement level by level in the various vacuum modules.

\subsection{The (E)TCS case}

The precise embedding of $(\text{Od}_3\oplus \text{Fr}^1)_-$ inside $\text{Od}_2\oplus \text{Fr}^3$ depends on the gluing angle of the ETCS construction we are considering. We have a circle worth of embeddings where any two of them can be mapped to each other by an automorphism leaving the underlying $\SVseven$ invariant. We first perform the computation for the TCS embedding corresponding to a gluing angle of $\vartheta=\pi/2$.

As we mentioned earlier, an analytic expression for the character of the $\SVseven$ algebra is unfortunately not available in the literature so far. We have however managed to compute it numerically up to level 9, and we obtained
\begin{align} \label{eq:svG2character}
    \chi(\SVseven)=&\,q^{-\frac{7}{16}}\bigg(1+2q^{3/2}+3 q^2+3 q^{5/2}+4 q^3+8 q^{7/2}+12 q^4+14 q^{9/2}+18 q^5+29 q^{11/2}\nonumber \\
    &+42 q^{6}+51 q^{13/2}+66 q^{7}+96 q^{15/2}+129 q^{8}+160 q^{17/2}+207 q^{9}+
    \ldots
    \bigg) \, .
\end{align}
Each coefficient in the expansion \eqref{eq:svG2character} corresponds to the number of independent fields of $\SVseven$ at the level given by the corresponding power of $q$. We have been able to verify numerically up to level 5 that these agree with the number of independent fields of $(\text{Od}_3\oplus \text{Fr})_+\cap(\text{Od}_3\oplus \text{Fr})_-$, thus giving evidence for our proposal in this case.

For the case of arbitrary gluing angle $\vartheta\in (0,\pi)$, the nontrivial trigonometric functions involved make the computation harder and we have only managed to verify the equality numerically up to level 3.

\subsection{The GCS case}

Now consider the GCS diagram, \cref{fig:GCS}(b). In this case we want to check that the intersection of the $\text{Od}_4$ and $\SVseven \oplus \text{Fr}^1$ subalgebras of $\text{Od}_3\oplus \text{Fr}^2$ is precisely the $\SVeight$ subalgebra.

The vacuum character of $\SVeight$ was recently computed in \cite{Benjamin:2014kna} and it is given by
\begin{align}
    \chi(\SVeight)=q^{-\frac{1}{2}}{\mathcal P}(\tau)\Bigg(1-\sum_{k=0}^\infty \Big(q^{\frac{15}{2}k^2+4k+\frac{1}{2}} +\frac{q^{\frac{15}{2}k^2+2k+\frac{1}{2}}}{1+q^\frac{6k+1}{2}} -\frac{q^{\frac{15}{2}k^2+7k+2}}{1+q^\frac{6k+3}{2}}- \nonumber
    \\
    -q^{\frac{15}{2}k^2+14k+\frac{13}{2}} +\frac{q^{\frac{15}{2}k^2+14k+\frac{11}{2}}}{1+q^\frac{6k+3}{2}} -\frac{q^{\frac{15}{2}k^2+19k+11}}{1+q^\frac{6k+5}{2}} \Big)\Bigg) \, ,
\end{align}
where
\begin{equation}
    {\mathcal P}(\tau)=\prod_{k=1}^\infty\left( \frac{1+q^{k-1/2}}{1-q^k} \right)^2 \, .
\end{equation}
Expanding in powers of $q$, the first few terms are
\begin{align} \label{eq:svSpin7character}
    \chi(\SVeight)=q^{-\frac{1}{2}}&\bigg(1+q^{3/2}+2 q^2+2 q^{5/2}+2 q^3+4 q^{7/2}+7 q^4 \nonumber \\
    &\qquad\qquad\qquad+8 q^{9/2}+9 q^5+14 q^{11/2}+21 q^{6}
    +\ldots
    \bigg) \, .
\end{align}
We have checked that the number of independent fields in the intersection algebra matches this expansion precisely up to level 6. Since we already know that the $\SVeight$ algebra is contained in the intersection, this is enough to show the equality between these algebras up to level 6 and provides a check to our proposal.

\section{Conclusion}

In this chapter we have explored the relationship between the geometry of connected sum manifolds $M$ of holonomies G$_2$ and Spin(7), and the chiral algebra of the associated sigma model. Starting from the geometric description of $M$ in terms of open patches, we have argued for a diamond of algebra inclusions in the worldsheet theory.
We have shown the validity of the diamond for Extra Twisted Connected Sum (ETCS) G$_2$-manifolds, and Generalized Connected Sum (GCS) Spin(7)-manifolds. We have checked numerically the agreement, at leading orders, between the Shatashvili-Vafa algebra at the bottom of the diamond and the intersection of the algebras at the lateral tips of the diamond, which suggests that these constructions provide generic special holonomy manifolds.

Additionally, we have {described all}
the {possible} automorphisms fixing the GCS diamond and we have interpreted them in terms of GCS mirror symmetry maps. {In the case of four different Joyce orbifolds, we have shown that every mirror map coming from T-dualities corresponds to one of these GCS mirror maps}.
This has lead us to propose new constructions of GCS mirror manifolds.

Our results set the ground for different future directions. A natural next step would be to study the diamond of algebras for other manifolds constructed by gluing two building blocks. Some Calabi--Yau manifolds can be obtained by this procedure; an example is the Schoen Calabi--Yau 3-fold \cite{Schoen1988}, see \cite{Braun:2017uku}. Further examples of Calabi--Yau 4-folds and G$_2$ manifolds built from gluing two ACyl copies can be found in \cite{Braun:2018joh}.

In our analysis of mirror maps for Spin(7) Joyce orbifolds we did not find any
that was acting as mirror symmetry on the ACyl CY$_4$ end while the ACyl G$_2$ end was dualized along a non-coassociative fibre. The existence of such a map is an intriguing possibility and one could try to look for an explicit realization in GCS manifolds beyond Joyce orbifolds such as the ones described in \cite{Braun:2018joh}. We know it would have to correspond to an $\textbf{A}_2$ automorphism.

It would also be interesting to perform a thorough study of the automorphisms available for the ETCS diamond in the same spirit as \cref{sec:Spin7Aut}. We could then try to give an interpretation of these as mirror maps for ETCS manifolds and look for realizations in explicit examples. In particular for the TCS case we should recover the maps $\mathcal{T}^3$ and $\mathcal{T}^4$ of \cite{Acharya:1997rh, Gaberdiel:2004vx}.

The general interpretation of our results given in \cref{sec:Idea} was in terms of worldsheet symmetries of local patches of the sigma model target space. Although this was sufficient for us, it is hard to miss the similarities with the so-called chiral de Rham complex, which is a sheaf of vertex operator algebras \cite{Malikov:1998dw}. In the context of sigma models, it is thought to describe localized fluctuations in the target space geometry, much in the spirit we have advocated \cite{Witten:2005px}. It would be interesting to understand precisely this connection.

Another tempting direction is to attempt to find modular invariant partition functions for exceptional holonomy manifolds by capitalizing on the variety of Calabi--Yau vacua under worldsheet control.
It would be particularly interesting to search for rational Od$_n$ CFTs which could be used as the ``theory on the neck region'' either for G$_2$ ($n=2$) or Spin(7) ($n=3$) backgrounds. They would also have to feature the corresponding diamond of subsymmetries we have presented, which constrains the possible choices.

\section*{Appendices}
\addcontentsline{toc}{section}{Appendices}

\setcounter{section}{0}
\renewcommand\theHsection{\thechapter.\Alph{section}}
\renewcommand\thesection{\thechapter.\Alph{section}}

\section{The Odake algebras}
\label{app:odakeOPEs}

In this appendix we collect our conventions for the OPEs of the Odake algebras \cite{Odake:1988bh} that we use in this chapter. We mostly follow \cite{Fiset:2018huv}, although we have added some subscripts to the different generators in order to distinguish them, and the operator $J^3$ was called $J$ in \cite{Fiset:2018huv}.
\begin{align*}
T_n(z)T_n(w)&=\frac{3\,n}{(z-w)^4}+\frac{2
   }{(z-w)^2}\,T_n(w)+\frac{1}{z-w}\,\partial T_n(w)+\cdots,\\
T_n(z)G_n(w)&=\frac{3}{2\, (z-w)^2}\, G_n(w)+\frac{1}{z-w}\,\partial G_n(w)+\cdots,\\
T_n(z)G^3_n(w)&=\frac{3}{2\, (z-w)^2}\, G^3_n(w)+\frac{1}{z-w}\, \partial G^3_n(w)+\cdots,\\
T_n(z)J^3_n(w)&=\frac{1}{(z-w)^2}\,J^3_n(w)+\frac{1}{z-w}\,\partial J^3_n(w)+\cdots,\\
G_n(z)G_n(w)&=\frac{4\,n}{(z-w)^3}+\frac{2}{z-w}\, T(w)+\cdots,\\
G_n(z)G^3_n(w)&=\frac{2}{(z-w)^2}\, J^3_n(w)+\frac{1}{z-w}\, \partial J^3_n(w)+\cdots,\\
G_n(z)J^3_n(w)&=\frac{1}{z-w}\, G^3_n(w)+\cdots,\\
G^3_n(z)G^3_n(w)&=\frac{4\, n}{(z-w)^3}+\frac{2}{z-w}\, T(w)+\cdots,\\
G^3_n(z)J^3_n(w)&=-\,\frac{1}{z-w}\,G_n(w)+\cdots,\\
J^3_n(z)J^3_n(w)&=-\,\frac{2\, n}{(z-w)^2}+\cdots,\\
T_n(z)A_n(w)&=\frac{n}{2\, (z-w)^2}\, A_n(w)+\frac{1}{z-w}\,\partial A_n(w)+\cdots,\\
T_n(z)B_n(w)&=\frac{n}{2\, (z-w)^2}\, B_n(w)+\frac{1}{z-w}\,\partial B_n(w)+\cdots,\\
T_n(z)C_n(w)&=\frac{(n+1)}{2\, (z-w)^2}\, C_n(w)+\frac{1}{z-w}\,\partial C_n(w)+\cdots,\\
T_n(z)D_n(w)&=\frac{(n+1) }{2\, (z-w)^2}\, D_n(w)+\frac{1}{z-w}\partial D_n(w)+\cdots,\\
G_n(z)A_n(w)&=\frac{1}{z-w}\,C_n(w)+\cdots,\\
G_n(z)B_n(w)&=\frac{1}{z-w}\,D_n(w)+\cdots,\\
G_n(z)C_n(w)&=\frac{n}{(z-w)^2}\, A_n(w)+\frac{1}{z-w}\,\partial A_n(w)+\cdots,\\
G_n(z)D_n(w)&=\frac{n}{(z-w)^2}\, B_n(w)+\frac{1}{z-w}\,\partial B_n(w)+\cdots,
\end{align*}
\begin{align*}
G^3_n(z)A_n(w)&=-\,\frac{1}{z-w}\,D_n(w)+\cdots,\\
G^3_n(z)B_n(w)&=\frac{1}{z-w}\,C_n(w)+\cdots,\\
G^3_n(z)C_n(w)&=\frac{n}{(z-w)^2}\, B_n(w)+\frac{1}{z-w}\, \partial B_n(w)+\cdots,\\
G^3_n(z)D_n(w)&=-\,\frac{n}{(z-w)^2}\, A_n(w)-\frac{1}{z-w}\,\partial A_n(w)+\cdots,\\
J^3_n(z)A_n(w)&=-\,\frac{n}{z-w}\, B_n(w)+\cdots,\\
J^3_n(z)B_n(w)&=\frac{n}{z-w}\, A_n(w)+\cdots,\\
J^3_n(z)C_n(w)&=-\,\frac{(n-1)}{z-w}\, D_n(w)+\cdots,\\
J^3_n(z)D_n(w)&=\frac{(n-1)}{z-w}\, C_n(w)+\cdots.
\end{align*}
The OPEs between the two superprimary multiplets $(A_n,C_n)$ and $(B_n,D_n)$ depend on $n$ in a non-trivial way. The explicit OPEs for $n=2$ are
\begin{align*}
A_2(z)A_2(w)&=-\,\frac{2}{(z-w)^2}+\cdots,\\
A_2(z)B_2(w)&=-\,\frac{2}{(z-w)}\,J_2(w)+\cdots,\\
A_2(z)C_2(w)&=\frac{1}{(z-w)}\,G_2(w)+\cdots,\\
A_2(z)D_2(w)&=-\,\frac{1}{(z-w)}\,G^3_2(w)+\cdots,\\
B_2(z)B_2(w)&=-\,\frac{2}{(z-w)^2}+\cdots,\\
B_2(z)C_2(w)&=\frac{1}{(z-w)}\,G^3_2(w)+\cdots,\\
B_2(z)D_2(w)&=\frac{1}{(z-w)}\,G_2(w)+\cdots,\\
C_2(z)C_2(w)&=\frac{4}{(z-w)^3}+\frac{2}{(z-w)}\,T_2(w)+\cdots,\\
C_2(z)D_2(w)&=\frac{2}{(z-w)^2}\,J_2(w)\,+\frac{1}{(z-w)}\,\partial J_2(w)+\cdots,\\
D_2(z)D_2(w)&=\frac{4}{(z-w)^3}\,+\frac{2}{(z-w)}\,T_2(w)+\cdots.
\end{align*}
The OPEs for the $n=3$ case are
\begin{align*}
A_3(z)A_3(w)&=-\,\frac{4}{(z-w)^3}+\frac{2}{z-w}\normord{J_3\, J_3}(w)+\cdots,\\
A_3(z)B_3(w)&=-\,\frac{4}{(z-w)^2}\,J_3(w)-\frac{2}{z-w}\,\partial J_3(w)+\cdots,\\
A_3(z)C_3(w)&=-\,\frac{2}{(z-w)^2}\,G_3(w)-\frac{2}{z-w}\normord{J_3\,G^3_3}(w)+\cdots,\\
A_3(z)D_3(w)&=\frac{2}{(z-w)^2}\,G^3_3(w)-\frac{2}{z-w}\normord{J_3\,G_3}(w)+\cdots,\\
B_3(z)B_3(w)&=-\,\frac{4}{(z-w)^3}+\frac{2}{z-w}\normord{J_3\,J_3}(w)+\cdots,\\
B_3(z)C_3(w)&=-\,\frac{2}{(z-w)^2}\,G^3_3(w)+\frac{2}{z-w}\normord{J_3\,G_3}(w)+\cdots,\\
B_3(z)D_3(w)&=-\,\frac{2}{(z-w)^2}\,G_3(w)-\frac{2}{z-w}\normord{J_3\,G^3_3}(w)+\cdots,\\
C_3(z)C_3(w)&=-\,\frac{12}{(z-w)^4}+\frac{1}{(z-w)^2}\left(2 \normord{J_3\,J_3}-\,4\, T_3\right)(w)+\\
&\quad+\frac{2}{z-w}\left(\normord{\partial J_3\,J_3}-\,\partial T_3\right)(w)+\cdots,\\
C_3(z)D_3(w)&=-\,\frac{8}{(z-w)^3}\,J_3(w)-\frac{4}{(z-w)^2}\,\partial J_3(w)+\\
&\quad+\frac{1}{z-w}\left(2\normord{G_3\, G^3_3}-\,4 \normord{T_3\,J_3}\right)(w)+\cdots,\\
D_3(z)D_3(w)&=-\,\frac{12}{(z-w)^4}+\frac{1}{(z-w)^2}\left(2 \normord{J_3\,J_3}-\,4\, T_3\right)(w)+\\
&\quad+\frac{2}{z-w}\left(\normord{\partial J_3\,J_3}-\,\partial T_3\right)(w)+\cdots.
\end{align*}
The OPEs for $n=4$ can be found in \cite{Figueroa-OFarrill:1996tnk}, although the reference contains some sign errors that we have corrected below
\begin{align*}
A_4(z)A_4(w)&=\frac{8}{(z-w)^4}-\frac{4}{(z-w)^2}\normord{J^3_4\,J^3_4}(w)-\frac{4}{z-w}\normord{\partial J^3_4\,J^3_4}(w)+\cdots,\\
A_4(z)B_4(w)&=\frac{8}{(z-w)^3}\, J^3_4(w)+\frac{4}{(z-w)^2}\,\partial J^3_4(w)+\\
&\quad +\frac{4}{3\,(z-w)}\left(-\normord{J^3_4\,J^3_4\,J^3_4}+\,\partial\partial J^3_4\right)(w)+\cdots,
\end{align*}
\begin{align*}
A_4(z)C_4(w)&=-\,\frac{4}{(z-w)^3}\,G_4(w)-\frac{4}{(z-w)^2}\left(\normord{G^3_4\, J^3_4}+\,\partial G_4\right)(w)+\\
&\quad+\frac{1}{z-w}\left(2\normord{G_4\,J^3_4\,J^3_4}-\,2\normord{G^3_4\, \partial J^3_4}-\,4\normord{\partial G^3_4\, J^3_4}-\,2\,\partial\partial G_4\right)(w)+\cdots,\\
A_4(z)D_4(w)&=\frac{4}{(z-w)^3}\, G^3_4(w)-\frac{4}{(z-w)^2}\left(\normord{G_4\, J^3_4}-\,\partial G^3_4\right)(w)+\\
&~~+\frac{1}{z-w}\left(-\,2\normord{G^3_4\,J^3_4\,J^3_4}-\,2\normord{G_4\, \partial J^3_4}-\,4\normord{\partial G_4\, J^3_4}+2\,\partial\partial G^3_4\right)(w)+\cdots,\\
B_4(z)B_4(w)&=\frac{8}{(z-w)^4}-\frac{4}{(z-w)^2}\normord{J^3_4\,J^3_4}(w)-\frac{4}{z-w}\normord{\partial J^3_4\,J^3_4}(w)+\cdots,\\
B_4(z)C_4(w)&=-\,\frac{4}{(z-w)^3}\,G^3_4(w)+\frac{4}{(z-w)^2}\left(\normord{G_4\, J^3_4}-\,\partial G^3_4\right)(w)+\\
&\quad+\frac{1}{z-w}\left(2\normord{G^3_4\,J^3_4\,J^3_4}+\,2\normord{G_4\, \partial J^3_4}+\,4\normord{\partial G_4\, J^3_4}-\,2\,\partial\partial G^3_4\right)(w)+\cdots,\\
B_4(z)D_4(w)&=-\,\frac{4}{(z-w)^3}\, G_4(w)-\frac{4}{(z-w)^2}\left(\normord{G^3_4\, J^3_4}+\,\partial G_4\right)(w)+\\
&\quad+\frac{1}{z-w}\left(2\normord{G_4\,J^3_4\,J^3_4}-\,2\normord{G^3_4\, \partial J^3_4}-\,4\normord{\partial G^3_4\, J^3_4}-2\,\partial\partial G_4\right)(w)+\cdots,\\
C_4(z)C_4(w)&=-\,\frac{32}{(z-w)^5}+\frac{8}{(z-w)^3}\left(\normord{J^3_4\,J^3_4}-\,T_4\right)(w)+\\
&\quad+\frac{1}{(z-w)^2}\left(8 \normord{\partial J^3_4\,J^3_4}-\,4\, \partial T_4\right)(w)+\\
&\quad+\frac{1}{z-w}\big(-4\normord{G_4\,G^3_4\,J^3_4}-\,2\normord{G_4\,\partial G_4}-\,2\normord{G^3_4\partial G^3_4}+\\
&\quad\qquad\qquad+2\normord{\partial J^3_4\,\partial J^3_4}+\,4\normord{T_4\,J^3_4\,J^3_4}\big)(w)+\cdots,\\
C_4(z)D_4(w)&=-\,\frac{24}{(z-w)^4}\, J^3_4(w)-\frac{12}{(z-w)^3}\,\partial J^3_4(w)+\\
&\quad+\frac{1}{(z-w)^2}\left(4\normord{G_4\, G^3_4}+\,\frac{4}{3}\normord{J^3_4\,J^3_4\,J^3_4}-\,8 \normord{T_4\,J^3_4}-\,\frac{4}{3}\,\partial\partial J^3_4\right)(w)+\\
&\quad+\frac{1}{z-w}\bigg(2\normord{\partial G_4\, G^3_4}+\,2\normord{G_4\,\partial G^3_4}+\,2\normord{\partial J^3_4\,J^3_4\,J^3_4}-\,4 \normord{T_4\,\partial J^3_4}-\\
&\quad\qquad\qquad-4 \normord{\partial T_4\,J^3_4}+\,\frac{1}{3}\,\partial\partial\partial J^3_4\bigg)(w)+\cdots,\\
D_4(z)D_4(w)&=-\,\frac{32}{(z-w)^5}+\frac{8}{(z-w)^3}\left(\normord{J^3_4\,J^3_4}-\,T_4\right)(w)+\\
&\quad+\frac{1}{(z-w)^2}\left(8 \normord{\partial J^3_4\,J^3_4}-\,4\, \partial T_4\right)(w)+\\
&\quad+\frac{1}{z-w}\big(-4\normord{G_4\,G^3_4\,J^3_4}-\,2\normord{G_4\,\partial G_4}-\,2\normord{G^3_4\,\partial G^3_4}+\\
&\quad\qquad\qquad+2\normord{\partial J^3_4\,\partial J^3_4}+\,4\normord{T_4\,J^3_4\,J^3_4}\big)(w)+\cdots.
\end{align*}

\section{Shatashvili-Vafa G${}_2$ algebra}
\label{app:SVG2OPEs}

In this appendix we collect our conventions for the OPEs of the Shatashvili-Vafa G${}_2$ algebra, which are taken from \cite{Shatashvili:1994zw}
\begin{align*}
T_7(z)T_7(w)&=\frac{21}{4\,(z-w)^4}+\frac{2}{(z-w)^2}\,
   T_7(w)+\frac{1}{z-w}\,\partial T_7(w)+\cdots,\\
T_7(z)G_7(w)&=\frac{3}{2\, (z-w)^2}\, G_7(w)+\frac{1}{z-w}\,\partial G_7(w)+\cdots,\\
G_7(z)G_7(w)&=\frac{21}{3\,(z-w)^3}+\frac{2}{z-w}\, T_7(w)+\cdots,\\
T_7(z)P(w)&=\frac{3}{2\, (z-w)^2}\, P(w)+\frac{1}{z-w}\,\partial P(w)+\cdots,\\
T_7(z)X(w)&=-\,\frac{7}{4\,(z-w)^4}+\frac{2}{(z-w)^2}X(w)+\frac{1}{(z-w)}\,\partial X(w)+\cdots,\\
T_7(z)K(w)&=\frac{2}{(z-w)^2}\,K(w)+\frac{1}{z-w}\,\partial K(w)+\cdots,\\
T_7(z)M(w)&=-\,\frac{1}{2\,(z-w)^3}+\frac{5}{2\,(z-w)^2}\,M(w)+\frac{1}{(z-w)}\,\partial M(w)+\cdots,\\
G_7(z)P(w)&=\frac{1}{(z-w)}\,K(w)+\cdots,\\
G_7(z)X(w)&=-\,\frac{1}{2\,(z-w)^2}\,G_7(w)+\frac{1}{(z-w)}\,M(w)+\cdots,\\
G_7(z)K(w)&=\frac{3}{2\,(z-w)^2}\,P(w)+\frac{1}{(z-w)}\,\partial P(w)+\cdots,\\
G_7(z)M(w)&=-\,\frac{7}{2\,(z-w)^4}+\frac{1}{(z-w)^2}\,(T_7+4X)(w)+\frac{1}{(z-w)}\,\partial X(w)+\cdots,\\
P(z)P(w)&=-\,\frac{7}{(z-w)^3}+\frac{6}{(z-w)}\,X(w)+\cdots,\\
P(z)X(w)&=-\,\frac{15}{2\,(z-w)^2}\, P(w)-\frac{5}{2\,(z-w)}\,\partial P(w)+\cdots,\\
P(z)K(w)&=-\,\frac{3}{(z-w)^2}\,G_7(w)-\frac{3}{(z-w)}\left(M+\frac{1}{2}\,\partial G_7\right)(w)+\cdots,\\
P(z)M(w)&=\frac{9}{2\,(z-w)^2}K(w)+\frac{1}{(z-w)}\left(3\normord{P\, G_7}-\,\frac{1}{2}\,\partial K\right)(w)+\cdots,\\
X(z)X(w)&=\frac{35}{4\,(z-w)^4}-\frac{10}{(z-w)^2}\,X(w)-\frac{5}{(z-w)}\,\partial X(w)+\cdots,\\
X(z)K(w)&=-\,\frac{3}{(z-w)^2}K(w)-\frac{3}{(z-w)}\normord{P\, G_7}(w)+\cdots,
\end{align*}
\begin{align*}
X(z)M(w)&=-\,\frac{9}{2\,(z-w)^3}G_7(w)-\frac{1}{(z-w)^2}\left(5M+\frac{9}{4}\,\partial G_7\right)(w)+\\
&\quad+\frac{1}{(z-w)}\left(4\normord{X\, G_7}+\frac{1}{2}\,\partial M+\frac{1}{4}\,\partial\partial G_7\right)(w)+\cdots,\\
K(z)K(w)&=-\,\frac{21}{(z-w)^4}+\frac{6}{(z-w)^2}\left(X-T_7\right)(w)+\frac{3}{(z-w)}\left(\partial X -\partial T_7\right)(w)+\cdots,\\
K(z)M(w)&=-\,\frac{15}{(z-w)^3}\,P(w)-\frac{11}{2\,(z-w)^2}\,\partial P(w)+\\
&\quad +\frac{1}{(z-w)}\left(3\normord{G_7\, K} -\,6 \normord{T_7\,P}\right)(w)+\cdots,\\
M(z)M(w)&=-\,\frac{35}{(z-w)^5}+\frac{1}{(z-w)^3}\,(20\,X-9\,T)(w)+\\
&\quad +\frac{1}{(z-w)^2}\left(10\,\partial X-\frac{9}{2}\,\partial T\right)(w)+\\
&\quad +\frac{1}{(z-w)}\left(\frac{3}{2}(\partial\partial X-\partial\partial T)-4\normord{G\, M}+\,8\normord{T\, X}\right)(w)+\cdots.
\end{align*}

\section{Shatashvili-Vafa Spin(7) algebra}
\label{app:SVSpin7OPEs}

In this appendix we collect our conventions for the OPEs of the Shatashvili-Vafa Spin(7) algebra, which are taken from \cite{Shatashvili:1994zw}
\begin{align*}
T_8(z)T_8(w)&=\frac{6}{(z-w)^4}+\frac{2}{(z-w)^2}\,
   T_8(w)+\frac{1}{z-w}\,\partial T_8(w)+\cdots,\\
T_8(z)G_8(w)&=\frac{3}{2\, (z-w)^2}\, G_8(w)+\frac{1}{z-w}\,\partial G_8(w)+\cdots,\\
G_8(z)G_8(w)&=\frac{24}{3\,(z-w)^3}+\frac{2}{z-w}\, T_8(w)+\cdots,\\
T_8(z)X(w)&=\frac{2}{(z-w)^4}+\frac{2}{(z-w)^2}\,X(w)+\frac{1}{(z-w)}\,\partial X(w)+\cdots,\\
T_8(z)M(w)&=-\,\frac{1}{2\,(z-w)^3}\,G_8(w)+\frac{5}{2\,(z-w)^2}\,M(w)+\frac{1}{(z-w)}\,\partial M(w)+\cdots,\\
G_8(z)X(w)&=-\,\frac{1}{2\,(z-w)^2}\,G_8(w)+\frac{1}{(z-w)}\,M(w)+\cdots,\\
G_8(z)M(w)&=\frac{4}{(z-w)^4}+\frac{1}{(z-w)^2}(-\,T_8+4\,X)(w)+\frac{1}{(z-w)}\,\partial X(w)+\cdots,
\end{align*}
\begin{align*}
X(z)X(w)&=\frac{16}{(z-w)^4}+\frac{16}{(z-w)^2}\,X(w)+\frac{8}{(z-w)}\,\partial X(w)+\cdots,\\
X(z)M(w)&=-\,\frac{15}{2\,(z-w)^3}\,G_8(w)+\frac{1}{(z-w)^2}\left(8\,M-\frac{15}{4}\,\partial G_8\right)(w)+\\
&\qquad+\frac{1}{(z-w)}\left(-\,6\normord{G_8\,X}+\frac{11}{2}\,\partial M-\frac{5}{4}\,\partial\partial G_8\right)(w)+\cdots,\\
M(z)M(w)&=-\frac{64}{(z-w)^5}+\frac{1}{(z-w)^3}\left(-15\,T-32\,X\right)(w)+\\
&+\frac{1}{(z-w)^2}\left(-\,\frac{15}{2}\,\partial T_8-16\,\partial X\right)(w)+\\
&+\frac{1}{(z-w)}\left(6\normord{G_8\,M}-12\normord{T_8\,X}-\frac{5}{2}\,\partial\partial T_8-\frac{5}{2}\,\partial\partial X\right)(w)+\cdots.
\end{align*}

\renewcommand\theHsection{\thechapter.\arabic{section}}
\renewcommand\thesection{\thechapter.\arabic{section}}

\chapter{Families of solutions of the heterotic G$_2$ system}
\label{chap:heterotic}

\section{Introduction}

In this chapter we adopt a different approach and study string compactifications from the supergravity point of view. As we explained in \cref{chap:introduction}, this means that we consider a low-energy limit of our string theory where a supergravity description is accurate. In this regime, the conditions to obtain a background solution can be spelled out in detail. Our goal in this chapter is to construct families of solutions for a specific type of compactifications.

We focus on heterotic string theory. It is worth noting that the heterotic string enjoys very particular features that are not present in type II string theories. First and foremost, the theory comes with a gauge sector at a perturbative level. This is not the case in type II superstrings, where non-abelian gauge fields appear only when considering the full non-perturbative effects due to the presence of D-branes.
This unfortunately comes at a cost: the presence of the Green-Schwarz anomaly cancellation condition imposes highly non-trivial relations between the gauge bundle, the flux and the geometry at first order in the string parameter $\alpha'$. This poses an additional challenge to the construction of heterotic background compactifications with maximally supersymmetric spacetime.

The backgrounds we consider are a particular case of the general ansatz we presented in \eqref{eq:compactificationansatz}. Specifically, we study heterotic compactifications on 7-dimensional manifolds giving 3-dimensional AdS$_3$ spacetime quantum field theories with minimal supersymmetry $\mathcal{N}=1$. The geometry of these compactifications is a warped product of AdS$_3$ and a 7-dimensional compact manifold together with a gauge bundle with a connection. The geometrical structure which preserves supersymmetry is called a \emph{heterotic} G$_2$ \emph{system}, and it includes an anomaly cancellation condition which ensures solutions of the system satisfy the equations of motion of the theory. These compactifications were first studied in references \cite{Gunaydin:1995ku,Gauntlett:2001ur, Friedrich:2001nh, Friedrich:2001yp, Gauntlett:2002sc, Gauntlett:2003cy, Ivanov:2003nd, Ivanov:2009rh, Kunitomo:2009mx, Lukas:2010mf, Gray:2012md, Beck:2015gqa}.

As we have already outlined in \cref{chap:introduction}, there are various properties that make these systems attractive. First and foremost, they allow not only Minkowski but also Anti-de Sitter spacetime solutions. This interesting feature is absent in the Hull--Strominger system \cite{Strominger:1986uh, Hull:1986kz} which gives supersymmetric four dimensional theories on Minkowski space when compactifying on 6-dimensional manifolds.

Secondly, these backgrounds have a deep connection with G$_2$-manifolds: the 7-dimensional compact manifolds must possess an \emph{integrable} G$_2$ \emph{structure}\footnote{A detailed description of these geometries can be found in \cref{sec:G2Structures}.} \cite{Gauntlett:2001ur, Friedrich:2001yp}---this guarantees the existence of a unique connection with totally antisymmetric torsion \cite{Friedrich:2001nh}.
In addition, the perturbative heterotic string  comes with a gauge bundle with a connection that must be a G$_2$-instanton\footnote{G$_2$-instantons are 7-dimensional analogues of 4-dimensional Anti Self-Dual (ASD) instantons \cite{Donaldson:1990}, a more detailed description is included in \cref{sec:instantons}. See also \cref{foot:g2references} for a non-exhaustive list of works on this topic.} \cite{Gauntlett:2002sc, Gauntlett:2003cy,Ivanov:2003nd,ReyesCarrion:1998si}.

Finally, there has been a recent surge in interest in the infinitesimal moduli space of the heterotic G$_2$ system \cite{Clarke:2016qtg, delaOssa:2016ivz, delaOssa:2017pqy, delaOssa:2017gjq, Fiset:2017auc,Clarke:2020erl}. The anomaly cancellation condition plays a major role here: the G$_2$-structure and the instanton connections on the compact manifold are intertwined with the flux rendering this (infinitesimal) moduli space finite-dimensional. One of our main goals is to construct solutions that are amenable to a deformation theory interpretation.

Unfortunately, not many explicit solutions to the heterotic G$_2$ system with minimal supersymmetry are available in the literature despite the efforts of the community in recent years \cite{Fernandez:2008wla, Nolle:2010nn, Fernandez:2014pfa, Clarke:2020erl, Lotay:2021eog}. In fact, some of these solutions have been shown \cite{delaOssa:2021cgd} to have more than one supersymmetry. In this chapter, we construct new families of solutions that present all the interesting features that we have mentioned: AdS$_3$ spacetime, a collection of G$_2$-instantons on the compact G$_2$-structure manifold, and the possibility of regarding the family as a finite version of the infinitesimal deformations of \cite{delaOssa:2017pqy}.

The geometry of the compact manifold constitutes a pivotal element in string compactifications and we dedicate \cref{sec:HomogeneousSasakianSolutions} to this aspect. We construct background solutions on certain 3-Sasakian manifolds---we review their general definition and mathematical properties in \cref{3Sasakianmanifolds}. These manifolds are naturally equipped with an integrable G$_2$-structure that can be deformed by rescaling the metric along certain SU(2) fibres. This process is known as \emph{squashing} and we call these manifolds with squashed metrics \emph{squashed 3-Sasakian} manifolds. 
In \cref{sec:Homogeneouscase} we specialize to homogeneous compact squashed 3-Sasakian manifolds, which are the ones we will use in our construction. To be precise, these are the squashed 7-sphere\footnote{Squashed 7-spheres first appeared in the physics literature in the context of compactifications of 11-dimensional supergravity down to 4 dimensions, see for example \cite{Awada:1982pk, Duff:1983gh, Duff:1986hr}.} and the squashed Aloff--Wallach space. 

In the heterotic setting the gauge background fields are described by instanton connections. These are the focus of \cref{sec:InstantonConnections}, where we present several G$_2$-instantons on bundles over squashed 3-Sasakian manifolds. We put a special emphasis on the tangent bundle and in the case of homogeneous manifolds.
In \cref{sec:CanonicalConnection} we review the construction of the canonical connection \cite{kobayashi1963foundations,Harland:2010ix}  and we check it is a G$_2$-instanton. In \cref{sec:ClarkeOliveiraConnection} we extend a G$_2$-instanton construction by Clarke and Oliveira \cite{Clarke:2019miv} to all squashed metrics and different representations.  
Lastly, in \cref{sec:tangentbundleinstantons} we describe explicitly a one-parameter family of instantons on the tangent bundle for both the squashed 7-sphere and the squashed Aloff--Wallach space.

We then proceed in \cref{sec:HeteroticBI} to set up the Bianchi identity for the anomaly cancellation condition, which is essential to obtain a vacuum solution. This identity consists in a rather complicated relation involving the torsion of a G$_2$ compatible connection on the manifold as well as the curvatures of the instanton connections on the gauge vector bundle and the tangent bundle. We collect the instanton curvature terms relevant for our discussion in \cref{sec:tracecanonicalconnection,sec:traceclarkeoliveira,sec:tracetangentbundle}.

Finally, we put all the previous data together to construct explicit solutions of the heterotic G$_2$ system. As anticipated, the highly non-trivial heterotic Bianchi identity significantly constrains the potential solutions: we can only find solutions for particular combinations of instantons. A fully detailed explanation is provided in \cref{sec:NewSolutions}.\footnote{We have also included at the end of the chapter various appendices collecting some of the relevant quantities for the computations we have performed.} We have included tables detailing the ranges of the solutions obtained as well as figures illustrating their behaviour. A brief comment on our results where several possible future directions are pointed out can be found in \cref{sec:conclusions}. 

In the remainder of this introductory section we provide the necessary background for the rest of the chapter, starting with a brief account of instanton connections in \cref{sec:instantons}. In \cref{sec:Heteroticg2Systems} we review the heterotic G${}_2$ system in detail, describing the geometric structures that are required to obtain supersymmetric vacua. In particular we briefly point out how Anti-de Sitter 3-dimensional spacetimes emerge.

\subsection{Instanton connections}
\label{sec:instantons}

As we have indicated, supersymmetric background solutions of the heterotic string support instanton connections on the gauge and tangent bundles. We now provide a summary of the most relevant features of such connections. We mostly follow \cite{Harland:2011zs}.

Instanton connections were first discovered in physics as field configurations satisfying the Yang--Mills equations \cite{Belavin:1975fg, Belavin:1979fb}. These equations are fundamental in gauge theory, yet finding solutions on a general manifold can become quite difficult. Instanton connections enjoy some particular symmetries that ensure they solve the Yang--Mills equations while simultaneously making them easier to find.

Instantons were originally studied in four dimensions, where a very particular phenomenon takes place: by dimensionality, the Hodge star operator maps two-forms to two-forms and one can study the eigenvalues of the operator. Since the Hodge star is an involution, this induces a decomposition of the space of two-forms into \emph{self-dual} (SD) and \emph{anti-self-dual} (ASD) forms, with eigenvalues $+1$ and $-1$ respectively
\begin{equation}
\label{eq:asdinstantondecomposition}
\Lambda^2=\Lambda^2_+\oplus\Lambda^2_-\, ,
\end{equation}
where $\Lambda^2_+$ denotes the space of SD forms and $\Lambda^2_-$ denotes the space of ASD forms. Given a connection $A$ on a vector bundle $V$ over a 4-dimensional manifold, the curvature $F_A$ of the connection is given by a two-form valued in the endomorphisms of $V$.  Thus, the curvature can be decomposed as in \eqref{eq:asdinstantondecomposition}.

A connection such that its curvature is in the ASD subspace is called an \emph{ASD instanton}. As we have already mentioned, such a connection satisfies the 4-dimensional Yang--Mills equations. In addition, Donaldson discovered that ASD instantons can be used to construct invariants of 4-dimensional manifolds and study their topology \cite{Donaldson:1983, Donaldson:1990kn}, see also the books \cite{Donaldson:1990, Freed:2012} for further details and references. This initiated the study of gauge theory in the mathematics community.

The self-duality of the Hodge star on two-forms is a very particular feature of four dimensions. It is therefore natural to search for a higher-dimensional generalization of the concept of instanton. This was first done by the physics community \cite{Corrigan:1982th, Ward:1983zm}, see also \cite{Dundarer:1983fe}, and it was mathematically formalized in \cite{ReyesCarrion:1998si}. The generalization relies on the existence of a $G$-structure on the manifold, hence these connections receive the name of $G$-\emph{instantons}.

First of all, note that locally the space of two-forms on a manifold can be naturally identified with the space of two-dimensional antisymmetric matrices. As a result, for an $n$-dimensional manifold we can locally write
\begin{equation}
\label{eq:2formsareson}
\Lambda^2=\mathfrak{so}(n)\, ,
\end{equation}
where $\mathfrak{so}(n)$ is the Lie algebra of SO($n$). When $n=4$ we have $\mathfrak{so}(4)\cong \mathfrak{su}(2)\oplus\mathfrak{su}(2)$ and this reproduces the decomposition \eqref{eq:asdinstantondecomposition}. We observe that the ASD instantons belong to a subspace corresponding to the Lie algebra of SU(2). This is the idea that can be generalized to higher dimensions.

Let $M$ be an $n$-dimensional manifold with a $G$-structure. Since $G\subset \text{SO}(n)$, the Lie algebra of SO($n$) can be written as
\begin{equation}
\label{eq:sonintermsofliealgebra}
\mathfrak{so}(n)=\mathfrak{g}\oplus\mathfrak{g}^\bot \, ,
\end{equation}
where $\mathfrak{g}$ denotes the Lie algebra of $G$ and $\mathfrak{g}^\bot$ denotes the orthogonal complement within $\mathfrak{so}(n)$. As we explained in \cref{chap:Gstructures}, two-forms on $M$ decompose into irreducible $G$ representations. Combining \eqref{eq:2formsareson} and \eqref{eq:sonintermsofliealgebra} we see this decomposition always includes a subspace corresponding to the Lie algebra of $G$ that we denote $\Lambda^2_{\mathfrak{g}}\,$. This is the subspace associated to the adjoint representation.

A connection $A$ on a vector bundle $V$ is a $G$-\emph{instanton} if its curvature is completely contained in the subspace corresponding to the Lie algebra of $G$, that is
\begin{equation}
F_A\in \Lambda^2_{\mathfrak{g}}(\text{End}(V))\, .
\end{equation}
We stressed in \cref{chap:Gstructures} that these subspaces can be described in terms of the $G$-forms associated to the $G$-structure. Thus, the instanton condition can be expressed as an equation involving the curvature of the connection and the $G$-forms.\footnote{It is possible to make a connection with the definition of ASD instantons in terms of the Hodge star using the $G$-forms. These forms together with the Hodge star can be used to construct an operator that takes two-forms to two-forms. For example, in the G$_2$ case one uses the associative form $\varphi$ to define
\begin{equation*}
\Lambda^2\ni\alpha\longmapsto *\left(\varphi\wedge\alpha\right)\in\Lambda^2\, .
\end{equation*}
This operator is not an involution in general, so the decomposition of the space of two-forms is more complicated than in \eqref{eq:asdinstantondecomposition}. In fact, it corresponds precisely to the decomposition in $G$ representations. Nevertheless, there exists a subspace with eigenvalue $-1$ and a connection is a $G$-instanton if its curvature lies in this subspace---which incidentally is the one associated to the Lie algebra of $G$.}

Higher-dimensional $G$-instantons satisfy interesting properties. When the $G$-structure is torsion-free they satisfy the Yang--Mills equation---as was the case for ASD instantons in four dimensions. Furthermore, $G$-instantons can be used to construct topological invariants on manifolds with a $G$-structure. This has been studied in physics via topological quantum field theories (TQFTs) following the pioneering work of Witten \cite{Witten:1988ze}, see also \cite{Baulieu:1997jx, Acharya:1997gp}. In the mathematics community this was started in refs. \cite{Donaldson:1996kp, Tian:2000fu, Donaldson:2009yq} and now there exists a vast literature in the subject.

In the context of this thesis the instanton condition appears as a consequence of the Killing spinor equations. This reflects the fact that instanton configurations preserve supersymmetry. The relevant connections appearing in this chapter are G$_2$-\emph{instantons} on 7-dimensional manifolds with a G$_2$-structure. In terms of the coassociative form $\psi$ these can be defined as connections whose curvature satisfies
\begin{equation}
F_A\wedge \psi=0\, .
\end{equation}

\subsection{Heterotic G$_2$ system}
\label{sec:Heteroticg2Systems}

We now introduce the heterotic G${}_2$ system following \cite{delaOssa:2017pqy}, see also \cite{Gauntlett:2001ur, Friedrich:2001nh, Friedrich:2001yp, Gauntlett:2002sc, Gauntlett:2003cy, Ivanov:2003nd, Ivanov:2009rh}. This system describes $\mathcal{N}=1$ supersymmetric vacuum solutions of heterotic string theory on a manifold of the form $\mathcal{M}_{3}\times Y$, where $\mathcal{M}_{3}$ is a maximally symmetric 3-dimensional Lorentzian space (the \emph{spacetime}), $Y$ is a compact 7-dimensional manifold and the metric is a warped product. 

A \emph{heterotic G${}_2$ system} is given by a quadruple $[(Y,\varphi),(V,A),(TY,\Theta),H]$ satisfying the following properties:
\begin{itemize}
\item $Y$ is a 7-dimensional manifold and $\varphi$ is a three-form on $Y$ defining an integrable G$_2$-structure on $Y$. We call $\psi=*\varphi $.
\item $V$ is a vector bundle on $Y$ with a connection $A$ that is a G$_2$-instanton, so that its curvature $F_A$ satisfies $F_A\wedge \psi=0$.
\item $TY$ is the tangent bundle of $Y$ and $\Theta$ is a connection on $TY$ which is a G$_2$-instanton, so that its curvature $R_\Theta$ satisfies $R_\Theta\wedge \psi=0$.\footnote{We denote the curvature of a gauge connection $A$ by $F_A$ and the curvature of a metric connection $\Theta$ by $R_\Theta \, $. We omit the subscript only if it is clear the connection we are making reference to.}
\item $H$ is a three-form on $Y$ defined by the formula
\begin{equation}
\label{eq:anomalycancellation}
    H=\dd B+\frac{\alpha'}{4}\left(\mathcal{CS}(A)-\mathcal{CS}(\Theta)\right) \, ,
\end{equation}
where $\mathcal{CS}$ denotes the Chern--Simons form of the corresponding connection, $B$ is the antisymmetric two-form and $\alpha'>0$ is the string parameter. In addition, $H$ is constrained to satisfy
\begin{equation}
\label{eq:fluxistorsion}
    H=T(\varphi)=\frac{1}{6}\,\tau_0\,\varphi-\tau_1\,\lrcorner\,\psi-\tau_3 \, .
\end{equation}
where $T(\varphi)$ denotes the torsion three-form of the G$_2$-structure as defined in \eqref{eq:torsiong2}.
\end{itemize}

\smallskip

These conditions ensure that we have a solution to the Killing spinor equations, thus preserving $\mathcal{N}=1$ supersymmetry. Together with the anomaly cancellation condition, they provide a solution of the equations of motion of the 10-dimensional supergravity action \cite{Ivanov:2009rh}.

Taking the exterior derivative of \eqref{eq:anomalycancellation} we obtain the \emph{heterotic Bianchi identity}
\begin{equation}
\label{eq:heteroticBianchiidentity}
\dd H=\frac{\alpha'}{4}\,(\tr F_A\wedge F_A -\tr R_\Theta\wedge R_\Theta) \, ,
\end{equation}
and finding a solution of \eqref{eq:heteroticBianchiidentity} automatically ensures that a solution to the anomaly cancellation condition \eqref{eq:anomalycancellation} exists.

As pointed out in \cite{delaOssa:2019cci}, the cosmological constant of the 3-dimensional spacetime is related to the external component of the flux $h$, which is determined by the torsion classes of the compact manifold: $h=\frac{1}{3}\,\tau_0\,$. Therefore,
\begin{equation}
\label{eq:tau0givescurvature}
\Lambda\sim -\,h^2 \implies \Lambda\sim -\,\tau_0^2 \, ,
\end{equation}
and we obtain that for $\tau_0\neq 0$ the non-compact spacetime is AdS$_3$. For a Minkowski spacetime, $\tau_0=0$ and, due to the dilaton condition $\tau_1=\frac{1}{2}\,\dd\phi $, the G$_2$-structure is cocalibrated of pure type as observed in \cite{Friedrich:2001yp}.

\section{Squashed 3-Sasakian manifolds}
\label{sec:HomogeneousSasakianSolutions}

\subsection{General aspects, squashing and G$_2$-structures}
\label{3Sasakianmanifolds}

In this section we introduce 3-Sasakian manifolds, which are one of the main elements in our solutions of the heterotic G${}_2$ system. A more detailed account of 3-Sasakian manifolds can be found in \cite{Boyer:1998sf} and \cite{Boyer:2007nr}. See also \cite{2010JGP....60..326A} for the 7-dimensional case.

Let $(Y,g)$ be a Riemannian manifold of dimension $n$, where $n=4k+3$ for $k\geq 1$. We say $(Y,g)$ is \emph{3-Sasakian} if its metric cone $(C(Y),\Bar{g})=(\mathbb{R}_+\times Y,\dd r^2+r^2 g)$
is a hyper-K\"{a}hler manifold. We are interested in the 7-dimensional case so from now on we fix $n=7 $.

Every 3-Sasakian manifold $Y$ has a triple of orthonormal Killing vector fields $(\xi_1, \xi_2, \xi_3)$  satisfying the relation $[\xi_i,\xi_j]=2\,\epsilon\indices{_{ij}^k}\,\xi_k \, $, where $\epsilon_{ijk}$ is the Levi-Civita symbol.  It then follows that these Killing vector fields form an integrable distribution and define a 3-dimensional foliation of $Y$. Moreover, it turns out that the space of leaves of this foliation is a compact orbifold and we can think of $Y$ as the total space of a bundle over an orbifold.\footnote{The simplest example of a 3-Sasakian manifold is the 7-sphere $\Sc^7$, which is the total space of an SU(2)-bundle over the 4-sphere as described by the Hopf fibration $\Sc^3\longrightarrow\Sc^7\longrightarrow\Sc^4$.}

We can locally complete $(\xi_1, \xi_2, \xi_3)$ to an orthonormal basis $\lbrace \xi_1, \dots , \xi_7\rbrace$ of the 3-Sasakian manifold $Y$ and work with the dual basis of one-forms $\lbrace \xi^1, \dots , \xi^7\rbrace $. We define for later convenience:
\begin{equation}
\label{defofomegas}
\omega^1=\xi^4\wedge\xi^5+\xi^6\wedge\xi^7 \, , \qquad
\omega^2=\xi^4\wedge\xi^6-\xi^5\wedge\xi^7 \, , \qquad
\omega^3=-\,\xi^4\wedge\xi^7-\xi^5\wedge\xi^6 \, .
\end{equation}
and it can be shown that the following formulas hold:
\begin{align}
\label{eq:derivativesofetas}
\dd \xi^i&=2 \, \omega^i-\epsilon\indices{^i_{jk}}\,\xi^j\wedge \xi^k \, , \\
\label{eq:derivativesofomegas}
\dd \omega^i&=-\,2 \, \epsilon\indices{^i_{jk}}\,\xi^j\wedge \omega^k \, ,
\end{align}
where $i,j,k\in\{1,2,3\}$.

We can deform the metric of $Y$ away from the 3-Sasakian metric by rescaling the metric along the fibres while keeping the base orbifold metric fixed. This process is known as \emph{squashing} and we obtain a one-parameter family of metrics
\begin{equation}
\label{eq:squashedmetric}
ds^2=\sum_{i=1}^3 s^2\, \xi^i\otimes \xi^i+\sum_{\alpha=4}^7 \xi^\alpha\otimes \xi^\alpha \, ,
\end{equation}
where $s>0$ is the \emph{squashing parameter} and we recover the original metric for $s=1 \, $. We call these manifolds \emph{squashed 3-Sasakian manifolds}. It will be convenient to define an orthonormal coframe $\{\eta^1,\dots,\eta^7\}$ for each value of $s$
\begin{equation}
\label{eq:coframe}
\eta^i=s \, \xi^i \qquad \text{for} \ i=1,2,3 \, ; \qquad \qquad
\eta^\alpha=\xi^\alpha \qquad \text{for} \ \alpha=4,5,6,7 \,.
\end{equation}
The two-forms $\omega^i$ from \eqref{defofomegas} have an analogous expression in this basis, and the formulas \eqref{eq:derivativesofetas} and \eqref{eq:derivativesofomegas} now take the form
\begin{align}
\label{eq:derivativesofetasimproved}
\dd \eta^i&=2 \, s\,\omega^i-\frac{1}{s}\,\epsilon\indices{^i_{jk}}\,\eta^j\wedge \eta^k \, , \\
\label{eq:derivativesofomegasimproved}
\dd \omega^i&=-\,\frac{2}{s}\,\epsilon\indices{^i_{jk}}\,\eta^j\wedge \omega^k \, ,
\end{align}
making manifest that the 3-Sasakian structure is lost by the squashing procedure.

We can define a G$_2$-structure\footnote{A detailed review of G$_2$-structures can be found in \cref{sec:G2Structures}.} on $Y$ for each value of the squashing parameter. The associative three-form is given by\footnote{For $s=1$ this is called the \emph{canonical} G$_2$-structure of the 3-Sasakian manifold in \cite{2010JGP....60..326A}.}
\begin{equation}
\label{eq:threeformforsquash}
\varphi_s=\eta^{123}+\eta^1\wedge\omega^1+\eta^2\wedge\omega^2+\eta^3\wedge\omega^3\, .
\end{equation}
It is important to remark that the structure depends on the parameter $s$. This can be seen explicitly writing \eqref{eq:threeformforsquash} in terms of the $\{\xi^\mu\}$ basis:
\begin{equation}
\label{threeforminotherbasis}
\varphi_s=s^3\,\xi^{123}+s\,\xi^1\wedge\omega^1+s\,\xi^2\wedge\omega^2+s\,\xi^3\wedge\omega^3 \, .
\end{equation}
Therefore we are defining a \emph{one-parameter family} of G$_2$-structures that change with the metric as the squashing parameter $s$ varies. Following \cite{Friedrich:1997}, we rewrite $\varphi_s=F_1+F_2$ with
\begin{equation}
F_1=\eta^{123} \, , \qquad
F_2=\eta^1\wedge\omega^1+\eta^2\wedge\omega^2+\eta^3\wedge\omega^3 \, ,
\end{equation}
we can then compute
\begin{equation}
\label{eq:dualF1andF2}
{*_s F_1}=\frac{1}{6}\sum_{i=1}^3\omega^i\wedge\omega^i \, , \qquad
{*_s F_2}=\frac{1}{2}\,\epsilon_{ijk}\,\eta^i\wedge \eta^j\wedge \omega^k \, ,
\end{equation}
and
\begin{equation}
\label{eq:exterdersF1andF2}
\dd F_1=2\,s\,{*_s F_2} \, , \qquad
\dd F_2=12\,s\,{*_s F_1}+2\,\frac{1}{s}\,{*_s F_2} \, ,
\end{equation}
where $*_s$ is the Hodge star with respect to the metric associated with the G$_2$-structure $\varphi_s \, $. This determines the coassociative four-form
\begin{equation}
\label{eq:fourformsquashed}
\psi_s=*_s \varphi_s=*_s F_1+*_s F_2 \, .
\end{equation}

We can then compute the exterior derivative of the G$_2$-forms
\begin{align}
\label{eq:extderG2squashed}
\begin{split}
\dd\varphi_s&=\frac{12}{7}\left(2 \, s+\frac{1}{s}\right)\psi_s+\left(10 \, s-\frac{2}{s}\right)\left( *_s F_1-\frac{1}{7}\,\psi_s \right),\\
\dd\psi_s&=0 \, ,
\end{split}
\end{align}
and the torsion classes of the G$_2$-structure can be extracted as in \eqref{eq:torsionclassequation}:
\begin{align}
\label{eq:tau0}
\tau_0(\varphi_s)&=\frac{12}{7}\left(2 \, s+\frac{1}{s}\right),\\
\label{eq:tau1}
\tau_1(\varphi_s)&=0 \, ,\\
\label{eq:tau2}
\tau_2(\varphi_s)&=0 \, ,\\
\tau_3(\varphi_s)&=\left(10 \, s-\frac{2}{s}\right)\left( F_1-\frac{1}{7}\,\varphi_s \right).
\label{eq:tau3}
\end{align}
For all values of the squashing parameter the G$_2$-structure is coclosed. In particular it is always integrable, $\tau_2=0 \, $, and can thus be used to construct solutions of the heterotic G${}_2$ system. Since $\tau_1$ vanishes, all these solutions will have constant dilaton. Note as well that $\tau_3$ vanishes if and only if $s=1/\sqrt{5} \, $, in this case the only nonzero torsion class is $\tau_0$ and we say that the G$_2$-structure is \emph{nearly parallel}.\footnote{We later apply these results to the 7-sphere and it is important to remark that the canonical 3-Sasakian G$_2$-structure we obtain from \eqref{eq:threeformforsquash} when $s=1$ is not nearly-parallel. It is therefore different from the \emph{standard} G$_2$-structure of the round 7-sphere, which is known to be nearly-parallel.}

Since the torsion class $\tau_0$ is nonzero for all values of the squashing parameter $s$, heterotic solutions constructed using these manifolds only give rise to AdS$_3$ spacetimes. This is interesting as the only solutions of this kind available in the literature so far are those of \cite{Lotay:2021eog}.

From \eqref{eq:torsiong2} we find for each value of the parameter $s$ the unique connection which is metric, compatible with the G$_2$-structure and has totally antisymmetric torsion
\begin{equation}
\label{eq:uniqueantitorsion}
T(\varphi_s)=2 \, s \, \varphi_s+\left(\frac{2}{s}-10 \, s\right) F_1 \, ,
\end{equation}
in agreement with \cite{Friedrich:2007}.

\subsection{Homogeneous 3-Sasakian manifolds}
\label{sec:Homogeneouscase}

We turn our attention to \emph{homogeneous} 3-Sasakian manifolds. These are described as coset spaces $G/K$ where $G$ is a Lie group and $K$ is a closed subgroup of $G$. Coset spaces are described in further detail in \cite{kobayashi1963foundations, Kapetanakis:1992hf, Harland:2010ix}. Homogeneous 3-Sasakian manifolds are fully classified, see \cite{Boyer:1998sf}, and in 7 dimensions we only have the 7-sphere $\Sc^7=\text{Sp}(2)/\text{Sp}(1)$ and the squashed Aloff--Wallach space $N_{1,1}=\text{SU}(3)/\text{U}(1)_{1,1} \, $. These are the only regular 3-Sasakian manifolds in dimension 7 \cite{Friedrich:1990zg} and we present them in sections \ref{sec:The7sphere} and \ref{sec:AloffWallach}.

The coset structure of these manifolds can be used to describe an orthonormal coframe realising the 3-Sasakian structure \eqref{eq:derivativesofetas} and \eqref{eq:derivativesofomegas}, as we now explain. Recall that a Lie group $G$ acts on itself by left translations and  left-invariant vector fields can be identified with elements of the Lie algebra $\mathfrak{g}$, so that left-invariant one-forms are identified with elements of the dual $\mathfrak{g}^*$.

Let $\lbrace I_1,\dots,I_{\rank(\mathfrak{g})}\rbrace$ be a basis of $\mathfrak{g}$ and let $\lbrace e^1,\dots,e^{\rank(\mathfrak{g})}\rbrace$ be the dual basis. The \emph{Maurer--Cartan form} is defined as the unique $\mathfrak{g}$-valued one-form that acts as the identity on the elements of $\mathfrak{g}$ and it can be written in a basis as
\begin{equation}
\label{eq:maurercartanform}
\theta=\sum_{\alpha=1}^{\rank(\mathfrak{g})} I_\alpha\otimes e^\alpha \, .
\end{equation}
The curvature of the Maurer--Cartan form vanishes identically: this is known as the \emph{Maurer--Cartan equation}. We can regard the Maurer--Cartan form as a metric on $G$, so the $\mathfrak{g}^*$ basis $\lbrace e^\alpha \rbrace$ is describing an orthonormal coframe of $G$. In terms of this coframe, the Maurer--Cartan equation can be written as a collection of structure equations for $G$
\begin{equation}
\label{eq:generalstructureequation}
\dd e^\alpha=-\,\frac{1}{2}f^\alpha_{\beta\gamma}\, e^\beta\wedge e^\gamma \, ,
\end{equation}
with $\alpha,\beta,\gamma=1,\dots,\rank(\mathfrak{g}) $, and where $f^\alpha_{\beta\gamma}$ are the structure constants of the Lie algebra $\mathfrak{g}$. 

We are interested in describing a coframe of the coset space $G/K$, to this end it is convenient to consider $G$ as a principal $K$-bundle over $G/K$. Let $\mathfrak{K}$ be the Lie algebra of $K$ and assume the coset is \emph{reductive}, that means there exists a subspace $\mathfrak{m}$ of $\mathfrak{g}$ such that $\mathfrak{g}=\mathfrak{K}\oplus\mathfrak{m}$ and $\mathfrak{m}$ is invariant under the adjoint action of $K$, $[\mathfrak{K},\mathfrak{m}]\subset\mathfrak{m} $. Under this assumption $\mathfrak{m}$ is a Lie subalgebra that can be identified with the tangent space of the coset space, see \cite{kobayashi1963foundations}.

We can split the basis of $\mathfrak{g}$ in generators of $\mathfrak{m}$, $\lbrace I_1 \, ,\dots,I_{\rank(\mathfrak{m})}\rbrace$, and generators of $\mathfrak{K}$, $\lbrace I_{\rank(\mathfrak{m})+1} \, ,\dots,I_{\rank(\mathfrak{g})}\rbrace$. The commutation relations can then be written as
\begin{equation}
\label{eq:structureconstants}
[I_\nu,I_\rho]=f^\mu_{\nu\rho}\,I_\mu + f^a_{\nu\rho}\, I_a \, , \qquad [I_b,I_\nu]=f^\mu_{b\nu}\, I_\mu \, , \qquad [I_b,I_c]=f^a_{bc}\, I_a \, ,
\end{equation}
where $\mu,\nu,\rho=1 \, ,\dots,\rank(\mathfrak{m})$ and $a,b,c=\rank(\mathfrak{m})+1 \, ,\dots,\rank(\mathfrak{g})$. The structure equations \eqref{eq:generalstructureequation} can be rewritten as
\begin{equation}
    \label{eq:forGstruceq}
\dd e^\mu=-f^\mu_{a\nu}\, e^a\wedge e^\nu-\frac{1}{2} f^\mu_{\nu\rho}\, e^\nu\wedge e^\rho \, , \qquad
\dd e^a=-\,\frac{1}{2} f^a_{\nu\rho}\, e^\nu\wedge e^\rho-\frac{1}{2} f^a_{bc}\, e^b\wedge e^c \, ,
\end{equation}
where again $\mu,\nu,\rho=1 \, ,\dots,\rank(\mathfrak{m})$ and $a,b,c=\rank(\mathfrak{m})+1 \, ,\dots,\rank(\mathfrak{g})$.

Take a local patch $U\subset G/K$ and a local section $L$ of the principal bundle $\pi :G\longrightarrow G/K$, that is, a map $L: U\longrightarrow G$ such that $\pi\circ L$ is the identity. We use $L$ to pull back the one-forms $\lbrace e^\alpha\rbrace$ to the coset space, $\lbrace L^*e^\alpha=\xi^\alpha\rbrace$. In particular the one-forms $\lbrace e^\mu\rbrace$ with $\mu=1 \, ,\dots,\rank(\mathfrak{m})$ are pulled back to a coframe $\{\xi^{\mu}\}$ of $G/K$.\footnote{The pulled-back one-forms $\xi^a$ with $a=\rank(\mathfrak{m})+1,\dots,\rank(\mathfrak{g})$ can be rewritten in terms of the coframe as $\xi^a=c^a_{\mu}\xi^{\mu}$ for some functions $c^a_{\mu}$. Nevertheless, it is convenient for computations to work with the forms $\xi^a$ directly and we will continue to do so for the rest of the chapter.} The pullback of the Maurer--Cartan form defines a metric for $G/K$ with respect to which $\{\xi^{\mu}\}$ is orthonormal. Note as well that the structure equations are still satisfied by the pulled-back forms
\begin{equation}
\label{eq:struceq}
\dd \xi^\mu=-f^\mu_{a\nu}\,\xi^a\wedge \xi^\nu-\frac{1}{2} f^\mu_{\nu\rho}\, \xi^\nu\wedge \xi^\rho \, , \qquad \dd \xi^a=-\,\frac{1}{2} f^a_{\nu\rho}\, \xi^\nu\wedge \xi^\rho-\frac{1}{2} f^a_{bc}\, \xi^b\wedge \xi^c \, .
\end{equation}

In the case of 7-dimensional homogeneous 3-Sasakian manifolds we can choose the coframe $\lbrace \xi^1 \, , \dots , \xi^7\rbrace$ so that equations \eqref{eq:derivativesofetas} and \eqref{eq:derivativesofomegas} are satisfied, making the 3-Sasakian structure explicit. Furthermore, we can consider a squashing of the metric as in \eqref{eq:squashedmetric}: define an orthonormal coframe as we did in \eqref{eq:coframe} and rename $\eta^a=\xi^a$ for $a=8 \, ,\dots,\rank(\mathfrak{g})$, then \eqref{eq:struceq} can be rewritten as
\begin{align}
\label{eq:squashedstruceq1}
\dd \eta^i&=-f^i_{aj}\,\eta^a\wedge \eta^j-s \, f^i_{a\alpha}\,\eta^a\wedge \eta^\alpha-f^i_{j\alpha}\, \eta^j\wedge \eta^\alpha-\frac{1}{2 \, s}f^i_{jk}\, \eta^j\wedge \eta^k-\frac{s}{2}f^i_{\alpha\beta}\, \eta^\alpha\wedge \eta^\beta \, , \\
\label{eq:squashedstruceq2}
\dd \eta^\alpha&=- \, \frac{1}{s}f^\alpha_{ai}\,\eta^a\wedge \eta^i-f^\alpha_{a\beta}\,\eta^a\wedge \eta^\beta-\frac{1}{s}f^\alpha_{j\beta}\, \eta^j\wedge \eta^\beta-\frac{1}{2 \, s^2}f^\alpha_{ij}\, \eta^i\wedge \eta^j-\frac{1}{2}f^\alpha_{\beta\gamma}\, \eta^\beta\wedge \eta^\gamma \, , 
\end{align}
\begin{align}
\label{eq:squashedstruceq3}
\dd \eta^a&=- \, \frac{1}{s}f^a_{i\alpha}\, \eta^i\wedge \eta^\alpha-\frac{1}{2 \, s^2}f^a_{ij}\, \eta^i\wedge \eta^j-\frac{1}{2}f^a_{\alpha\beta}\, \eta^\alpha\wedge \eta^\beta-\frac{1}{2}f^a_{bc}\, \eta^b\wedge \eta^c \, ,
\end{align}
where $i,j,k\in\lbrace 1,2,3\rbrace\,$, $\alpha,\beta,\gamma\in\lbrace 4 \, ,\dots,7\rbrace\,$, $a,b,c\in\lbrace 8 \, ,\dots,\rank(\mathfrak{g})\rbrace$ and $s$ is the squashing parameter. We now particularize this result to the two manifolds we are interested in.

\subsubsection{The squashed 7-sphere}
\label{sec:The7sphere}

The 7-sphere is a homogeneous 3-Sasakian manifold given by the coset
\begin{equation}
\Sc^7=\text{Sp}(2)/\text{Sp}(1) \, ,
\end{equation}
where Sp($n$) denotes the quaternionic unitary groups\footnote{For $n=1,2$ we have alternative characterizations due to some accidental Lie algebra isomorphisms: $\text{Sp}(1)\cong\text{SU}(2)$ and $\text{Sp}(2)\cong\text{Spin}(5)$.}
\begin{equation}
\text{Sp}(n)=\{M\in \text{Mat}_{n\times n}(\mathbb{H}) \text{ such that } MM^\dagger=1\}.
\end{equation}
There are two diagonal $\text{Sp}(1)\cong\text{SU}(2)$ subgroups inside $\text{Sp}(2)$. Quotienting one of them gives the 7-sphere, and the quotient map $\text{Sp}(2)\longrightarrow\Sc^7$ describes a principal SU(2)-bundle over $\Sc^7$. If we further quotient $\Sc^7$ by the remaining $\text{Sp}(1)$, the resulting coset manifold is the 4-sphere and we recover the Hopf fibration $\Sc^3\longrightarrow\Sc^7\longrightarrow\Sc^4$, which makes the 3-Sasakian structure manifest.

As explained above for the general case, we need to specify a convenient basis of the Lie algebra of Sp(2) in order to describe a coframe of the 7-sphere. See also \cite{Geipel:2017tmp} for an alternative equivalent method to obtain a local section using the Hopf fibration. Our conventions can be found in Appendix~\ref{sec:Sp2struceq}. The coframe structure equations are obtained by substituting the structure constants in \eqref{eq:squashedstruceq1}, \eqref{eq:squashedstruceq2} and \eqref{eq:squashedstruceq3}, obtaining
\begin{align}
\label{s7structureequations}
\begin{split}
\dd\eta^1&= - \, \frac{2}{s}\eta^2\wedge\eta^3 +2 \, s \, \eta^4\wedge\eta^5 +2 \, s \, \eta^6\wedge\eta^7 \, , \\
\dd\eta^2&= - \, \frac{2}{s}\eta^3\wedge\eta^1 +2 \, s \, \eta^4\wedge\eta^6 -2 \, s \, \eta^5\wedge\eta^7 \, , \\
\dd\eta^3&= - \, \frac{2}{s}\eta^1\wedge\eta^2 -2 \, s \, \eta^4\wedge\eta^7 -2 \, s \, \eta^5\wedge\eta^6 \, , \\
\dd\eta^4&= - \, \frac{1}{s}\eta^1\wedge\eta^5 -\frac{1}{s}\eta^2\wedge\eta^6 +\frac{1}{s}\eta^3\wedge\eta^7 -\eta^5\wedge\eta^ 8-\eta^6\wedge\eta^9 +\eta^7\wedge\eta^{10} \, , \\
\dd\eta^5&= + \, \frac{1}{s}\eta^1\wedge\eta^4 +\frac{1}{s}\eta^2\wedge\eta^7 +\frac{1}{s}\eta^3\wedge\eta^6 +\eta^4\wedge\eta^8 -\eta^7\wedge\eta^9 -\eta^6\wedge\eta^{10} \, , \\
\dd\eta^6&= - \, \frac{1}{s}\eta^1\wedge\eta^7 +\frac{1}{s}\eta^2\wedge\eta^4 -\frac{1}{s}\eta^3\wedge\eta^5 +\eta^7\wedge\eta^8 +\eta^4\wedge\eta^9 +\eta^5\wedge\eta^{10} \, , \\
\dd\eta^7&= + \, \frac{1}{s}\eta^1\wedge\eta^6 -\frac{1}{s}\eta^2\wedge\eta^5 -\frac{1}{s}\eta^3\wedge\eta^4 -\eta^6\wedge\eta^8 +\eta^5\wedge\eta^9 -\eta^4\wedge\eta^{10} \, , \\
\dd\eta^8&= -2 \, \eta^9\wedge\eta^{10} -2 \, \eta^4\wedge\eta^5 +2 \, \eta^6\wedge\eta^7 \, , \\
\dd\eta^9&= -2 \, \eta^{10}\wedge\eta^8 -2 \, \eta^4\wedge\eta^6 -2 \, \eta^5\wedge \eta^7 \, , \\
\dd \eta^{10}&= -2 \, \eta^8\wedge\eta^9 +2 \, \eta^4\wedge\eta^7 -2 \, \eta^5\wedge\eta^6 \, .
\end{split}
\end{align}
It can be checked the structure equations satisfy \eqref{eq:derivativesofetasimproved} and \eqref{eq:derivativesofomegasimproved} so the coframe describes the squashed 3-Sasakian structure of the 7-sphere.

\subsubsection{The squashed Aloff--Wallach space}
\label{sec:AloffWallach}

The Aloff--Wallach spaces were first described in \cite{bams/1183536240}, we introduce them following \cite{Ball:2016xte}. Consider the matrix group SU(3), let $k,l\in\Z$ and let U(1)${}_{k,l}$ be the circle subgroup of SU(3) whose elements are of the form:
\begin{equation}
\begin{pmatrix}
e^{ik\theta} &0 &0 \\
0 &e^{il\theta} &0 \\
0 &0 &e^{im\theta}
\end{pmatrix},
\end{equation}
where $k+l+m=0$ and $\theta\in\R$. The \emph{Aloff--Wallach space} $N_{k,l}$ is the coset
\begin{equation}
N_{k,l}=\text{SU}(3)/\text{U}(1)_{k,l} \, .
\end{equation}
The only 3-Sasakian Aloff--Wallach space is $N_{1,1}$ and this is the only case that we will study in this chapter.
Our choice of coframe for SU(3) differs slightly from the ones in \cite{Ball:2016xte} and \cite{Geipel:2016hpk} and can be found in Appendix~\ref{SU3struceq}. After squashing we obtain the following structure equations
\begin{align}
\label{aloffwallachstructureequations}
\begin{split}
\dd\eta^1&= - \, \frac{2}{s}\eta^2\wedge\eta^3 +2 \, s \, \eta^4\wedge\eta^5 +2 \, s \, \eta^6\wedge\eta^7 \, , \\
\dd\eta^2&= - \, \frac{2}{s}\eta^3\wedge\eta^1 +2 \, s \, \eta^4\wedge\eta^6 -2 \, s \, \eta^5\wedge\eta^7 \, , \\
\dd\eta^3&= - \, \frac{2}{s}\eta^1\wedge\eta^2 -2 \, s \, \eta^4\wedge\eta^7 -2 \, s \, \eta^5\wedge\eta^6 \, , \\
\dd\eta^4&= - \, \frac{1}{s}\eta^1\wedge\eta^5 -\frac{1}{s}\eta^2\wedge\eta^6 +\frac{1}{s}\eta^3\wedge\eta^7 -\eta^5\wedge\eta^ 8 \, , \\
\dd\eta^5&= + \, \frac{1}{s}\eta^1\wedge\eta^4 +\frac{1}{s}\eta^2\wedge\eta^7 +\frac{1}{s}\eta^3\wedge\eta^6 +\eta^4\wedge\eta^8 \, , \\
\dd\eta^6&= - \, \frac{1}{s}\eta^1\wedge\eta^7 +\frac{1}{s}\eta^2\wedge\eta^4 -\frac{1}{s}\eta^3\wedge\eta^5 +\eta^7\wedge\eta^8 \, , \\
\dd\eta^7&= + \, \frac{1}{s}\eta^1\wedge\eta^6 -\frac{1}{s}\eta^2\wedge\eta^5 -\frac{1}{s}\eta^3\wedge\eta^4 -\eta^6\wedge\eta^8 \, , \\
\dd\eta^8&= - \, 6 \, \eta^4\wedge\eta^5 + \, 6 \, \eta^6\wedge\eta^7 \, .
\end{split}
\end{align}
It is again easy to check that \eqref{eq:derivativesofetasimproved} and \eqref{eq:derivativesofomegasimproved} are satisfied and the coframe describes the squashed 3-Sasakian structure of the Aloff--Wallach space.

\bigskip

\section{Instanton connections}
\label{sec:InstantonConnections}

The construction of solutions to the heterotic G$_2$ system requires the existence of G$_2$-instanton connections on certain vector bundles. In this section we introduce instanton connections on 3-Sasakian manifolds and their squashed deformations, which will play a role in the new solutions we present later.

\subsection{Canonical connection}
\label{sec:CanonicalConnection}

Homogeneous manifolds are equipped with a natural bundle and connection, as explained for example in \cite{Harland:2010ix}. Let $G$ be a Lie group, $K$ a closed Lie subgroup and let $\mathfrak{g}$ and $\mathfrak{K}$ be their Lie algebras. Theorem 11.1 from \cite{kobayashi1963foundations} states that if the coset $G/K$ is reductive then the $\mathfrak{K}$-component of the Maurer--Cartan one-form of $G$ \eqref{eq:maurercartanform} defines a connection on the principal $K$-bundle $G\longrightarrow G/K$ which is left-invariant under the $G$-action. We call this the  \emph{canonical connection on the principal bundle} and write it as
\begin{equation}
\label{eq:bundlecanonicalcon}
A_\text{can}=\sum_a I_a\otimes e^a \, ,
\end{equation}
where $\{e^a\}$ are the vertical one-forms and $\{I_a\}$ is the associated dual basis of $\mathfrak{K}$. Note the canonical connection does not depend on the coset metric so for the case of squashed 3-Sasakian manifolds it is well-defined for all values of the squashing parameter $s$.

For applications to the heterotic G$_2$ system we are interested in considering a representation of the group $K$ and work with the canonical connection \emph{on the associated vector bundle}. The reductiveness of the coset provides a natural representation of $K$ as follows: write $\mathfrak{g}=\mathfrak{K}\oplus\mathfrak{m}$ where $\mathfrak{m}$ is identified with the tangent space at the identity of $G/K$. The elements of $\mathfrak{K}$ act on $\mathfrak{m}$ via the adjoint action, and the condition $[\mathfrak{K},\mathfrak{m}]\subset\mathfrak{m}$ ensures that the action defines a representation of $\mathfrak{K}$ in $\mathfrak{m}$. This provides a representation of $K$ whose associated vector bundle is the tangent bundle of the coset $G/K$. 

The existence of a canonical connection on the tangent bundle is not exclusive to homogeneous manifolds. As shown in \cite{Harland:2011zs} all 3-Sasakian manifolds have a canonical connection constructed using the forms\footnote{Note our conventions differ slightly from \cite{Harland:2011zs}.} preserved by the Sp(1)-structure of the 3-Sasakian manifold
\begin{equation}
\label{formP}
P=\frac{1}{3}\, \xi^{123}-\frac{1}{3}\sum_{i=1}^3 \xi^i\wedge\omega^i \, , \qquad
Q=*F_1=\frac{1}{6}\sum_{i=1}^3\omega^i\wedge\omega^i \, .
\end{equation}
The Christoffel symbols of the canonical connection on the tangent bundle are
\begin{equation}
\label{eq:eqsofcanonical}
\Gamma\indices{^i_{\mu\nu}}={}^{LC}\Gamma\indices{^i_{\mu\nu}}+3 \, P_{i\mu\nu} \, , \qquad
\Gamma\indices{^\nu_{\mu i}}=-\,{}^{LC}\Gamma\indices{^i_{\mu\nu}}-3 \, P_{i\mu\nu} \, , \qquad
\Gamma\indices{^\beta_{\mu\alpha}}={}^{LC}\Gamma\indices{^\beta_{\mu\alpha}} \, ,
\end{equation}
where $\mu,\nu\in\{1 \, ,\dots,7\}\,$, $i\in\{1,2,3\}\,$, $\alpha,\beta\in\{4,5,6,7\}\,$, ${}^{LC}\Gamma$ denotes the Christoffel symbols of the Levi-Civita connection and $P$ is the form defined in \eqref{formP}. The torsion is given by
\begin{equation}
\label{eq:torsionHarland}
    T^i=3 \, P_{i\mu\nu}\,\xi^\mu\wedge \xi^\nu \, , \qquad T^\alpha=\frac{3}{2} P_{\alpha\mu\nu}\,\xi^\mu\wedge \xi^\nu \, .
\end{equation}
This connection is also compatible with all the squashed metrics. In the case of squashed homogeneous 3-Sasakian manifolds, we can regard the structure equations \eqref{eq:struceq} of the coset as Cartan structure equations for the canonical connection
\begin{equation}
\label{cartanstructureeqs}
\dd \xi^\mu=-\,\omega\indices{^\mu_\nu} \wedge \xi^\nu+ \xi^\mu\lrcorner \,  T \, .
\end{equation}
From this we can read the connection one-form of the canonical connection
\begin{equation}
\label{eq:connectiononeformintermsofstrucconst}
\omega\indices{^\mu_\nu}=f^\mu_{a\nu}\, \xi^a \, ,
\end{equation}
as well as the torsion
\begin{equation}
\label{eq:torsionintermsofstrucconst}
\xi^\mu\lrcorner \,  T=-\,\frac{1}{2} f^\mu_{\nu\rho}\, \xi^\nu\wedge \xi^\rho \, ,
\end{equation}
where $\mu,\nu,\rho=1 \, ,\dots,7$ and $a=8 \, ,\dots,\rank(\mathfrak{g})$. This agrees with \eqref{eq:torsionHarland},
and we note that with our conventions the canonical connection has totally antisymmetric torsion only when $s=1/\sqrt{2}\,$.\footnote{This is because for $s=1/\sqrt{2}$ the coset metric we choose is proportional to the Killing form.}

\bigskip

We now turn to the question of whether this connection is a G$_2$-instanton. It is shown in \cite{Harland:2011zs} that the canonical connection on the tangent bundle is an Sp(1)-instanton. This is done by proving that the indices of the curvature of the connection have interchange symmetry and then observing that it has $\text{Sp}(1)$-holonomy.
Since the Sp(1)-structure is given by the forms $\lbrace(\xi^i,\omega^i)_{i=1,2,3}\rbrace$ and the G$_2$-structure is determined by \eqref{eq:threeformforsquash}, the $G$-structures are such that $\mathfrak{sp}(1)\subset\mathfrak{g}_2 \, $, see \cite{delaOssa:2021cgd}. As a result, the canonical connection on the tangent bundle is in fact a G$_2$-instanton, and this remains true for all values of the squashing parameter. 

Let us show that the G$_2$-instanton condition also holds for the canonical connection on the principal bundle \eqref{eq:bundlecanonicalcon}: consider a squashed 3-Sasakian homogeneous manifold.\footnote{Our approach is similar to \cite{Harland:2010ojo} and we show it case-by-case rather than giving a general derivation. Even though canonical connections seem to always satisfy an instanton condition, to the best of our knowledge there is not a general argument for this in the full general case.} The curvature
\begin{equation}
\label{eq:curvatureofgaugebundle}
F_\text{can}=\dd A_\text{can}+\frac{1}{2}[A_\text{can}\wedge A_\text{can}] \, ,
\end{equation}
of the canonical connection \eqref{eq:bundlecanonicalcon} can be computed using the structure equations \eqref{eq:squashedstruceq3}, obtaining
\begin{equation}
\label{eq:curvaturecanonicalgeneral}
F_\text{can}=\sum_a I_a\otimes \left( -\,\frac{1}{s}f^a_{j\alpha}\, \eta^j\wedge \eta^\alpha-\frac{1}{2\,s^2} f^a_{jk} \,\eta^j\wedge \eta^k-\frac{1}{2}f^a_{\alpha\beta}\, \eta^\alpha\wedge \eta^\beta \right),
\end{equation}
where $j,k\in\lbrace 1,2,3\rbrace\,$, $\alpha,\beta\in\lbrace 4 \, ,\dots,\rank(\mathfrak{m})\rbrace\,$, $a\in\lbrace \rank(\mathfrak{m})+1 \, ,\dots,\rank(\mathfrak{g})\rbrace$ and $s$ is the squashing parameter. Comparing \eqref{eq:derivativesofetasimproved} and \eqref{eq:squashedstruceq1} we can rewrite the two-forms $\omega^i$ in terms of the structure constants
\begin{equation}
    \label{eq:omegaintermsofstructureconstants}
    \omega^i=-\,\frac{1}{2 \, s}f^i_{aj}\,\eta^a\wedge \eta^j-\frac{1}{2}f^i_{a\alpha}\,\eta^a\wedge \eta^\alpha-\frac{1}{2 \, s}f^i_{j\alpha}\, \eta^j\wedge \eta^\alpha-\frac{1}{4}f^i_{\alpha\beta}\, \eta^\alpha\wedge \eta^\beta \, .
\end{equation}
In fact, these expressions are greatly simplified in the 7-dimensional case since most of the structure constants vanish, see \eqref{eq:structureconstantssp2} and \eqref{eq:structureconstantssu3}. We have
\begin{equation}
    \label{eq:curvatureandomega}
    F_\text{can}=\sum_a I_a\otimes \left( -\,\frac{1}{2}f^a_{\alpha\beta}\, \eta^\alpha\wedge \eta^\beta \right), \qquad \omega^i=-\,\frac{1}{4}f^i_{\alpha\beta}\, \eta^\alpha\wedge \eta^\beta \, ,
\end{equation}
this can be used together with \eqref{eq:dualF1andF2} and \eqref{eq:fourformsquashed} to express the coassociative form in terms of the structure constants
\begin{equation}
\label{eq:dualF1andF2structureconstants}
\psi_s=\frac{1}{96}\sum_{i=1}^3f^i_{\alpha\beta}\,f^i_{\gamma\rho}\, \eta^{\alpha\beta\gamma\rho}-\frac{1}{8}\epsilon_{ijk}\,f^k_{\alpha\beta}\, \eta^{ij\alpha\beta} \, ,
\end{equation}
where $i,j,k\in\lbrace 1,2,3\rbrace\,$, $\alpha,\beta,\gamma,\rho\in\lbrace 4,\dots,7\rbrace$. The instanton condition $F_\text{can}\wedge\psi=0$ for the canonical connection can then be written in terms of the structure constants as
\begin{equation}
    \label{eq:instantonconditionstructureconstants}
    f^a_{[\alpha\beta}\,f^k_{\gamma\rho]}=0 \, ,
\end{equation}
for all $k\in\lbrace 1,2,3\rbrace\,$, $\alpha,\beta,\gamma,\rho\in\lbrace 4 \, ,\dots,7\rbrace\,$, $a\in\lbrace 8 \, ,\dots,\rank(\mathfrak{g})\rbrace$. It is immediate to check this is satisfied for both the squashed 7-sphere and the squashed Aloff--Wallach space, so the canonical connection is a G$_2$-instanton for both of them, for all values of the squashing parameter $s$.

\subsection{Clarke--Oliveira connection}
\label{sec:ClarkeOliveiraConnection}

An additional G${}_2$-instanton connection on squashed 3-Sasakian manifolds is described in \cite{Clarke:2019miv}. We briefly review the construction and adapt it to our purposes. This involves a straightforward generalization to all values of the squashing parameter $s$.

Any 3-Sasakian manifold $Y$ is described as the total space of an SU(2)-bundle over a Riemannian orbifold $Z$. This bundle can be identified with a bundle of self-dual antisymmetric two-forms on $Z$, and as a result, the Levi-Civita connection of $Z$ induces a connection on the bundle $Y\longrightarrow Z$,
\begin{equation}
\label{eq:connectiononZbeforelift}
a_Z=\sum_{i=1}^3 I_i\otimes \xi^i \, ,
\end{equation}
where $I_i\in\mathfrak{su}(2)$ with $[I_i,I_j]=2 \, \epsilon\indices{_{ij}^k} \, I_k$ for $i,j,k\in\lbrace 1,2,3\rbrace$, and $\xi^i$ denotes the pullback of the 3-Sasakian one-forms to $Z$.\footnote{When the 4-dimensional orbifold $Z$ can be described as a coset, this is just the canonical connection on the $Y\longrightarrow Z$ bundle as described in \cref{sec:CanonicalConnection}.} Consider now the trivial SU(2) bundle over $Y$, and pull back the connection \eqref{eq:connectiononZbeforelift} to the bundle $Y\times\text{SU}(2)\longrightarrow Y$ to obtain\footnote{The main focus of \cite{Clarke:2019miv} is the study of Spin(7)-instantons on cones over squashed 3-Sasakian manifolds, which makes convenient to choose the vector bundle associated to the fundamental SU(2) representation and to introduce a radial dependence in the connection. We do not make a choice of representation so there is no radial factor present in our version of the construction.}
\begin{equation}
a(x_1,x_2,x_3)=\sum_{i=1}^{3} x_i \,  I_i\otimes\xi^i \, .
\end{equation}
Here $x_i\in\R$ are free parameters to be fixed later by the instanton condition. As we squash the metric, the connection in terms of the orthonormal coframe \eqref{eq:coframe} is given by
\begin{equation}
\label{eq:clarkeoliveiraconnection}
a(x_1,x_2,x_3)=\sum_{i=1}^{3} \frac{x_i}{s}\,I_i\otimes\eta^i \, .
\end{equation}
The curvature of the connection
\begin{equation}
\label{eq:curvatureofgaugebundle2}
F=\dd a+\frac{1}{2}\,[a\wedge a]
\end{equation}
changes with the squashing and can be computed using \eqref{eq:derivativesofetasimproved}, obtaining
\begin{equation}
\label{eq:ClarkeOliveiracurvaturebeforeinstanton}
F(x_1,x_2,x_3)=\sum_{i=1}^{3} \left(2 \, x_i \, \omega^i+\sum_{j,k=1}^{3}\epsilon\indices{^i_j_k}\,\frac{1}{s^2}\,\left(-\,x_i+x_j\,x_k\right)\eta^j\wedge \eta^k\right)\otimes I_i \, .
\end{equation}
The G${}_2$-instanton condition $F\wedge\psi=0$ for this connection reduces to three equations
\begin{equation}
\left(2-\frac{1}{s^2}\right)x_i+\frac{1}{s^2}\,x_j \, x_k=0\,, \qquad \text{where } \ i,j,k\in\lbrace 1,2,3\rbrace \ \text{ and } \ i\neq j\neq k \, .
\end{equation}
Imposing these equations leads to two different non-zero kind of instantons. For a fixed squashing of $s=1/\sqrt{2}$ setting any two of the $x_i$ parameters to zero gives a solution for any value of the remaining $x_i \, $. Therefore, there are three one-parameter families of G$_2$-instantons for $s=1/\sqrt{2}\,$. Unfortunately, the curvature of these connections is such that they do not provide solutions of the heterotic G$_2$ system.

The other group of instantons is defined for any value of $s$ and can be found by setting the $x_i$ parameters to $\pm(1-2\,s^2)$ with the following restriction: either we choose all signs positive,\footnote{For the nearly-parallel case $s=1/\sqrt{5}$ this connection has $x_i=3/5$ for $i=1,2,3$ and it is precisely the instanton connection described in Example 2 of \cite{Clarke:2019miv}.} or we choose one of them positive and the rest negative. This provides four different G$_2$-instantons for all values of $s$ which we will use to construct new solutions of the heterotic system. Note that all of them reduce to the trivial flat connection for $s=1/\sqrt{2}\,$.

\subsection{Tangent bundle instantons}
\label{sec:tangentbundleinstantons}

In this section we focus on the tangent bundle and describe a one-parameter family of G$_2$-instantons. We briefly review how to compute the connection one-form and fix our notation, which agrees with \cite{Nakahara:2003nw}. Given a connection on the tangent bundle, consider the Cartan structure equations in the orthonormal coframe
\begin{equation}
\label{eq:cartanfororthonormalbasis}
\dd \eta^\mu=-\,\omega\indices{^\mu_\nu} \wedge \eta^\nu+ \eta^\mu\lrcorner T \, ,
\end{equation}
where $\omega\indices{^\mu_\nu}=\Gamma\indices{^\mu_{\alpha\nu}} \, \eta^\alpha$ is the connection one-form and $T$ is the torsion tensor, with $\mu,\nu,\rho=1 \, ,\dots,7$ and $\alpha=1 \, ,\dots,\rank(\mathfrak{g})$.
The torsion is antisymmetric in the last two indices, $T\indices{^\mu_{\nu\rho}}=-\,T\indices{^\mu_{\rho\nu}}$, and $\eta^\mu\lrcorner  \, T=T^\mu=\frac{1}{2}\,T\indices{^\mu_{\nu\rho}}\eta^\nu\wedge\eta^\rho$.

Manipulating the Cartan equations \eqref{eq:cartanfororthonormalbasis} we can write the connection one-form as
\begin{equation}
\label{eq:oneformfinalformula}
\omega\indices{^\mu_{\nu}}={}^{LC}\omega\indices{^\mu_{\nu}}+K\indices{^\mu_{\rho\nu}}\eta^\rho \, .
\end{equation}
where ${}^{LC}\omega$ is the connection one-form of the Levi-Civita connection, given by
\begin{equation}
\label{eq:levicivitaformula}
{}^{LC}\Gamma_{\mu\nu\rho}=-\frac{1}{2}\left[(\dd \eta_\mu)_{\nu\rho}-(\dd \eta_\nu)_{\rho\mu}+(\dd \eta_\rho)_{\mu\nu}\right] \, ,
\end{equation}
and $K$ is the \emph{contorsion tensor}
\begin{equation}
\label{eq:contorsioneq}
K_{\mu\nu\rho}=\frac{1}{2}\left(T_{\mu\nu\rho}+T_{\nu\mu\rho}+T_{\rho\mu\nu}\right) \, ,
\end{equation}
which is antisymmetric in the first and last indices, ensuring the connection we define is compatible with the metric. Indices are raised and lowered with the orthonormal metric. Note that since the Levi-Civita connection is determined by the structure equations, \eqref{eq:oneformfinalformula} shows that connections are completely specified by the choice of torsion tensor.

Every manifold with a G$_2$-structure \cite{Bryant:2005mz} has a two-parameter family of metric connections preserving the G$_2$-structure. The torsion can in fact be written explicitly and it is given by \cite{delaOssa:2017pqy}
\begin{align*}
\frac{1}{2}\,T_{\mu\nu\rho}(a,\beta)=\frac{1}{12}\,\tau_0 \, \varphi_{\mu\nu\rho}-\frac{1}{6}\,\tau_{2}\phantom{}_{\sigma[\nu} \, \varphi\phantom{}_{\rho]\mu}{}^\sigma+a \, \tau_{3\mu\nu\rho}+\frac{1}{4}(1+2 \, a)S\indices{_\mu^\sigma} \, \varphi_{\nu\rho\sigma}+\\
+\frac{1}{3}(\beta-1)(\tau_1\lrcorner \, \psi)_{\mu\nu\rho}+\frac{2}{3}(1+2 \, \beta)\tau_{1[\nu} \, g_{\rho]\mu} \, ,
\end{align*}
where $S_{\mu\nu}=\frac{1}{4}\, \varphi^{\rho\sigma}{}_{(\mu} \, (\tau_3){}_{\nu)\rho\sigma}$ and $a,\beta\in\R$. For squashed 3-Sasakian manifolds the torsion classes $\tau_2$ and $\tau_1$ vanish, see \eqref{eq:tau1} and \eqref{eq:tau2}, so the family reduces to one parameter\footnote{The choice of parameter $a=-1/2$ corresponds the unique G$_2$-compatible connection with totally antisymmetric torsion \eqref{eq:uniqueantitorsion}.}
\begin{equation}
\label{eq:torsiongeneralg2}
\frac{1}{2}\,T_{\mu\nu\rho}(a)=\frac{1}{12}\,\tau_0 \, \varphi_{\mu\nu\rho}+a \, \tau_{3\mu\nu\rho}+\frac{1}{4}\,(1+2 \, a)\,S\indices{_\rho^\sigma}\varphi_{\sigma\mu\nu}.
\end{equation}
Using the torsion classes computed in \eqref{eq:tau0} and \eqref{eq:tau3} we obtain the torsion of the family\footnote{For the nearly-parallel case $s=1/\sqrt{5}$ the vanishing of $\tau_3$ reduces the family to a single connection which is the one with totally antisymmetric torsion.} of G${}_2$-compatible connections for squashed 3-Sasakian manifolds in terms of $s$ and $a$
\begin{align}
T_{ijk} &=\left[(2+20 \, a)\,s-4 \, a\,\frac{1}{s}\right]\varphi_{ijk}\,, \\
T_{iab} &=2 \, s \, \varphi_{iab}\,, \\
T_{aib} &=\left[-\left(\frac{1}{2}+5 \, a\right)s+\left(\frac{1}{2}+a\right)\frac{1}{s}\right]\varphi_{aib}\,,
\end{align}
where $i,j,k=1,2,3$ and $a,b=4,5,6,7$. The curvature of this family of connections has to be computed in a case-by-case basis since it depends on the Levi-Civita connection of the manifold, recall \eqref{eq:oneformfinalformula}. We have done this for the 7-dimensional squashed homogeneous 3-Sasakian manifolds in Appendix~\ref{sec:ConnectionCurvatureg2compatible} and used it to check the G${}_2$-instanton equations $F\wedge\psi=0$.

We find that the connections with torsion \eqref{eq:torsiongeneralg2} are G${}_2$-instantons for all values of the parameters $s$ and $a$ for both the squashed 7-sphere and the squashed Aloff--Wallach space.\footnote{\label{foot:canisinfamily} This family includes the canonical connection with the representation on the tangent bundle we described in \cref{sec:CanonicalConnection}, which is recovered for the choice of parameter $a(s)=\frac{1+s^2}{2-10s^2}$. On the other hand, the Clarke--Oliveira connection does not belong to this one-parameter family.}

We finish with a comment on other possible instantons: note the heterotic G${}_2$-system introduced in \cref{sec:Heteroticg2Systems} does not require the tangent bundle instanton to be compatible with the G$_2$-structure. Therefore, we could look for instantons outside the family \eqref{eq:torsiongeneralg2}. We have explored this possibility for the homogeneous cases and several families of instantons can be found. Nevertheless, the curvatures of these connections are quite involved and imposing the heterotic Bianchi identity \eqref{eq:heteroticBianchiidentity} becomes difficult. Therefore, we will not use these instantons in our solutions. It would be interesting to verify if it is indeed possible to use these other connections to obtain solutions.

\bigskip

\section{The heterotic Bianchi identity}\label{sec:HeteroticBI}

As described in \cref{sec:Heteroticg2Systems}, the heterotic G$_2$ system includes an instanton connection $A$ on a vector bundle $V$ as well as an instanton connection $\Theta$ on the tangent bundle. The curvatures of these connections must satisfy the heterotic Bianchi identity
\begin{equation}
\dd H=\frac{\alpha'}{4}\big(\tr (F_A\wedge F_A) -\tr (R_\Theta\wedge R_\Theta)\big) \, ,
\qquad H = T \, .
\end{equation}
In \cref{sec:InstantonConnections} we have presented several instanton connections on different bundles over 7-dimensional squashed homogeneous 3-Sasakian manifolds. In order to use these instantons to construct solutions to the heterotic G$_2$ system and verify that they satisfy the heterotic Bianchi identity, in this section we compute the curvature terms $\tr (F\wedge F)$ and $\tr (R\wedge R)$.

Considering the curvature as a Lie algebra-valued two-form, the trace is taken over a product of Lie algebra generators $\tr (I_a I_b)$. The value of this trace will depend on the Lie algebra representation associated to the vector bundle where the connection is defined. Some of the connections introduced in \cref{sec:InstantonConnections} are defined on principal bundles, so a representation of the gauge group has to be chosen to obtain an associated vector bundle. We will explicitly see how the value of $\tr (F\wedge F)$ depends on this choice.

\subsection{Canonical connection}
\label{sec:tracecanonicalconnection}

Recall the canonical connection of a homogeneous manifold was described in \cref{sec:CanonicalConnection}
\begin{equation}
A_\text{can}=\sum_a I_a\otimes e^a \, .
\end{equation}
Using the structure equations \eqref{eq:generalstructureequation}, we computed the curvature of the canonical connection for 7-dimensional squashed homogeneous 3-Sasakian manifolds
\begin{equation}
    F_\text{can}=\sum_a I_a\otimes \left( -\frac{1}{2}f^a_{\alpha\beta}\, \eta^\alpha\wedge \eta^\beta \right) \, ,
\end{equation}
and now it is immediate to compute
\begin{equation}
\label{eq:tracegeneralcanonical}
    \tr( F_\text{can}\wedge F_\text{can})=\sum_{a,b} \tr(I_aI_b) \otimes \left( \frac{1}{4}f^a_{\alpha\beta}\,f^b_{\gamma\rho} \ \eta^{\alpha\beta\gamma\rho} \right) \, ,
\end{equation}
where $\alpha,\beta,\gamma,\rho\in\lbrace 4,\dots,7\rbrace\,$, $a\in\lbrace 8 \, ,\dots,\rank(\mathfrak{g})\rbrace$. As we have said before, a representation of the gauge group has to be chosen to compute explicitly $\tr(I_aI_b)$, so we need 
to distinguish between the 7-sphere and the Aloff--Wallach space.

\subsubsection{The squashed 7-sphere}

\label{sec:canonicalconnforthe7sphere}

The canonical connection on the squashed 7-sphere is defined on a principal SU(2)-bundle. Representations of SU(2) are in bijective correspondence with representations of $\mathfrak{su}(2)$ and are well understood, see for example \cite{Fulton1991RepresentationTA} or \cite{Yamatsu:2015npn}. Complex representations of $\mathfrak{su}(2)$ are classified by a single Dynkin label $(m)$, so that $(m)$ corresponds to the $(m+1)$-dimensional representation.

The $\mathfrak{su}(2)$ generators $\lbrace I_8 \, , I_9 \, , I_{10} \rbrace$ satisfy the commutation relations $[I_a,I_b]=2 \, \epsilon\indices{_{ab}^c} \, I_c$. The trace of a product of generators of a representation is directly related to the \emph{Dynkin index} of the representation. With our conventions, the index of the representation $(m)$ is given by
\begin{equation}
    c(m)=\frac{1}{3}\,m(m+1)(m+2)
\end{equation}
and the trace of the  representation $(m)$ is
\begin{equation}
    \tr(I_a I_b)=-\,c(m) \, \delta_{ab} \, .
\end{equation}
Consider now a general finite-dimensional representation of SU(2) and the canonical connection on the associated vector bundle $V$. The corresponding $\mathfrak{su}(2)$ representation is a direct sum of $k$ irreducible representations with Dynkin labels $m_1,\dots,m_k$. Since the Dynkin index is additive and using the structure constants \eqref{eq:structureconstantssp2}, we obtain from \eqref{eq:tracegeneralcanonical}
\begin{equation}
\label{eq:tracecanonicalsphere}
    \tr( F_\text{can}\wedge F_\text{can})=24\left( c(m_1)+\cdots+c(m_k) \right){*_sF_1} \, ,
\end{equation}
where $*_sF_1=\eta^{4567}$ was defined in \eqref{eq:dualF1andF2}.

Some representations deserve a special mention. First of all, note that the smallest value of the trace is obtained for the fundamental representation, where
\begin{equation}
    \tr( F_\text{can}\wedge F_\text{can})=48\,{*_sF_1} \, .
\end{equation}
whereas the adjoint representation gives
\begin{equation}
    \tr( F_\text{can}\wedge F_\text{can})=192\,{*_sF_1} \, .
\end{equation}
We stressed in \cref{sec:CanonicalConnection} that the canonical connection has a natural representation on the tangent bundle due to the coset $G/K$ being reductive. The representation is described by the adjoint action of $K$ on $G/K$, so the matrix representation of the generators is given by the structure constants
\begin{equation}
    \left( I_a \right)_{\mu\nu}=f^\mu_{a\nu} \, ,
\end{equation}
the explicit matrices can be found in Appendix~\ref{sec:ExpRepMatSp1AdjAct} and they satisfy
\begin{equation}
    \tr(I_a I_b)=-4\,\delta_{ab} \, .
\end{equation}
Therefore, the canonical connection in the tangent bundle representation has
\begin{equation}
\label{eq:tracecanonicaltangentsphere}
    \tr(F_\text{can}\wedge F_\text{can})=96\,{*_sF_1} \, .
\end{equation}
Equivalently, the tangent bundle representation of the canonical connection can be described using Cartan structure equations as we showed in \eqref{eq:connectiononeformintermsofstrucconst}, obtaining the Christoffel symbols \eqref{eq:eqsofcanonical} of \cite{Harland:2011zs}.

\subsubsection{The squashed Aloff--Wallach space}

In the case of the squashed Aloff--Wallach space, the canonical connection is on a principal U(1) bundle.
Since U(1) is abelian, all irreducible complex representations are one-dimensional and described by maps from the circle to itself classified by an integer $q$
\begin{equation}
    \text{U}(1)\ni e^{i\theta} \longmapsto e^{iq\theta}\in\text{U}(1) \, .
\end{equation}
The Lie algebra generator is then given by
\begin{equation}
    I_8=i \, q \, ,
\end{equation}
which results in
\begin{equation}
    \tr(I_8 I_8)=-q^2 \, .
\end{equation}
On the other hand, we can consider irreducible (non-trivial) real representations. These are all two-dimensional, taking values in SO(2) and classified by a positive integer $p$
\begin{equation}
    \text{U}(1)\ni e^{i\theta} \longmapsto \begin{pmatrix}
    \cos{(p\theta)} & -\sin{(p\theta)} \\
    \sin{(p\theta)} & ~\cos{(p\theta)}
    \end{pmatrix} \in\text{SO}(2) \, .
\end{equation}
The Lie algebra generator is then given by
\begin{equation}
    I_8=\begin{pmatrix}
0 &-p \\
p & ~0 
\end{pmatrix},
\end{equation}
which results in
\begin{equation}
    \tr(I_8 I_8)=-2 \, p^2 \, .
\end{equation}
Let us consider now a general finite-dimensional representation of U(1) and the associated vector bundle $V$ together with the corresponding canonical connection. The representation is given by a direct sum of irreducible representations. Suppose we have $k$ irreducible complex representations with indices $p_1,\dots,p_k$ and $\ell$ irreducible real representations with indices $q_1,\dots,q_\ell$. Using the structure constants \eqref{eq:structureconstantssu3}, we obtain from \eqref{eq:tracegeneralcanonical}
\begin{equation}
\label{eq:tracecanonicalaloffwallach}
    \tr( F_\text{can}\wedge F_\text{can})=72\left( q_1^2+\cdots+q_k^2 \right)\,{*_sF_1}+144\left( p_1^2+\cdots+p_\ell^2 \right)\,{*_sF_1} \, ,
\end{equation}
where $*_sF_1=\eta^{4567}$ was defined in \eqref{eq:dualF1andF2}.

In this case, the smallest value possible for the trace term is obtained for a single complex representation
\begin{equation}
\label{eq:traceAloffcanonicalgeneral}
    \tr( F_\text{can}\wedge F_\text{can})=72\,{*_sF_1} \, .
\end{equation}
Interestingly, a real representation with index $q$ gives the same trace value as a direct sum of two complex representations with indices $\pm q$. This means different representations can be used in exactly the same way to obtain solutions. 

Consider now the natural representation of the canonical connection on the tangent bundle described in \cref{sec:CanonicalConnection}. The matrix representation of the generator is given by
\begin{equation}
    \left( I_8 \right)_{\mu\nu}=f^\mu_{8\nu} \, ,
\end{equation}
the matrix is explicitly written in Appendix~\ref{sec:ExpRepMatSp1AdjAct} and satisfies
\begin{equation}
    \tr(I_8 I_8)=-4 \, .
\end{equation}
The canonical connection in the tangent bundle representation gives
\begin{equation}
\label{eq:tracecanonicaltangentAloffWallach}
    \tr(F_\text{can}\wedge F_\text{can})=288\,{*_sF_1} \, .
\end{equation}
It is clear from the explicit matrix form that the tangent bundle representation is a direct sum of the trivial representation on the coordinates $1,2,3$ and two real representations with index $p=1$ on the coordinates $4,5$ and $7,6$. This can be used to recover the trace term from the general formula \eqref{eq:traceAloffcanonicalgeneral}. Alternatively, one can describe the canonical connection using Cartan structure equations as in \eqref{eq:connectiononeformintermsofstrucconst}, and the computation of the curvature becomes very simple since U(1) is abelian.

\subsection{Clarke--Oliveira connection}
\label{sec:traceclarkeoliveira}

As found in \cref{sec:ClarkeOliveiraConnection} the Clarke--Oliveira connection
\begin{equation}
a(x_1,x_2,x_3)=\sum_{i=1}^{3} \frac{x_i}{s}\, I_i\otimes\eta^i \, ,
\end{equation}
is an instanton for all values of the squashing parameter $s$ if we set the $x_i$ parameters to $\pm(1-2\,s^2)$ with the following restriction: either we choose all signs positive, or we choose one of them positive and the rest negative.

Since the Clarke--Oliveira connection is defined on an SU(2)-principal bundle, we can choose representations in the same way as for the canonical connection on the squashed 7-sphere in \cref{sec:canonicalconnforthe7sphere}: the $(m+1)$-dimensional complex representations of $\mathfrak{su}(2)$ is denoted by the Dynkin label $(m)$, and the trace of its generators is given by
\begin{equation}
    \tr(I_i I_j)=-\,c(m)\,\delta_{ij} \, ,
\end{equation}
where $c(m)$ is the Dynkin index of the representation, given in our conventions by
\begin{equation}
    c(m)=\frac{1}{3}\,m(m+1)(m+2) \, .
\end{equation}
For every representation the nonzero contribution to $\tr(I_i I_j)$ comes from the terms with $i=j$, so from the curvature \eqref{eq:ClarkeOliveiracurvaturebeforeinstanton} we obtain, for all four choices of $x_i$ we have previously indicated
\begin{equation}
\label{eq:clarkeoliveiratracewedgeunfinished}
    \tr(F\wedge F)=\sum_{i=1}^3 8\,(1-2 \, s^2)^2\left( *_s F_1-\epsilon\indices{^i_j_k}\,\omega^i\wedge\eta^j\wedge\eta^k \right)\tr(I_i I_i) \, ,
\end{equation}
Take a finite-dimensional representation of SU(2), the corresponding $\mathfrak{su}(2)$ representation is a direct sum of $k$ irreducible representations with Dynkin labels $m_1,\dots,m_k$. Substituting $\tr(I_i I_i)$ for this representation in \eqref{eq:clarkeoliveiratracewedgeunfinished} we obtain the general formula
\begin{equation}
\label{eq:traceclarkeoliveira}
    \tr( F\wedge F)=-\,8\,(1-2 \, s^2)^2\left( c(m_1)+\cdots+c(m_k) \right)\left( 3\,{*_s F_1}-2\,{*_s F_2} \right) \, ,
\end{equation}
where $*_s F_1$ and $*_s F_2$ were defined in \eqref{eq:dualF1andF2}.

The Clarke--Oliveira connection is obtained via pullback by regarding the squashed 3-Sasakian manifold as the total space of an SU(2)-bundle. Therefore, the $\mathfrak{su}(2)$ algebra has a natural adjoint action on the tangent bundle of the 3-Sasakian manifold which can be used to construct a representation of the Clarke--Oliveira connection. For homogeneous manifolds this is analogous to the tangent bundle representation of the canonical connection and the explicit matrices are given by the structure constants
\begin{equation}
    \left( I_i \right)_{\mu\nu}=f^\mu_{i\nu} \, .
\end{equation}
The matrices are the same for both the squashed 7-sphere and the squashed Aloff--Wallach space and they can be found in Appendix~\ref{sec:ExpRepMatCOconn}. They satisfy
\begin{equation}
\tr(I_i I_j)=-12 \, \delta_{ij} \, ,
\end{equation}
so that the Clarke--Oliveira connection on the tangent bundle satisfies
\begin{equation}
\label{eq:traceclarkeoliveiratangent}
    \tr( F\wedge F)=-\,96\,(1-2 \, s^2)^2\left( 3\,{*_s F_1}-2\,{*_s F_2} \right).
\end{equation}

\subsection{Tangent bundle instantons}
\label{sec:tracetangentbundle}

In \cref{sec:tangentbundleinstantons} we introduced a one-parameter family of instantons on the tangent bundle of the squashed 7-sphere and the squashed Aloff--Wallach space consisting on the most general metric connections compatible with the G$_2$-structures. The explicit expressions of the connections and curvatures can be found in Appendix~\ref{sec:ConnectionCurvatureg2compatible} and can be used to compute the term $\tr(F\wedge F)$ for both manifolds. For simplicity, let us denote
\begin{equation}
    \kappa(a,s)=(1+10 \, a)\,s+(1-2 \, a)\,\frac{1}{s} \, ,
\end{equation}
for the squashed 7-sphere we then find
\begin{equation}
\label{eq:traceofgeneralg2connection}
    \tr(F\wedge F)= -\,72 \, s^2 \left( \kappa(a,s)^2 -\frac{4}{3\,s^2} \right)\,{*_s F_1} -12 \, s \ \kappa(a,s)^2\left(\kappa(a,s)-\frac{2}{s}\right)\,{*_s F_2} \, .
\end{equation}
whereas for the squashed Aloff--Wallach space we obtain
\begin{equation}
\label{eq:traceofgeneralg2connectionAloffWallach}
    \tr(F\wedge F)= -\,72 \, s^2 \left( \kappa(a,s)^2 -\frac{4}{s^2} \right)\,{*_s F_1} -12 \, s \ \kappa(a,s)^2\left(\kappa(a,s)-\frac{2}{s}\right)\,{*_s F_2} \, .
\end{equation}

\bigskip

\section{New solutions}
\label{sec:NewSolutions}

In this section we provide new solutions to the heterotic G$_2$ system introduced in \cref{sec:Heteroticg2Systems}. This means we have to specify a quadruple
\begin{equation}
    [(Y,\varphi),(V,A),(TY,\Theta),H] \, .
\end{equation}
The 7-dimensional manifold with an integrable G$_2$-structure $(Y,\varphi)$ is a squashed homogeneous 3-Sasakian manifold with squashed metric \eqref{eq:squashedmetric} and G$_2$-structure given by \eqref{eq:threeformforsquash}, for a certain value of the squashing parameter $s$ that we do not fix for now. The flux $H$ is then completely determined and equal to the unique totally antisymmetric torsion given by \eqref{eq:uniqueantitorsion}. Using \eqref{eq:exterdersF1andF2} and \eqref{eq:extderG2squashed} we can compute
\begin{equation}
\label{eq:exteriorderivativeflux}
    \dd H=24 \, s^2\,{*_s F_1}+8\,(1-2 \, s^2)\,{*_s F_2} \, .
\end{equation}

For an instanton connection $\Theta$ on the tangent bundle $TY$, we consider either the one-parameter family we introduced in \cref{sec:tangentbundleinstantons}, or the tangent bundle representations of the canonical connection or the Clarke--Oliveira connection.

For the vector bundle $V$ and the instanton connection $A$, one option is to choose the canonical connection or the Clarke--Oliveira connection with a representation of the gauge bundle where they were defined. Another option is to take the gauge vector bundle to be the tangent bundle and use the one-parameter family of instantons. We would like to emphasize from the beginning that this is not the so-called standard embedding \cite{Candelas:1985en} because we will choose different connections on each bundle.

The final step to solve the heterotic G$_2$ system is to impose the heterotic Bianchi identity \eqref{eq:heteroticBianchiidentity}
\begin{equation}
    \dd H=\frac{\alpha'}{4}\left(\tr(F\wedge F)-\tr(R_\Theta\wedge R_\Theta)\right) \, ,
\end{equation}
with positive string parameter $\alpha'>0$.

The elements $\dd H$, $\tr(F\wedge F)$ and $\tr(R_\Theta\wedge R_\Theta)$ consist of a sum of two terms proportional to $*_s F_1$ and $*_s F_2\,$, as defined in \eqref{eq:dualF1andF2}. By grouping the coefficients of $*_s F_1$ and $*_s F_2$ in the heterotic Bianchi identity we obtain two independent equations for $s$, $\alpha'$ and any additional coefficients of the connections. These equations impose non-trivial relations between the G$_2$-structure and the curvature of the instantons. In fact, for some choices of connections it is not possible to find a solution, whereas for the rest, the valid ranges of $s$ and $\alpha'$ are restricted. We summarize our findings in \cref{tab:solutions}, which apply to both the squashed 7-sphere and the squashed Aloff--Wallach space.

\begin{table}[h]
{
\centering
\begin{tabular}{|C{0.22\textwidth}||C{0.22\textwidth}|C{0.22\textwidth}|C{0.22\textwidth}|}
\hline
\diagbox[innerwidth=0.22\textwidth,height=2.9\line]{$(V,A)$}{$(TY,\Theta)$} &
Canonical \newline connection &
Clarke--Oliveira connection &
One-parameter family \\
\hline
\hline
Canonical \newline connection &
Solutions only for $s=1/\sqrt{2}\,$, fixed $\alpha'$ &
Solutions for isolated values of $s$ and $\alpha'$ &
Solutions in different ranges of $s$ with $\alpha'$ determined \\
\hline
Clarke--Oliveira connection &
No solution &
No solution &
Solutions in different ranges of $s$ with $\alpha'$ determined \\
\hline
One-parameter family &
No solution &
Solutions in different ranges of $s$ with $\alpha'$ determined &
Solutions with arbitrary $s$ and $\alpha'$ within a certain range \\
\hline
\end{tabular}
\caption{Summary of solutions obtained for squashed homogeneous 7-dimensional 3-Sasakian manifolds in terms of the choice of instanton connections.
}
\label{tab:solutions}
}
\end{table}

We give details of these solutions for both the squashed 7-sphere and the squashed Aloff--Wallach space in the following sections, following the order of \cref{tab:solutions} row by row.

Let us first comment on some general features of our solutions. First of all, we are not able to find solutions for the nearly-parallel squashed metric $s=1/\sqrt{5}\,$. One of the reasons is that the one-parameter family of connections on the tangent bundle collapses to a single connection which is in fact the canonical connection on the tangent bundle. Therefore, the set of instanton connections at our disposal for this particular value of $s$ is much smaller.

On the other hand, we do find solutions for all other values of $s>0$. It is interesting to analyze the behaviour of the solutions close to the limits $s=0$ and $s=\infty$. Since the nonzero elements of the Ricci tensor \cite{2010JGP....60..326A} are given by
\begin{equation}
    \mathcal{R}_{ii}=6\,(2-s^2)\,, \qquad \mathcal{R}_{\alpha\alpha}=\frac{2+4 \, s^4}{s^2} \, ,
\end{equation}
where $i=1,2,3$ and $\alpha=4,5,6,7$, we see the Ricci tensor blows up both for $s=0$ and $s=\infty$. Therefore, these are singular limits and we do not have well-defined solutions for them. Nevertheless, we have a geometrical understanding of the origin of the singularities.

When $s\rightarrow 0$ the $\text{SU}(2)\cong\Sc^3$ fibres of the squashed 3-Sasakian manifold are shrunk to zero size, obtaining a singular space. In this limit $\dd H\rightarrow 0$ and we always find $\alpha'\rightarrow 0$. This means that the string parameter vanishes as the fibres shrink. We stress that we obtain well-defined solutions only while $s>0$.

The case $s\rightarrow\infty$ corresponds to a large volume limit. The radius of the $\text{SU}(2)\cong\Sc^3$ fibres of the squashed 3-Sasakian manifold increases and in the limit the manifold decompactifies. Since the G$_2$-structure becomes singular our solutions are only well-defined while $s<\infty$. In this limit $\dd H$ blows up and typically so do the curvatures of the connections involved in the heterotic Bianchi identity \eqref{eq:heteroticBianchiidentity}. Hence, the asymptotic behaviour of $\alpha'$ depends on the particular solution and we find that $\alpha'$ can tend to $0$, $\infty$ or a fixed constant. We give more details about this behaviour in \cref{Summary}.

As mentioned earlier, another aspect all the solutions we have constructed have in common is that the $\tau_0$ torsion class \eqref{eq:tau0} is always nonzero. Therefore, the 3-dimensional spacetime emerging from these compactifications is Anti-de Sitter space. Note that the AdS$_3$ curvature is proportional to $\tau_0^2$ and therefore tends to $\infty$ both in the limits $s\rightarrow 0$ and $s\rightarrow\infty$. The minimum value of the curvature is achieved for $s=1/\sqrt{2}\,$.

Finally, we explain how the solutions can be described when the one-parameter family of instantons of \cref{sec:tangentbundleinstantons} is chosen. These solutions are richer, as was to be expected since we are introducing an extra parameter $a$ in our equations. As we see from \eqref{eq:traceofgeneralg2connection} and \eqref{eq:traceofgeneralg2connectionAloffWallach}, the contribution of this one-parameter family of connections to the heterotic Bianchi identity is cubic in $a$. Thus, the heterotic Bianchi identity can be rewritten as a cubic equation for $a$ with coefficients depending on $s$. We list the coefficients of the cubic equations in Appendix~\ref{sec:cubiceqs}.

These can be solved explicitly: for each value of $s$ we have up to three different real solutions $a(s)$ of the cubic equation, depending on the sign of the discriminant. For later convenience we define the following functions, motivated by the discriminants in the squashed 7-sphere and the squashed Aloff--Wallach case respectively
\begin{equation}
\label{eq:fandg}
    f(s)=-\frac{4}{9} \, \frac{32 \, s^6-96 \, s^4+78s \, ^2-23}{2 \, s^2-1} \, , \qquad g(s)=-\frac{4}{27} \, \frac{32 \, s^6-96 \, s^4+42 \, s^2-5}{2 \, s^2-1} \, .
\end{equation}
From $a(s)$ we can then obtain the value of $\alpha'$ that solves the heterotic Bianchi identity and ensure it is positive. In most cases the precise expressions of the solutions are not particularly illuminating and we will not show them explicitly.

From the point of view of physics, we are interested in solutions for which the string parameter $\alpha'$ is small. We will highlight these solutions along our presentation.

\subsection{Canonical connection on the vector bundle}

If we want to use the canonical connection on the vector bundle we need to make a choice of representation for the principal bundle. This differs slightly depending on the homogeneous space we choose as we now recall, see \cref{sec:tracecanonicalconnection} for further details.

The canonical connection for the squashed 7-sphere is defined on an SU(2)-bundle. We choose an arbitrary $\mathfrak{su}(2)$ representation, which is given as a direct sum of $k$ irreducible representations with Dynkin labels $m_1,\dots,m_k\,$. Let us denote for simplicity
\begin{equation}
    c=\left( c(m_1)+\cdots+c(m_k) \right),
\end{equation}
note $c$ can take any natural even value. Then the contribution of the canonical connection to the heterotic Bianchi identity \eqref{eq:tracecanonicalsphere} is given by
\begin{equation}
    \tr( F_\text{can}\wedge F_\text{can})=24 \, c\,{*_s F_1} \, .
\end{equation}
The canonical connection on the squashed Aloff--Wallach space is defined on a U(1) bundle. An arbitrary representation is given as a direct sum of $k$ irreducible complex representations with indices $p_1,\dots,p_k$ and $\ell$ irreducible real representations with indices $q_1,\dots,q_\ell$. Let us denote for simplicity
\begin{equation}
    q=\left( q_1^2+\cdots+q_k^2 \right)+2\left( p_1^2+\cdots+p_\ell^2 \right),
\end{equation}
note $q$ can take any natural value. Then the contribution of the canonical connection to the heterotic Bianchi identity \eqref{eq:tracecanonicalaloffwallach} is given by
\begin{equation}
    \tr( F_\text{can}\wedge F_\text{can})=72 \, q\,{*_sF_1} \, .
\end{equation}
We study the available solutions depending on the choice of tangent bundle instanton.

\subsubsection{Canonical connection on the tangent bundle}
\label{subsec:cancan}

We only find a single solution in each case. For the squashed 7-sphere, using \eqref{eq:tracecanonicaltangentsphere} we obtain a solution of the heterotic G$_2$ system if and only if we set
\begin{equation}
    s=\frac{1}{\sqrt{2}}\,, \qquad \alpha'=\frac{2}{c-4}\,.
\end{equation}
Note the vector bundle representation has to be chosen such that $c>4$ in order to have a solution, which discards the SU(2) representations on $\C$ and $\C\oplus\C$.

For the squashed Aloff--Wallach space, using \eqref{eq:tracecanonicaltangentAloffWallach} the solution of the heterotic G$_2$ system is obtained if and only if we set
\begin{equation}
    s=\frac{1}{\sqrt{2}}\,, \qquad \alpha'=\frac{2}{3(q-4)}\,.
\end{equation}
Note that the representation has to be such that $q>4$, which rules out several low-dimensional representations.

Even though these solutions are isolated, they present the interesting feature that solutions with arbitrary small string parameter $\alpha'$ can be found by choosing a gauge bundle representation of sufficiently large dimension.

\subsubsection{Clarke--Oliveira connection on the tangent bundle}
\label{subsec:canCO}

Using \eqref{eq:traceclarkeoliveiratangent} we have two isolated solutions for each representation, both of them with the same value of $\alpha'$. One is always obtained for $s=1/\sqrt{2}\,$, where the Clarke--Oliveira connection reduces to the trivial flat connection on the tangent bundle. For the squashed 7-sphere the solutions are
\begin{equation}
    s=\frac{1}{\sqrt{2}} \ \ \text{or} \ \ s=\frac{\sqrt{12+c}}{2\sqrt{6}} \, , \qquad \alpha'=\frac{2}{c} \, ,
\end{equation}
whereas for the squashed Aloff--Wallach space we have
\begin{equation}
    s=\frac{1}{\sqrt{2}} \ \ \text{or} \ \ s=\frac{\sqrt{4+q}}{2\sqrt{2}}, \qquad \alpha'=\frac{2}{3 \, q} \, .
\end{equation}
In both cases all bundle representations are allowed. The string parameter $\alpha'$ can be made arbitrarily small by choosing a bundle representation of arbitrary large dimension.

\subsubsection{One-parameter family of connections on the tangent bundle}
\label{subsec:canfam}

The contribution of the one-parameter family to the heterotic Bianchi identity can be found in \eqref{eq:traceofgeneralg2connection} for the squashed 7-sphere or in \eqref{eq:traceofgeneralg2connectionAloffWallach} for the squashed Aloff--Wallach space. The solutions we find follow three different behaviours depending on the value of $c$ or $q$. The range where the solution is defined is controlled by the discriminant of the cubic equation for $a$, whose coefficients can be found in Appendix~\ref{sec:cubiceqs}.

\begin{figure}[h]
\centering
\includegraphics[scale=0.7]{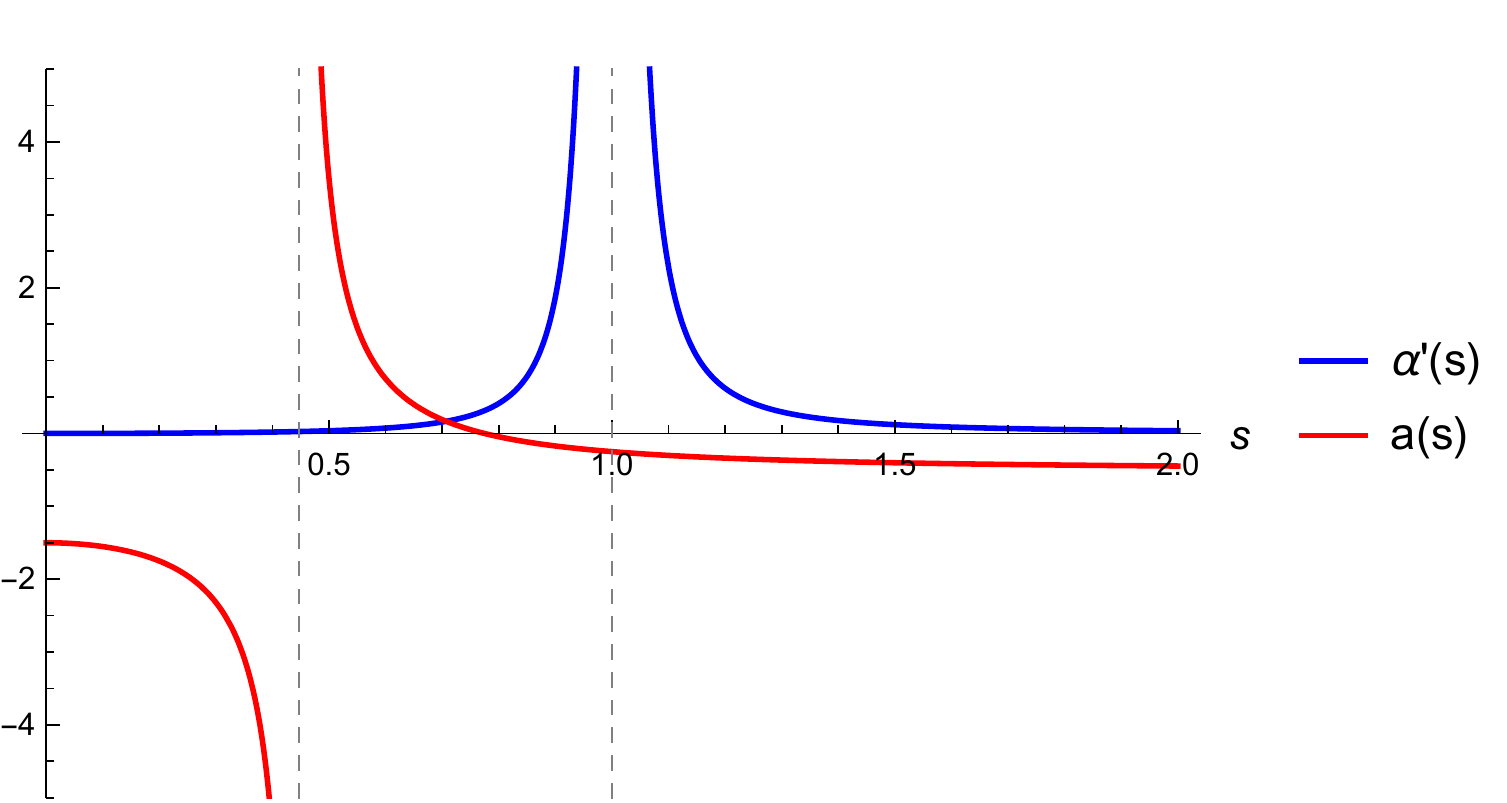}
\caption{Values of $\alpha'(s)$ and $a(s)$ in terms of $s$ for the squashed 7-sphere and the squashed Aloff--Wallach space solution \eqref{eq:easysolution}. The instanton on the tangent bundle belongs to the one-parameter family, with parameter $a(s)$. The instanton on the gauge bundle is the canonical connection in a representation with $c=4$. The dashed grey lines indicate the values of $s=1/\sqrt{5}\,, \, 1\,$, where no solutions exist.}
\label{fig:easysolution}
\end{figure}

The first case corresponds to representations with $c=4$ for the squashed 7-sphere or $q=4$ for the squashed Aloff--Wallach space,\footnote{Note these representations include in particular the tangent bundle representation of the canonical connection. Since that particular representation is part of the one-parameter family---see \cref{foot:canisinfamily}---these solutions also appear when the one-parameter family is chosen in the vector bundle.} in this case the discriminant vanishes identically and we obtain a unique solution for all $s$ except for the nearly parallel $s=1/\sqrt{5}$ and round $s=1$ values. The solution takes the same form for both the squashed 7-sphere and the squashed Aloff--Wallach space
\begin{equation}
\label{eq:easysolution}
    a(s)=-\,\frac{5 \, s^2-3}{10 \, s^2-2}\,, \qquad \alpha'(s)=\frac{s^2}{12\,(s^2-1)^2}\,.
\end{equation}
We show this solution in \cref{fig:easysolution}. Note the string parameter tends to 0 when $s\rightarrow 0$ and when $s\rightarrow\infty$, whereas it blows up whenever $s\rightarrow 1$. This means that the most interesting solutions from the point of view of physics are away from the round metric $s=1$.

For other values of $c$ or $q$ we find three different solutions which are well-defined only in certain ranges. We summarize them in \cref{tab:rangescanpar}, where we have introduced the quantity $s_1\in\left( 1/\sqrt{2},1 \right)$ given implicitly by
\begin{equation}
    f(s_1)=c \qquad \text{or} \qquad g(s_1)=q \, ,
\end{equation}
for the squashed 7-sphere or the squashed Aloff--Wallach space, respectively. Recall $f(s)$ and $g(s)$ were defined in \eqref{eq:fandg}.

\begin{table}[h]
{
\centering
\begin{tabular}{|c|c||c|c|c|c|}
\hline 
\multicolumn{2}{|c||}{Range of $s$} &
$\left(0,\frac{1}{\sqrt{5}}\right)$ \rule{0pt}{3ex} &
$\left(\frac{1}{\sqrt{5}},\frac{1}{\sqrt{2}}\right]$ &
$\left[\frac{1}{\sqrt{2}},s_1\right]$ &
$\left[s_1,\infty\right)$\\[1ex]
\hline
\hline
\multirow{3}{*}{$c>4$ or $q>4$} 
& Solution 1 & \checkmark &\checkmark &\checkmark &\checkmark \\ \cline{2-6}
& Solution 2 & & &\checkmark & \\ \cline{2-6}
& Solution 3 & & &\checkmark & \\
\hline
\multirow{3}{*}{$c<4$ or $q<4$} 
& Solution 1 &\checkmark & &\checkmark &\checkmark \\ \cline{2-6}
& Solution 2 & & & &\checkmark \\ \cline{2-6}
& Solution 3 & &\checkmark & &\checkmark \\
\hline
\end{tabular}
\caption{Ranges of solutions obtained for squashed homogeneous 7-dimensional 3-Sasakian manifolds with one-parameter family of connections on the tangent bundle and canonical connection on the vector bundle.
}
\label{tab:rangescanpar}
}
\end{table}

\begin{figure}[h]
\centering
\includegraphics[scale=0.7]{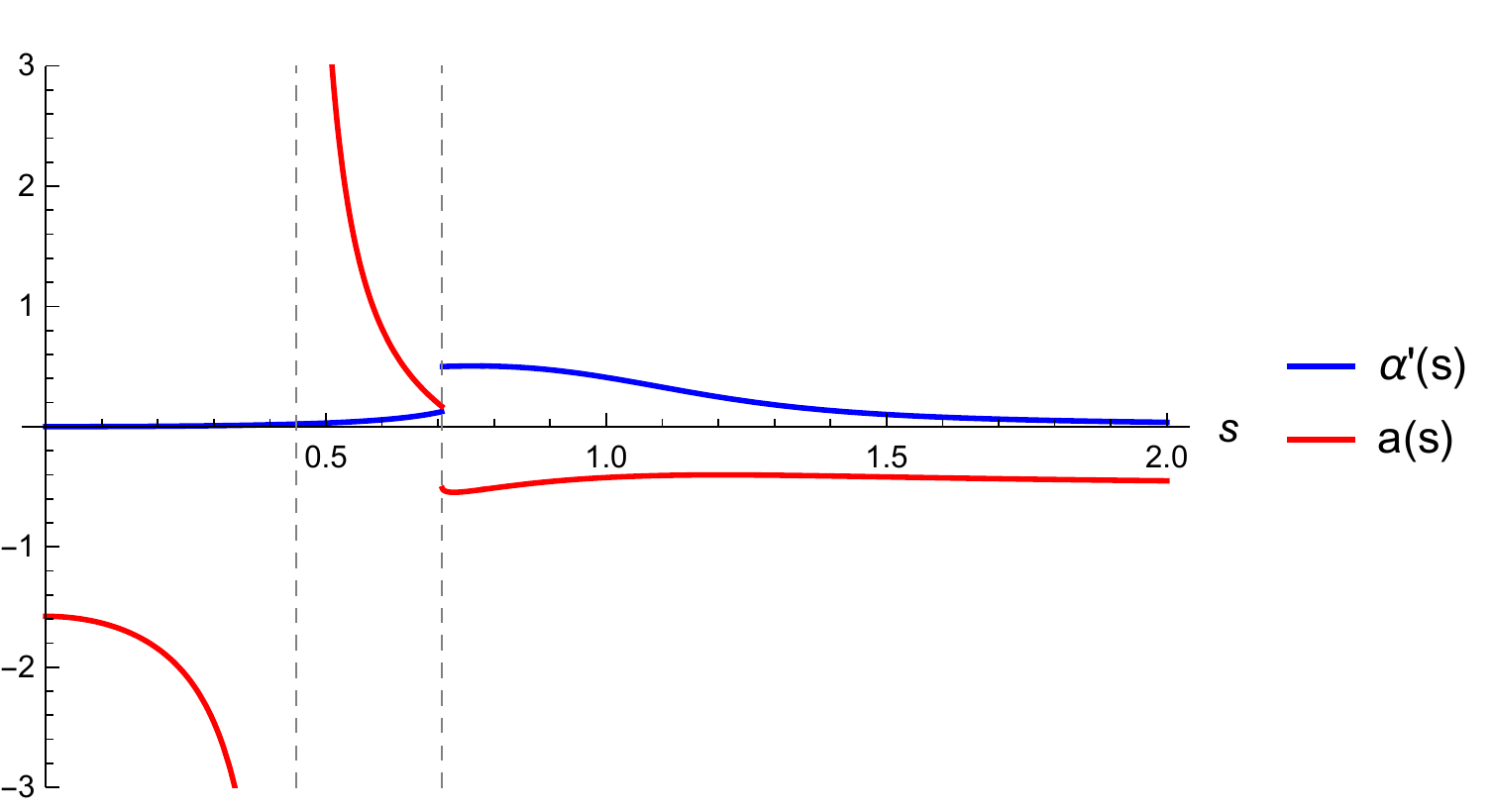}
\caption{Values of $\alpha'(s)$ and $a(s)$ in terms of $s$ for solution 1 in the squashed 7-sphere case. The instanton on the tangent bundle belongs to the one-parameter family, with parameter $a(s)$. The instanton on the gauge bundle is the canonical connection in a representation with $c=8$. The dashed grey lines indicate the values of $s=1/\sqrt{5}\,, \,1/\sqrt{2}\,$. At $s=1/\sqrt{5}$ no solution exists whereas at $s=1/\sqrt{2}\,$ there is a discontinuity.}
\label{fig:oneparcanc8}
\end{figure}

Solution 1 for $c>4$ or $q>4$ is well defined for all $s$ except for $s=1/\sqrt{5}\,$. It is continuous at $s=s_1\,$, whereas at $s=1/\sqrt{2}$ there is a discrete jump: the left and right limits correspond to two different valid solutions
\begin{equation}
\label{eq:pairofsols1sqrt2}
    \left(a,\alpha'\right)=\left(\frac{1}{6} \, ,\frac{2}{z+8}\right), \qquad \left(a,\alpha'\right)=\left(- \, \frac{1}{2} \, ,\frac{2}{z-4}\right),
\end{equation}
where $z$ represents either $c$ or $q$. Similarly to the $c,q=4$ case, we have $\alpha'\rightarrow 0$ for both $s\rightarrow 0$ and $s\rightarrow\infty$. Nevertheless, $\alpha'$ does not blow up for any value of $s$. This provides a large number of solutions with a small value of the string parameter $\alpha'$. We show this solution for the squashed 7-sphere with the choice $c=8$ in \cref{fig:oneparcanc8}.
As for solutions 2 and 3 in the case $c>4$ or $q>4$, they are only defined in a very small range and they remain very close to the $s=1/\sqrt{2}$ solutions described in \eqref{eq:pairofsols1sqrt2}.

\begin{figure}[h]
\centering
\includegraphics[scale=0.7]{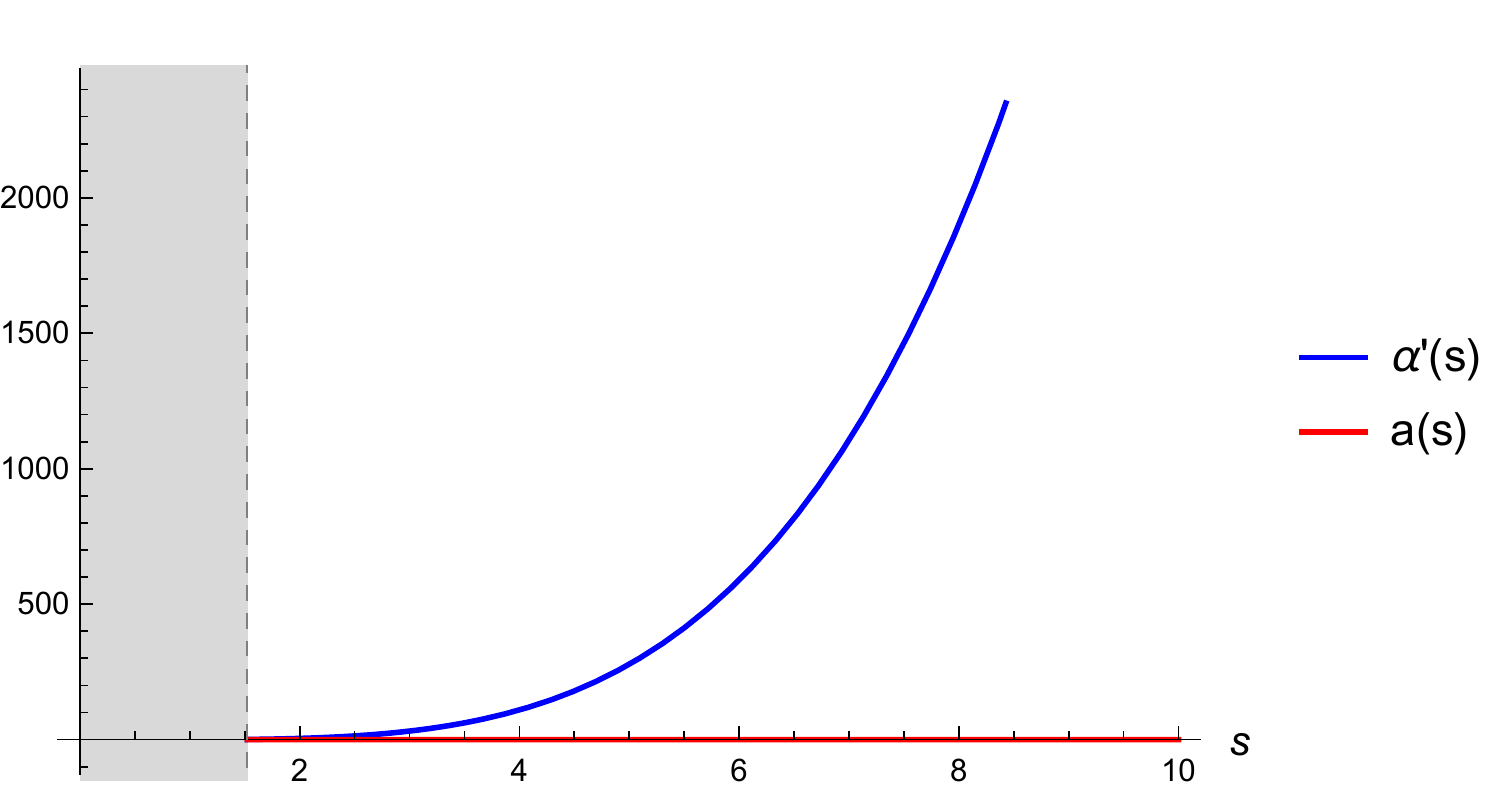}
\caption{Values of $\alpha'(s)$ and $a(s)$ in terms of $s$ for solution 2 in the squashed Aloff--Wallach case. The instanton on the tangent bundle belongs to the one-parameter family, with parameter $a(s)$. The instanton on the gauge bundle is the canonical connection in a representation with $q=1$. The grey region indicates values of $s<s_1$ where no solutions exist.}
\label{fig:oneparcanq1}
\end{figure}

When $c<4$ or $q<4$, solution 1 now presents a discrete jump at $s=s_1\,$. This solution still presents the physically interesting behaviour $\alpha'\rightarrow 0$ when $s\rightarrow 0$ or $s\rightarrow\infty$. On the other hand, solutions 2 and 3 satisfy $\alpha'\rightarrow\infty$ as $s\rightarrow\infty$. Even though these solutions are perfectly valid from a mathematical point of view, they are less attractive for physical purposes---we should think of $\alpha'$ as a small perturbative parameter, a property which is clearly not satisfied in this case. We illustrate this behaviour for solution 2 for the squashed Aloff--Wallach space case with $q=1$ in \cref{fig:oneparcanq1}.

\subsection{Clarke--Oliveira connection on the vector bundle}

In order to use the Clarke--Oliveira connection on the vector bundle we need to make a choice of representation for the SU(2) principal bundle as explained in \cref{sec:traceclarkeoliveira}. Take an arbitrary $\mathfrak{su}(2)$ representation, which is given as a direct sum of $k$ irreducible representations with Dynkin labels $m_1 \, ,\dots,m_k\,$. Let us denote for simplicity
\begin{equation}
    c=\left( c(m_1)+\cdots+c(m_k) \right).
\end{equation}
Note $c$ can take any natural even value. Then the contribution of the Clarke--Oliveira connection to the heterotic Bianchi identity \eqref{eq:traceclarkeoliveira} is given by
\begin{equation}
\label{eq:abridgedtraceclarkeoliveira}
    \tr( F\wedge F)=-8 \, (1-2 \, s^2)^2 \, c\,\left( 3\,{*_sF_1}-2\,{*_sF_2} \right),
\end{equation}
We study the available solutions depending on the choice of tangent bundle instanton.

\subsubsection{Canonical connection on the tangent bundle}

This particular choice of connections contributes with a negative coefficient to the $*_sF_1$ term in the Bianchi identity, as can be seen from \eqref{eq:tracecanonicaltangentsphere} or \eqref{eq:tracecanonicaltangentAloffWallach} and \eqref{eq:abridgedtraceclarkeoliveira}. On the other hand, the coefficient of $\dd H$ is positive, see \eqref{eq:exteriorderivativeflux}. Therefore, there are no solutions with positive string parameter $\alpha'$.

\subsubsection{Clarke--Oliveira connection on the tangent bundle}

In this case the $*_sF_1$ and $*_sF_2$ terms corresponding to the connections in the heterotic Bianchi identity are proportional to each other, and they can not be equal to the $\dd H$ contribution while keeping $\alpha'>0 \, $.

\subsubsection{One-parameter family of connections on the tangent bundle}
\label{subsec:COfam}

For each value of $c$ we find three different solutions. The domain of these solutions depends on the value of $c$ itself and on the chosen homogeneous space, so we describe the solutions for the squashed 7-sphere and the squashed Aloff--Wallach space separately.

For the squashed 7-sphere, the contribution of the one-parameter family to the heterotic Bianchi identity can be found in \eqref{eq:traceofgeneralg2connection}. The domain of solutions is determined by the discriminant of the cubic equation for $a$ (see Appendix~\ref{sec:cubiceqs}) and the positive sign of $\alpha'$, as explained around equation \eqref{eq:fandg}. This is controlled by three quantities that depend on the chosen value of $c$. We denote them by $s_2\,$, $s_3$ and $s_4\,$. We define $s_2$ and $s_4$ as the positive roots of the equation
\begin{equation}
\label{eq:equationfor-fs}
    -f(s)=c \, ,
\end{equation}
where $f(s)$ was defined in \eqref{eq:fandg}, $1/\sqrt{2}<s_4$ and $s_2$ is only present in the case $c>10$, determined by $s_2<1/\sqrt{2}$. We define $s_3$ as the positive real root of the polynomial
\begin{equation}
    -\left( c \, (1-2 \, s^2) \right)^3-48 \, s^2(1-s^2)\left( c \, (1-2 \, s^2) \right)^2+48 \, (1-4 \, s^2) \,  c \, (1-2 \, s^2) +128 \, .
\end{equation}
We show the ranges of the solutions in \cref{tab:rangesclarpar1}.

\begin{table}[h]
{
\centering
\begin{tabular}{|c|c||c|c|c|c|c|c|}
\hline 
\multicolumn{2}{|c||}{Range of $s$} &
$\left(0,\frac{1}{\sqrt{5}}\right)$ \rule{0pt}{3ex} &
$\left(\frac{1}{\sqrt{5}},s_2\right]$ &
$\left(s_2,s_3\right)$ &
$\left[s_3,\frac{1}{\sqrt{2}}\right]$ &
$\left[\frac{1}{\sqrt{2}},s_4\right]$ &
$\left[s_4,\infty\right)$\\[1ex]
\hline
\hline
\multirow{3}{*}{$c<8$} 
& Solution 1 & \checkmark & & & &\checkmark &\checkmark \\ \cline{2-8}
& Solution 2 & & & & & &\checkmark \\ \cline{2-8}
& Solution 3 & &\checkmark &\checkmark &\checkmark & &\checkmark \\
\hline
\multirow{3}{*}{$10<c<40$} 
& Solution 1 & & & & &\checkmark &\checkmark \\ \cline{2-8}
& Solution 2 & & &\checkmark & & &\checkmark \\ \cline{2-8}
& Solution 3 & & &\checkmark &\checkmark & &\checkmark \\
\hline
\multirow{3}{*}{$40\le c$} 
& Solution 1 & & & & &\checkmark &\checkmark \\ \cline{2-8}
& Solution 2 & & & & & &\checkmark \\ \cline{2-8}
& Solution 3 & & & &\checkmark & &\checkmark \\
\hline
\end{tabular}
\caption{Ranges of solutions obtained for the squashed 7-sphere with one-parameter family of connections on the tangent bundle and Clarke--Oliveira connection on the vector bundle.
}
\label{tab:rangesclarpar1}
}
\end{table}

Only two cases are absent in the table. For the case $c=10$ the equation \eqref{eq:equationfor-fs} has a third positive root that we denote $\tilde{s_2}$, with $\tilde{s_2}<s_2<s_4 \, $. The behaviour in this case is as in $10<c<40$ with the addition that solutions 1 and 2 are also well-defined in the interval $(0,\tilde{s_2}] \, $.

\begin{figure}[h]
\centering
\includegraphics[scale=0.7]{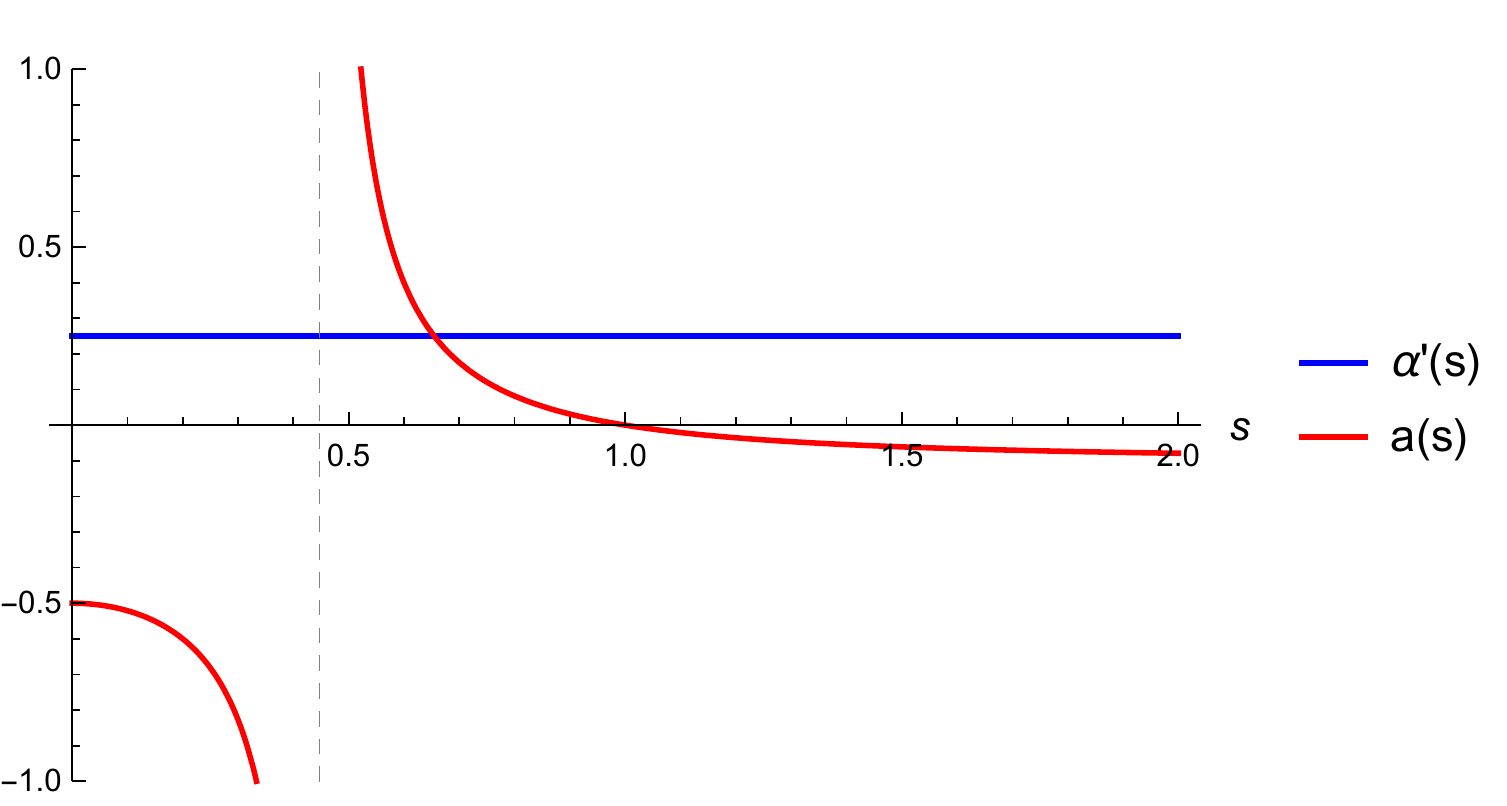}
\caption{Values of $\alpha'(s)$ and $a(s)$ in terms of $s$ for solution 3 in the squashed 7-sphere case. The instanton on the tangent bundle belongs to the one-parameter family, with parameter $a(s)$. The instanton on the gauge bundle is the Clarke--Oliveira connection in a representation with $c=8$. The dashed grey line indicates the value $s=1/\sqrt{5}$ where no solution exists.}
\label{fig:oneparCOc8}
\end{figure}

For the case $c=8$ something remarkable happens. Solution 1 has the same range as for $c<8$ and solution 2 incorporates the interval $\left(1/\sqrt{5}\,,s_2\right]$ to the range we find for $c<8$. Solution 3, on the other hand, is very special: it is defined everywhere except for the nearly-parallel value $s=1/\sqrt{5}$ and the string parameter is fixed to a constant value $\alpha'=1/4 \, $. We show this solution in \cref{fig:oneparCOc8}. This is, to the best of our knowledge, the first example of a family of solutions of the heterotic G$_2$ system with a fixed value of $\alpha'<1$. Keeping the string parameter $\alpha'$ fixed, we can deform the G$_2$-structure and the instanton connection on the tangent bundle from a fixed value of $s$. We comment further on this in \cref{sec:conclusions}.

Now we discuss \cref{tab:rangesclarpar1} more generally. For all values of $c$, solution 1 is discontinuous at $s=s_4\,$. The value of $\alpha'$ remains small for all the range of $s$ and in fact $\alpha'\rightarrow 0$ when $s\rightarrow\infty$. This means for every choice of $c$ we can obtain solutions with arbitrary small string parameter $\alpha'$.

\begin{figure}[h]
\centering
\includegraphics[scale=0.7]{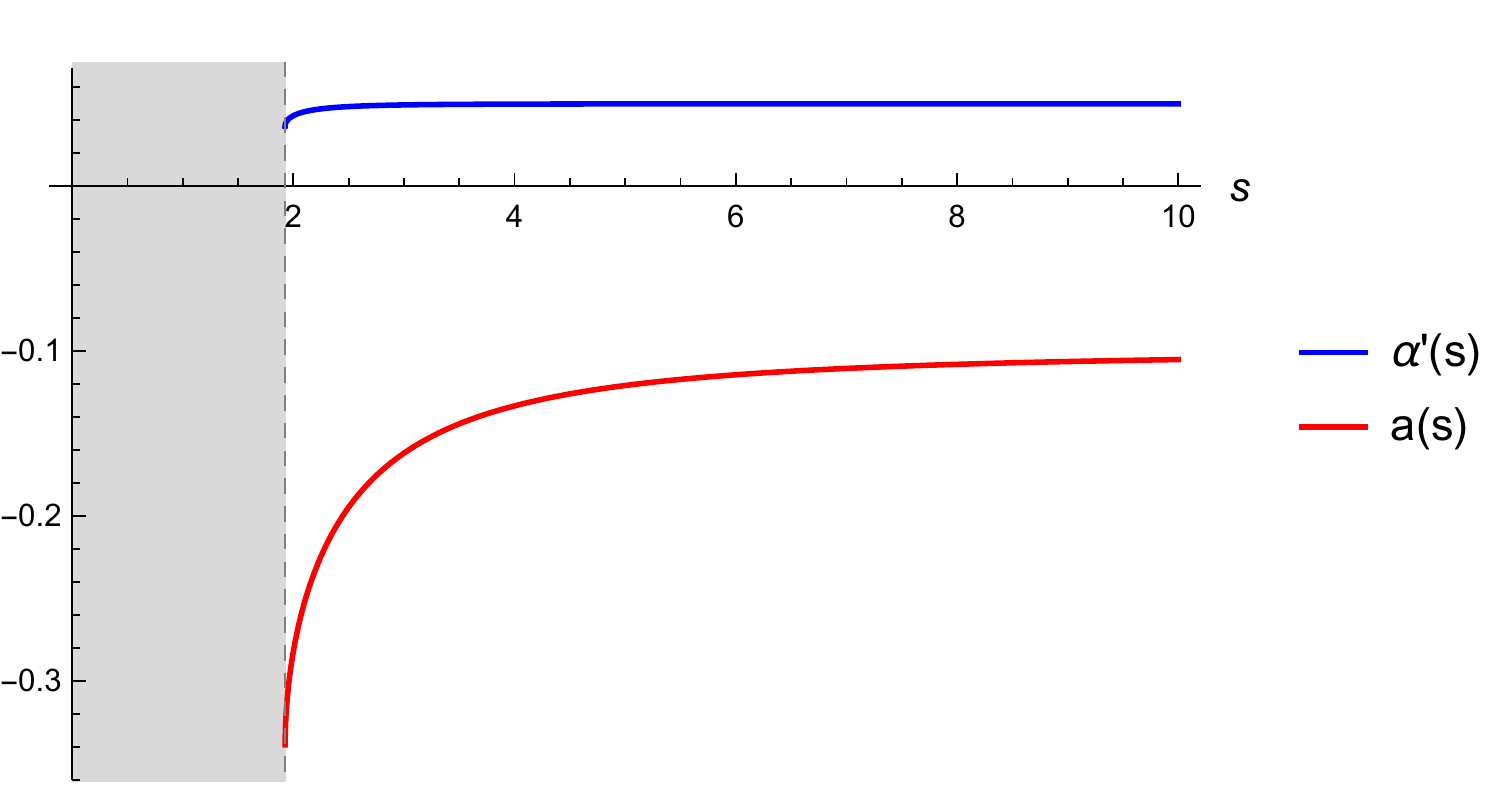}
\caption{Values of $\alpha'(s)$ and $a(s)$ in terms of $s$ for solution 2 in the squashed 7-sphere case. The instanton on the tangent bundle belongs to the one-parameter family, with parameter $a(s)$. The instanton on the gauge bundle is the Clarke--Oliveira connection in a representation with $c=40$. The grey region indicates values of $s<s_4$ where no solutions exist.}
\label{fig:oneparCOc20}
\end{figure}

Solutions 2 and 3 present a slightly different behaviour. First of all, they do not present discontinuities in their domains. Within the interval $\left(1/\sqrt{5},1/\sqrt{2}\right)$, the solutions are typically defined over a very small range and the value of $\alpha'$ blows up. Therefore, that part of the solutions is less amenable to an interesting physical interpretation. On the other hand, we always find the limiting behaviour $\alpha'\rightarrow 2/c$ when $s\rightarrow\infty$. In fact, the solutions soon stabilize in values close to $\alpha'=2/c \, $. This means we find solutions with an almost constant value of $\alpha'$ which can be made arbitrarily small by choosing different representations of the gauge bundle. We illustrate this for the choice $c=40$ in \cref{fig:oneparCOc20}.

\bigskip

For the squashed Aloff--Wallach space, the contribution of the one-parameter family to the heterotic Bianchi identity can be found in \eqref{eq:traceofgeneralg2connectionAloffWallach}. The domain of the solutions is simpler and we just distinguish two cases. These are shown in \cref{tab:rangesclarpar2}: all solutions are well defined beyond a certain value $s_4\,$, which is given by the highest root of the equation
\begin{equation}
\label{eq:equationfor-3gs}
    -3\,g(s)=c \, ,
\end{equation}
for $g(s)$ as defined in \eqref{eq:fandg}. When $c=2$ we have an additional interval in the domain determined by the smallest positive root of equation \eqref{eq:equationfor-3gs}, which we denote by $s_2\,$.

\begin{table}[h]
{
\centering
\begin{tabular}{|c|c||c|c|c|c|}
\hline 
\multicolumn{2}{|c||}{Range of $s$} &
$\left(0,s_2\right]$ \rule{0pt}{3ex} &
$\left(s_2,\frac{1}{\sqrt{2}}\right]$ &
$\left[\frac{1}{\sqrt{2}},s_4\right]$ &
$\left[s_4,\infty\right)$\\[1ex]
\hline
\hline
\multirow{3}{*}{$c=2$} 
& Solution 1 & \checkmark & &\checkmark &\checkmark \\ \cline{2-6}
& Solution 2 & \checkmark & & &\checkmark \\ \cline{2-6}
& Solution 3 & & & &\checkmark \\
\hline
\multirow{3}{*}{$c>2$} 
& Solution 1 & & &\checkmark &\checkmark \\ \cline{2-6}
& Solution 2 & & & &\checkmark \\ \cline{2-6}
& Solution 3 & & & &\checkmark \\
\hline
\end{tabular}
\caption{Ranges of solutions obtained for the squashed Aloff--Wallach space with one-parameter family of connections on the tangent bundle and Clarke--Oliveira connection on the vector bundle.
}
\label{tab:rangesclarpar2}
}
\end{table}

\begin{figure}[h]
\centering
\includegraphics[scale=0.7]{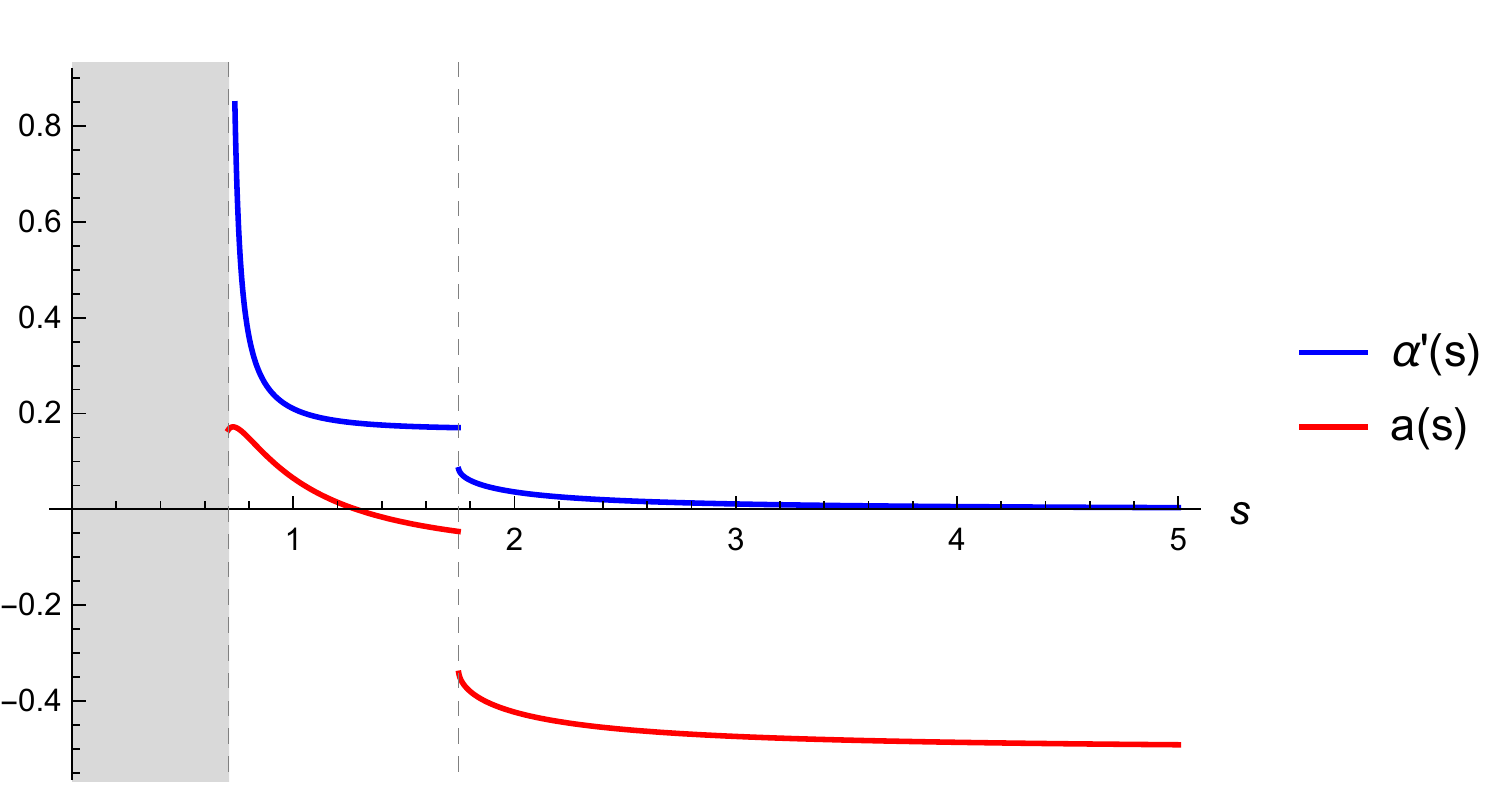}
\caption{Values of $\alpha'(s)$ and $a(s)$ in terms of $s$ for solution 1 in the squashed Aloff--Wallach case. The instanton on the tangent bundle belongs to the one-parameter family, with parameter $a(s)$. The instanton on the gauge bundle is the Clarke--Oliveira connection in a representation with $c=12$. The grey region indicates values of $s<1/\sqrt{2}$ where no solutions exist, and the dashed line indicates a discontinuity of the solution at $s=s_4\,$.}
\label{fig:oneparCOc12}
\end{figure}

The behaviour of the solutions is very similar to the squashed 7-sphere case we have just described. For all values of $c$, solution 1 is discontinuous at $s=s_4$ and we have $\alpha'\rightarrow 0$ when $s\rightarrow\infty$. We present an illustrative example with $c=20$ in \cref{fig:oneparCOc12}. As for solutions 2 and 3, we find again the same behaviour $\alpha'\rightarrow 2/c$ when $s\rightarrow\infty$ and solutions stay almost constant for the whole range of $s$.

\newpage

\subsection{One-parameter family of connections on the vector bundle}

For this last set of solutions we choose the tangent bundle $TY$ as the vector bundle $V$, and a connection from the one-parameter family of \cref{sec:tangentbundleinstantons} as the vector bundle G$_2$-instanton. As we will see, our solutions have different connections on each bundle. We denote the parameter of this family as $b$, and the contribution to the heterotic Bianchi identity is given by \eqref{eq:traceofgeneralg2connection} for the squashed 7-sphere
\begin{equation}
    \tr(F\wedge F)= -\,72 \,  s^2 \left( \kappa(b,s)^2 -\frac{4}{3 \, s^2} \right)\,{*_sF_1} -12 \, s \ \kappa(b,s)^2\left(\kappa(b,s)-\frac{2}{s}\right)\,{*_sF_2} \, .
\end{equation}
and \eqref{eq:traceofgeneralg2connectionAloffWallach} for the squashed Aloff--Wallach space
\begin{equation}
    \tr(F\wedge F)= -\,72 \,  s^2 \left( \kappa(b,s)^2 -\frac{4}{s^2} \right)\,{*_sF_1} -12 \, s \ \kappa(b,s)^2\left(\kappa(b,s)-\frac{2}{s}\right)\,{*_sF_2} \, .
\end{equation}
where $\kappa(b,s)=(1+10 \, b)s+(1-2 \, b)/s$. We find different solutions in terms of the choice of tangent bundle instanton.

\subsubsection{Canonical connection on the tangent bundle}

The equations for $*_sF_1$ and $*_sF_2$ in this case have a unique solution that is exactly as the one presented in \eqref{eq:easysolution} but with opposite $\alpha'$ sign. As a result, we can not impose $\alpha'>0$ and there are no solutions for this choice of instantons.

\subsubsection{Clarke--Oliveira connection on the tangent bundle}
\label{subsec:famCO}

We treat the squashed 7-sphere and the squashed Aloff--Wallach space separately. For each of them we have three solutions with domains given by the discriminant of the cubic equation for the parameter $b$ together with the condition $\alpha'>0$, as explained around equation \eqref{eq:fandg}. The coefficients of the cubic equation are listed in Appendix~\ref{sec:cubiceqs}.

For the squashed 7-sphere, there are three relevant numbers $s_5$, $s_6$ and $s_7$ that we define as follows: $s_5$ and $s_7$ are the positive roots of the polynomial
\begin{equation}
    16s^6+6s^4-15s^2+2 \, ,
\end{equation}
with $s_5<s_7$. On the other hand, $s_6$ is the only positive root of the polynomial
\begin{equation}
    -36s^6+36s^4-15s^2+2 \, .
\end{equation}
We have detailed the range of the solutions in \cref{tab:rangesparclar1}. In addition to the ranges depicted, solutions 1 and 2 can be extended to $s=\sqrt{2/3} \, $.

\begin{table}[h]
{
\centering
\begin{tabular}{|c||c|c|c|c|c|c|c|}
\hline 
Range of $s$ &
$\left(0,s_5\right)$ \rule{0pt}{3.5ex} &
$\left[s_5,\frac{1}{\sqrt{5}}\right)$ &
$\left(\frac{1}{\sqrt{5}},s_6\right]$ &
$\left(s_6,\frac{1}{\sqrt{2}}\right]$ &
$\left(\frac{1}{\sqrt{2}},\sqrt{\frac{2}{3}}\right)$ &
$\left(\sqrt{\frac{2}{3}},s_7\right]$ &
$\left(s_7,\infty\right)$\\[1ex]
\hline
\hline
 Solution 1 & \checkmark & & \checkmark & \checkmark & &\checkmark & \checkmark \\ \hline
 Solution 2 & & & & \checkmark & &\checkmark & \\ \hline
 Solution 3 & &\checkmark & & & &\checkmark & \\
\hline
\end{tabular}
\caption{Ranges of solutions obtained for the squashed 7-sphere with Clarke--Oliveira connection on the tangent bundle and one-parameter family of connections on the vector bundle.
}
\label{tab:rangesparclar1}
}
\end{table}

\begin{figure}[h]
\centering
\includegraphics[scale=0.7]{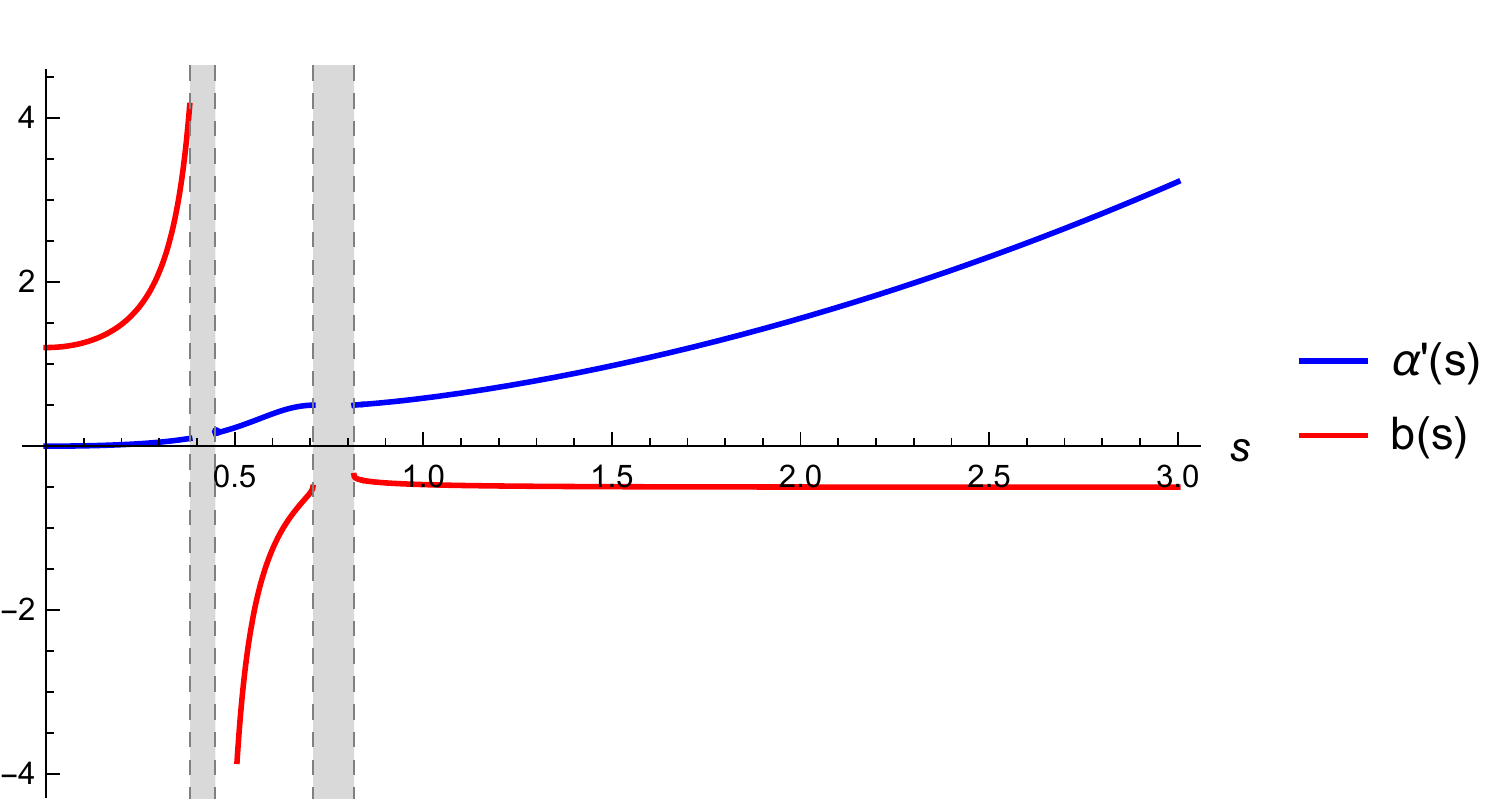}
\caption{Values of $\alpha'(s)$ and $b(s)$ in terms of $s$ for solution 1 in the squashed 7-sphere case. The instanton on the tangent bundle is the Clarke--Oliveira connection in the tangent bundle representation. The instanton on the gauge bundle belongs to the one-parameter family, with parameter $b(s)$. The grey regions indicate values of $s_5<s<1/\sqrt{5}$ and $1/\sqrt{2}<s<\sqrt{2}/\sqrt{3}$ where no solutions exist.}
\label{fig:COonepar}
\end{figure}

Note that solution 1 is the only one defined for $s\rightarrow\infty$, in this limit we have $\alpha'\rightarrow\infty$. It is also the only solution defined for $s\rightarrow 0$ and in this case we find $\alpha'\rightarrow 0$. It is continuous at $s_6$ and $s_7$. These features can be observed in \cref{fig:COonepar}. Solutions 2 and 3 are defined in two small intervals and the string parameter $\alpha'$ blows up for one of them, making the solutions less interesting.

\bigskip

For the squashed Aloff--Wallach space, the ranges are simpler. We introduce the number
\begin{equation}
    s_8=\frac{\sqrt{3\sqrt{33}-11}}{4} \, ,
\end{equation}
and we list the domain of each solution in \cref{tab:rangesparclar2}. In addition to the ranges depicted, solutions 2 and 3 can be extended to $s=1/\sqrt{2}$ preserving continuity from the left.

\smallskip

\begin{table}[h]
{
\centering
\begin{tabular}{|c||c|c|c|c|}
\hline 
Range of $s$ &
$\left(0,\frac{1}{\sqrt{5}}\right)$ &
$\left(\frac{1}{\sqrt{5}},s_8\right]$ &
$\left[s_8,\frac{1}{\sqrt{2}}\right)$ \rule{0pt}{3ex} &
$\left(\frac{1}{\sqrt{2}},\infty\right)$\\[1ex]
\hline
\hline
 Solution 1 & \checkmark & \checkmark &\checkmark & \checkmark \\ \hline
 Solution 2 & & & \checkmark &  \\ \hline
 Solution 3 & & & \checkmark  & \\
\hline
\end{tabular}
\caption{Ranges of solutions obtained for the squashed Aloff--Wallach space with Clarke--Oliveira connection on the tangent bundle and one-parameter family of connections on the vector bundle.
}
\label{tab:rangesparclar2}
}
\end{table}

\smallskip

\begin{figure}[h]
\centering
\includegraphics[scale=0.7]{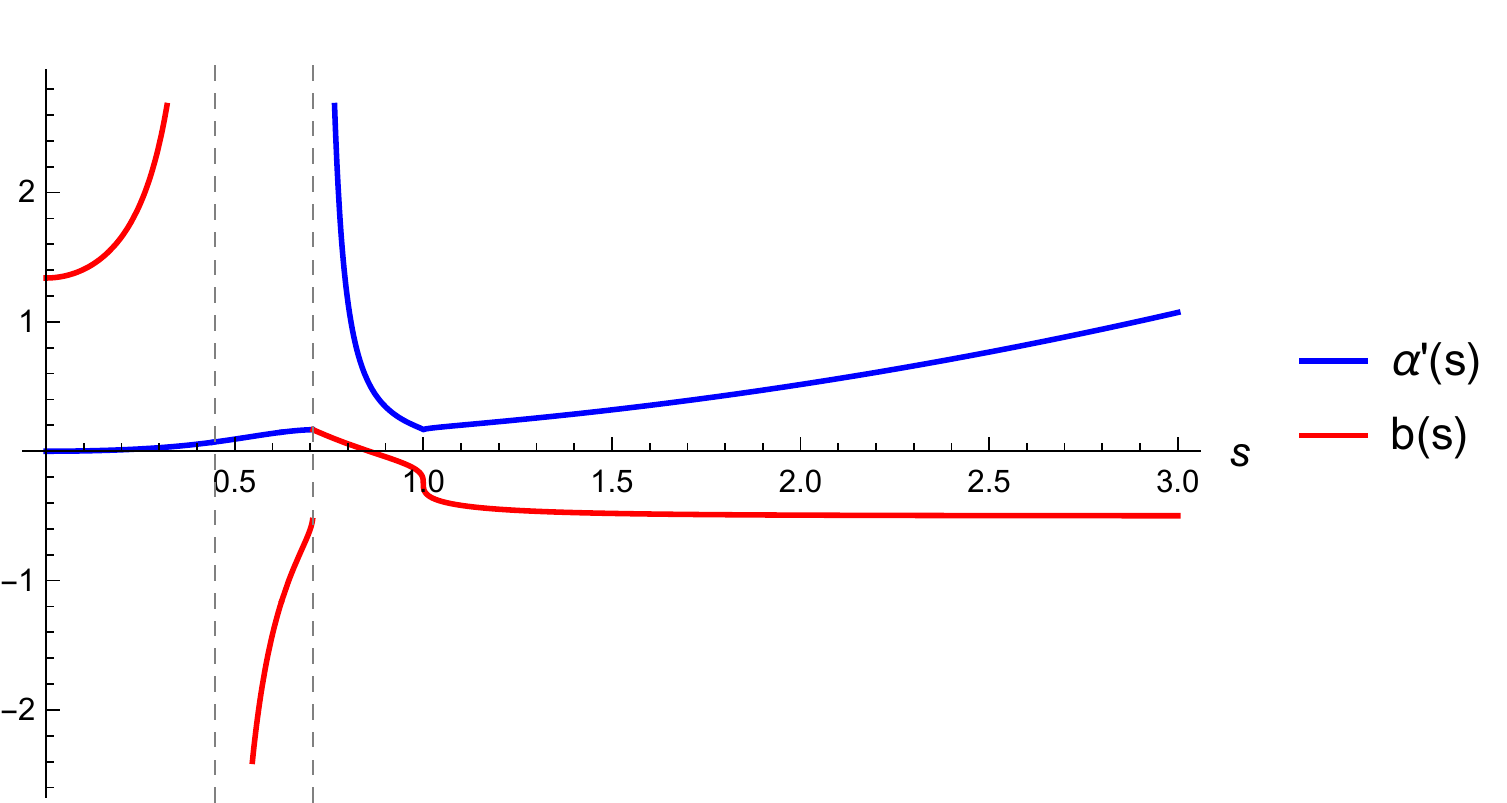}
\caption{Values of $\alpha'(s)$ and $b(s)$ in terms of $s$ for solution 1 in the squashed Aloff--Wallach case. The instanton on the tangent bundle is the Clarke--Oliveira connection in the tangent bundle representation. The instanton on the gauge bundle belongs to the one-parameter family, with parameter $b(s)$. The dashed grey lines indicate the values of $s=1/\sqrt{5}\,, \, 1/\sqrt{2}\,$, where no solutions exist.}
\label{fig:COoneparAW}
\end{figure}

Solution 1 is continuous at $s_8$ and discontinuous at $1/\sqrt{5}$ and $1/\sqrt{2}\,$, where no solutions exist. We find $\alpha'\rightarrow\infty$ as $s\rightarrow\infty$ and $\alpha'\rightarrow 0$ as $s\rightarrow 0$. We plot the solution in \cref{fig:COoneparAW}. Solutions 2 and 3 are again defined in a very small interval and $\alpha'$ blows up.

\bigskip

\subsubsection{One-parameter family of connections on the tangent bundle}
\label{subsec:famfam}

In this case we choose connections from the one-parameter family both for the vector bundle and the tangent bundle, and we denote the parameters by $b$ and $a$ respectively. Having two continuous parameters allows for a wider range of solutions, as we now explain.

Both instantons contribute in a similar fashion to the heterotic Bianchi identity but with opposite signs. This means the equations for $*_sF_1$ and $*_sF_2$ have the same coefficients for $a$ and $b$ up to the sign. The $*_sF_1$ equation is quadratic in $a$ and $b$ whereas the $*_sF_2$ equation is cubic in $a$ and $b$. We impose these equations together with the condition $\alpha'>0$.

\begin{figure}[h]
\centering
\includegraphics[scale=1]{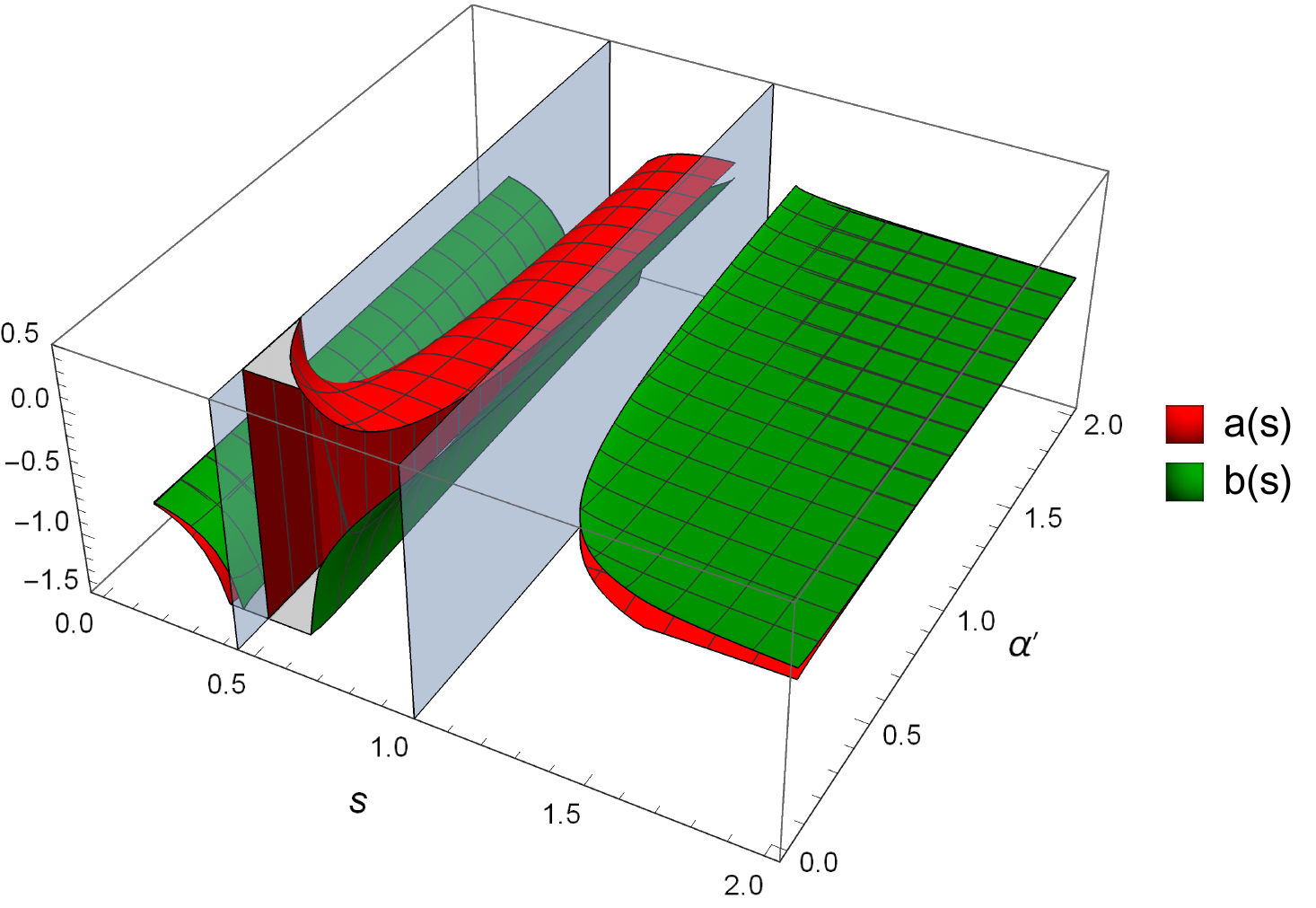}
\caption{First set of values of $a(s,\alpha')$ and $b(s,\alpha')$ in terms of $s$ and $\alpha'$. These provide solutions for both the squashed 7-sphere and the squashed Aloff--Wallach space. The instantons on both the tangent bundle and the gauge bundle belong to the one-parameter family, with parameters $a(s,\alpha')$ and $b(s,\alpha')$ respectively. The grey planes indicate values of $s=1/\sqrt{5}\,, \, 1\,$, where no solutions exist.}
\label{fig:FamFam1}
\end{figure}

\begin{figure}[h]
\centering
\includegraphics[scale=1]{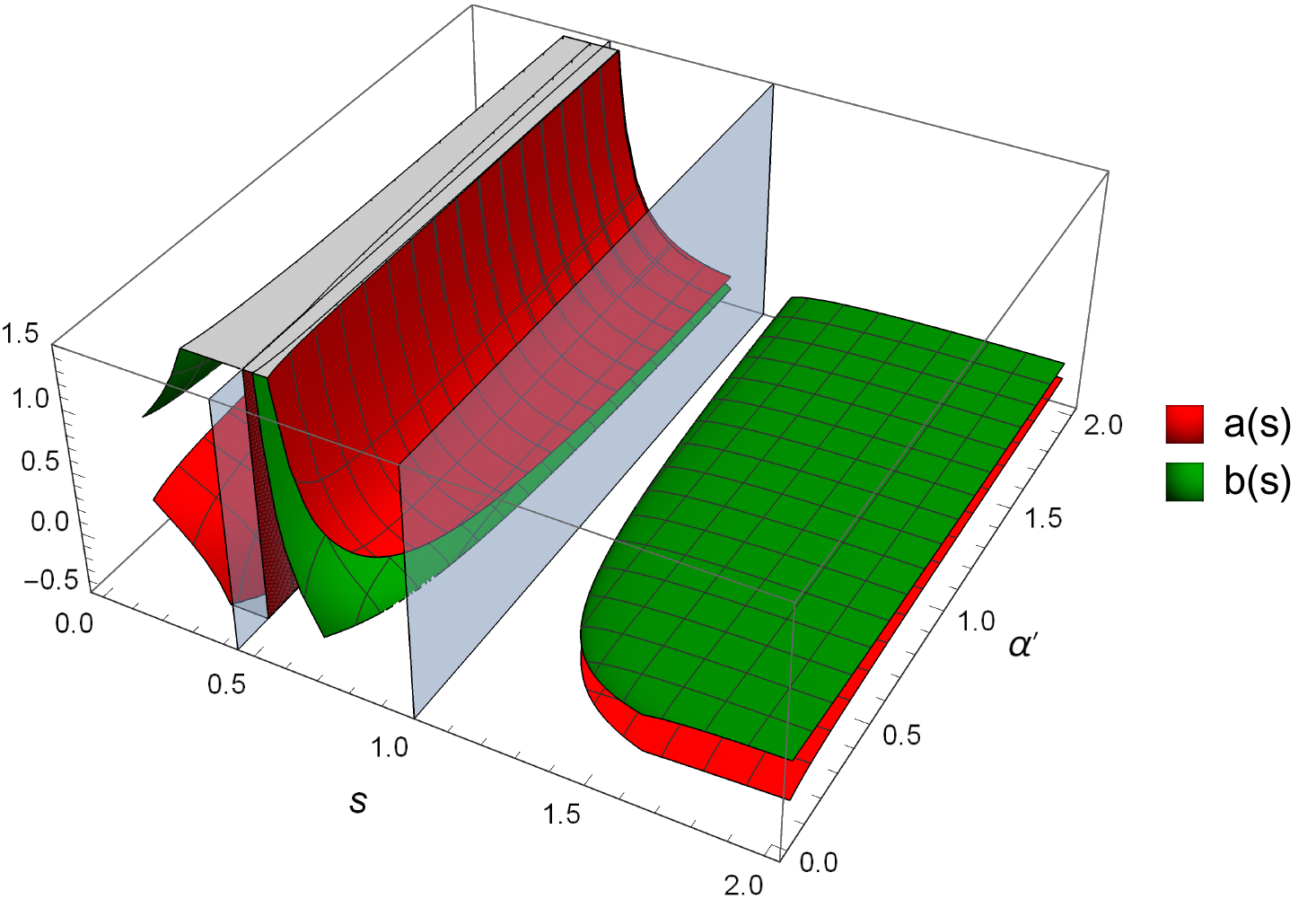}
\caption{Second set of values of $a(s,\alpha')$ and $b(s,\alpha')$ in terms of $s$ and $\alpha'$. These provide solutions for both the squashed 7-sphere and the squashed Aloff--Wallach space. The instantons on both the tangent bundle and the gauge bundle belong to the one-parameter family, with parameters $a(s,\alpha')$ and $b(s,\alpha')$ respectively. The grey planes indicate values of $s=1/\sqrt{5}\,, \, 1\,$, where no solutions exist.}
\label{fig:FamFam2}
\end{figure}

Both for the squashed 7-sphere and the squashed Aloff--Wallach space, we find that for every value of the squashing parameter $s$ except for $s=1$ and $s=1/\sqrt{5}\,$, and for every value of the string parameter $\alpha'$ such that
\begin{equation}
\label{eq:existsolcond}
    \alpha'> \frac{s^2}{12\,(s^2-1)^2} \, ,
\end{equation}
we have two sets of values for $a(s,\alpha')$ and $b(s,\alpha')$ that solve the heterotic Bianchi identity \eqref{eq:heteroticBianchiidentity} and provide solutions of the heterotic G$_2$ system. We present plots of the solutions in \cref{fig:FamFam1,fig:FamFam2}.

If we consider the limit value $\alpha'=\frac{s^2}{12\,(s^2-1)^2}\,$, we find a unique solution of the form
\begin{equation}
\label{eq:limitsolution}
    a(s)=-\,\frac{5 \, s^2-3}{10\,s^2-2} \, , \qquad b(s)=-\,\frac{s^2+1}{10 \, s^2-2}\,.
\end{equation}
Note this is precisely the same solution we found in \eqref{eq:easysolution}. This is because for this specific choice of parameter $b(s)$ the instanton coincides with the tangent bundle representation of the canonical connection, see \cref{foot:canisinfamily}.

The solutions we find present a very interesting profile. As is the case with other solutions, the parameters of the connections blow up at $s=1/\sqrt{5}$ reflecting the fact that the family of instantons collapses to a single connection. For $s=1$ it is not possible to obtain a solution with finite $\alpha'$, analogously to the solution depicted in \cref{fig:easysolution}. In fact, a whole region around $s=1$ is excluded by the condition \eqref{eq:existsolcond}.

When $s\rightarrow 0$ or $s\rightarrow\infty$, the parameters of both connections tend to the same value independently of the value of $\alpha'$.\footnote{Note in these limits the connections tend to a configuration analogous to the standard embedding, where the curvatures from both connections cancel each other.} For the first set (\cref{fig:FamFam1}) we find $a,b\rightarrow-5/6$ as $s\rightarrow 0$, and $a,b\rightarrow-11/30$ as $s\rightarrow \infty$. For the second set (\cref{fig:FamFam2}) we find $a,b\rightarrow 1/2$ as $s\rightarrow 0$, and $a,b\rightarrow-1/10$ as $s\rightarrow \infty$. Interestingly, for $s>1$ the parameters of the connections remain almost constant. Analogously, varying the string parameter $\alpha'$ has a very mild effect on the parameters.

One of the most remarkable features of these solutions is that they can be regarded as finite deformations from a given solution. For a fixed value of $\alpha'$, the solutions describe a deformation with the squashing parameter $s$ as the deformation parameter. Together with the solution depicted in \cref{fig:oneparCOc8}, these are the first examples of finite deformations of the heterotic G$_2$ system. See \cite{delaOssa:2017pqy} for a description of the infinitesimal deformations.

\begin{figure}[h]
\centering
\includegraphics[scale=0.7]{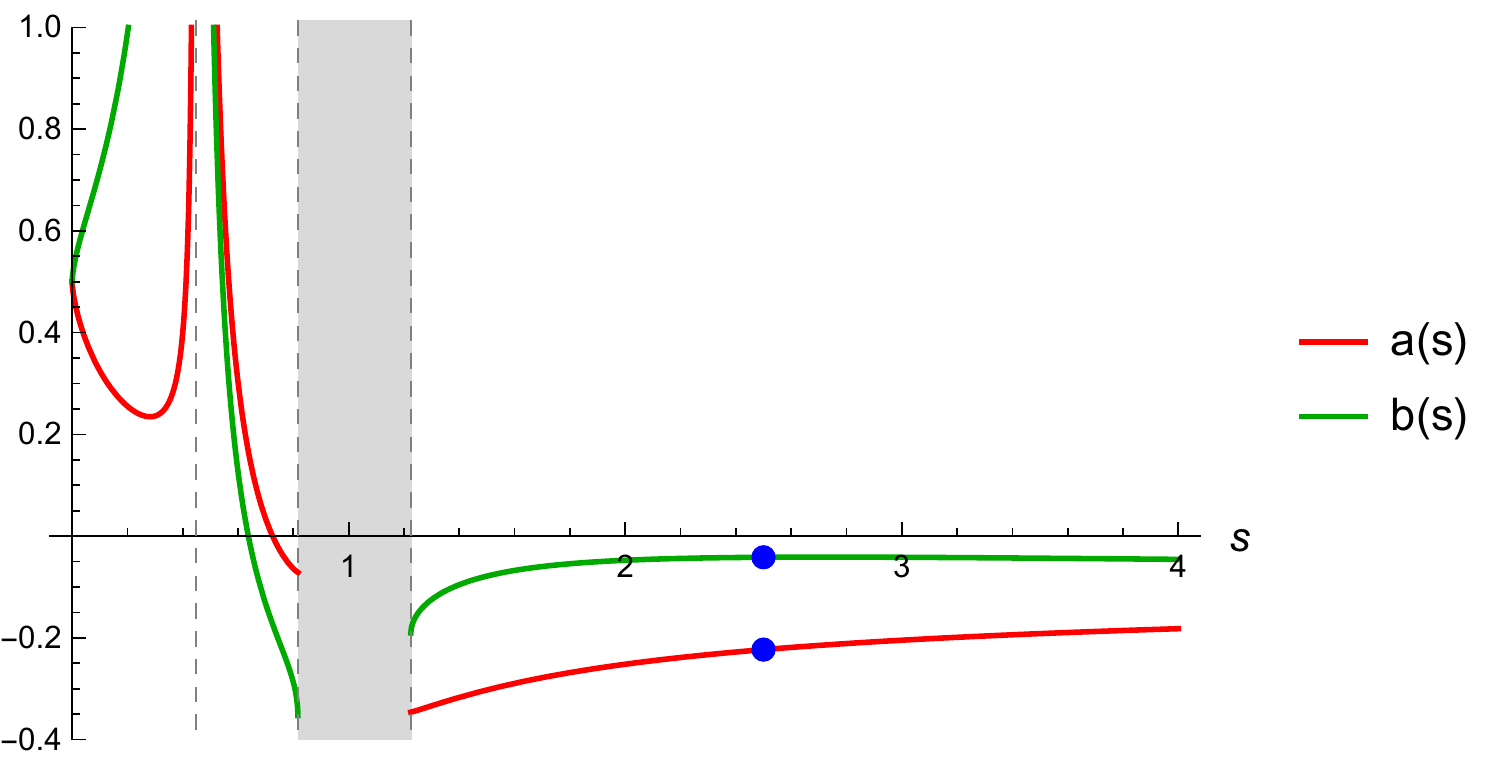}
\caption{One set of values of $a(s,\alpha')$ and $b(s,\alpha')$ in terms of $s$ and with $\alpha'=1/2$ fixed. These provide solutions for both the squashed 7-sphere and the squashed Aloff--Wallach space. The instantons in both the tangent bundle and the gauge bundle belong to the one-parameter family, with parameters $a(s,\alpha')$ and $b(s,\alpha')$ respectively. The blue dots refer to the starting point for the deformation $s=5/2$. The grey region indicates values of $\sqrt{2/3}<s<\sqrt{3/2}$ where no solutions exist, and the dashed line indicates the value $s=1/\sqrt{5}$ where no solution exists.}
\label{fig:DefEx}
\end{figure}

We show an example in \cref{fig:DefEx}. Taking $\alpha'= 1/2$ and starting from the solution with $s=5/2$, we can perform a finite deformation by increasing or decreasing the squashing parameter. We keep a solution of the heterotic G$_2$ system if we deform the instantons in a very specific way as we change $s$. Note that deformation towards higher $s$ is unrestricted, whereas when decreasing $s$ we find that deformations beyond $s=\sqrt{3/2}$ are not allowed. therefore, the solutions with $s<1$ can not be accessed through a continuous deformation.

\subsection{Summary of results}\label{Summary}

We now summarize the results of this section, emphasizing the differences between the solutions obtained.

In \cref{subsec:cancan,subsec:canCO} we have described some isolated solutions where the string parameter $\alpha'$ can be made arbitrarily small by choosing a convenient representation of the gauge bundle. Nevertheless, from the point of view of heterotic string theory the gauge bundle representation must be contained in the adjoint representation of the gauge group $\text{E}_8\times\text{E}_8$. This imposes some restrictions on the representations we can actually choose, in particular the dimension of the representation can not be arbitrarily large.

In \cref{subsec:canfam,subsec:COfam,subsec:famCO} we obtained solutions for a variety of ranges of the squashing parameter $s$, with the string parameter $\alpha'$ depending on the value of $s$. Some of the solutions are well-defined only in very small intervals for which the string parameter quickly blows up. These solutions are less interesting from a physical point of view, since $\alpha'$ should be understood as a small perturbation parameter.

For solutions defined on a wider range, it is interesting to analyze the behaviour in the limit where the squashing parameter becomes very large $s\rightarrow\infty$. Recall that this limit corresponds to the case where the $\text{SU}(2)\cong\Sc^3$ fibres have large radius. Equivalently, the volume of the compact manifold blows up and the space decompactifies. In \cref{subsec:canfam,subsec:famCO} we find solutions where $\alpha'\rightarrow\infty$ in this limit. These solutions, although completely valid from a mathematical point of view, are not satisfactory for string theory as the parameter $\alpha'$ can not be made small.

In \cref{subsec:COfam} we find solutions where $\alpha'$ tends to a constant value as $s\rightarrow\infty$. The exact value of this constant depends on the gauge bundle representation and can be chosen as small as desired. This gives raise to plenty of solutions where the string parameter is small. Furthermore, in \cref{subsec:canfam,subsec:COfam} we find solutions where $\alpha'\rightarrow 0$ as $s$ increases, providing further examples relevant to physics. These solutions correspond to a large volume limit: as the string parameter vanishes the volume of the compact manifold blows up.

It is also worth exploring the limit $s\rightarrow 0$. Geometrically, this corresponds to a situation where the $\text{SU}(2)\cong\Sc^3$ fibres are shrunk to a point and the compact manifold becomes singular. For the solutions in \cref{subsec:canfam,subsec:COfam,subsec:famCO} where this limit can be considered, we always encounter $\alpha'\rightarrow 0$. Hence, the string parameter vanishes in the limit of maximum squashing.

At this point we would like to compare our results with the interesting solutions of Lotay and S\'a Earp \cite{Lotay:2021eog}. They construct solutions to the heterotic G$_2$ system on contact Calabi-Yau 7-folds, which are manifolds obtained from a circle fibration over a Calabi-Yau 3-fold. These present some common features with our solutions: first of all, whereas we rescale the metric along SU(2) fibres and our G$_2$-structure depends on the squashing parameter $s$, they perform a rescaling along the U(1) fibre and their G$_2$-structure depends on the corresponding parameter---which they denote $\varepsilon$.

The authors of \cite{Lotay:2021eog} obtain connections which depend on the parameter $\varepsilon$, in the same way as our connections depend on the parameter $s$. However, their tangent bundle connections satisfy the instanton condition only at first order in $\alpha'$, so they are approximate G$_2$-instantons. Our tangent bundle connections are exact G$_2$-instantons. We use them to solve the heterotic G$_2$ system given in \cref{sec:Heteroticg2Systems}, which is itself valid at first order in $\alpha'$.\footnote{In \cite{delaOssa:2014msa} it was argued that these equations remain valid to all orders in $\alpha'$ after suitable field redefinitions.}

In \cite{Lotay:2021eog}, the authors find solutions with an AdS$_3$ spacetime where the string parameter $\alpha'(\varepsilon)$ depends on the parameter $\varepsilon$, as is the case for most of our solutions where $\alpha'(s)$ depends on $s$. The behaviour $\alpha'\rightarrow 0$ that we find when $s\rightarrow 0$ is also present in their solutions when $\varepsilon\rightarrow 0$. Nevertheless, the curvature of the AdS$_3$ tends to zero in their solutions whereas it blows up in our case. As we have mentioned before, our solutions are such that that spacetime can never be  Minkowski.

Finally, let us stress that in all the solutions of \cite{Lotay:2021eog} there is a mutual dependence between the parameters $\alpha'$ and $\varepsilon$. Our solutions in \cref{subsec:canfam,subsec:COfam,subsec:famCO} present an analogous feature in terms of $\alpha'$ and $s$. Nevertheless, we have also managed to construct solutions where $\alpha'$ does not depend on the squashing parameter $s$.

This is the case of the solutions of \cref{subsec:famfam}, which deserve special attention. By using two different instanton connections from the one-parameter family, we construct a family of solutions where $s$ and $\alpha'$ are two parameters that can be chosen independently within a certain region. This provides a large family of solutions. Furthermore, we can regard these backgrounds as finite deformations: taking any solution as the starting point, we can keep the string parameter $\alpha'$ fixed while we modify the squashing parameter $s$. This results in a deformation of the G$_2$-structure as well as the instanton connections.

These are the first explicit examples of families of finite deformations (together with a very particular solution described in \cref{subsec:COfam}) and they provide a finite version of the infinitesimal deformations discussed in \cite{delaOssa:2017pqy}. Making a thorough study of the relation between infinitesimal and finite deformations will be the subject of future work.

\section{Conclusion}
\label{sec:conclusions}

In this chapter we have constructed new solutions of the heterotic G$_2$ system on squashed homogeneous 3-Sasakian manifolds. We have given a detailed description of the specific family of G$_2$-structures used in the construction as well as the instanton connections employed. These solutions are grouped into different families in terms of the choice of instanton connections on the gauge and tangent bundles. The resulting spacetime is AdS$_3$ for all the solutions found and, together with those of \cite{Lotay:2021eog}, constitute the only few explicit examples of this type.

Moreover, these families of solutions include the first examples of heterotic G$_2$ backgrounds that can be described as finite deformations from a given solution. For any particular solution of those presented in \cref{subsec:famfam}---as well as for the solution depicted in \cref{fig:oneparCOc8} in \cref{subsec:COfam}---we can keep the string parameter $\alpha'$ fixed and obtain new, deformed solutions by performing a squashing of the metric and deforming the bundle instantons.

In \cite{delaOssa:2021cgd} it is explained how heterotic G$_2$ solutions can secretly present enhanced $\mathcal{N}=2$ supersymmetry. The authors showed that this was indeed the case for one of the solutions constructed in \cite{Fernandez:2008wla}. Nevertheless, the supersymmetry enhancement requires the 3-dimensional spacetime to be Minkowski. This is not the case for our solutions, so we can guarantee these are proper $\mathcal{N}=1$ background solutions of the heterotic G$_2$ system.

There are several possible future directions to pursue. First of all, it would be very interesting to use our solutions to improve our understanding of the moduli space of the heterotic G$_2$ system. An easy first step would be to compute the infinitesimal version of the deformation described by our solutions and relate it to the formalism of \cite{delaOssa:2017pqy}. One could also try to explicitly compute the cohomologies encoding the infinitesimal deformations.

We have obtained solutions that follow one particular deformation direction, but we expect further deformations to be allowed. A possibility would be to look into alternative deformations of the G$_2$-structure, such as the ones described for the Aloff--Wallach space in \cite{Ball:2016xte}, and try to find solutions extending the ones we have constructed.

Alternatively, we could explore further choices of instanton connections and study if these provide new solutions of the heterotic G$_2$ system. One possibility would be to use the instantons on the tangent bundle we mentioned at the end of \cref{sec:tangentbundleinstantons}. Further options are available in the literature: for example a 15-dimensional family of G$_2$-instantons for the 7-sphere is described in \cite{Waldron:2020zjo}. Additionally, a detailed description of infinitesimal deformations of G$_2$-instantons for nearly-parallel G$_2$-manifolds can be found in \cite{Singhal:2021fsa}. Finally, connections on homogeneous 3-Sasakian manifolds are described in some generality in \cite{draper2020affine}.

A natural direction---beyond the study of deformations---would be to generalize our construction to obtain completely different solutions. A first option would be to consider alternative G$_2$-structures (not necessarily connected to our family via deformations) such as the standard nearly parallel G$_2$-structures of \cite{2010JGP....60..326A}. Going further, one could try to extend our construction to squashed non-homogeneous 3-Sasakian manifolds, or to generalized 3-($\alpha$,$\delta$) Sasakian manifolds: \cite{agricola2020generalizations,draper2021holonomy}. It would also be interesting to construct solutions involving Sasaki or Sasaki-Einstein manifolds, or more general Aloff--Wallach spaces $N_{p,q}$ making use of the connections of \cite{Ball:2016xte}. Another option would be to generalize the solutions of \cite{Fernandez:2008wla, Fernandez:2014pfa} to other nilmanifolds using the detailed study of \cite{delBarco:2020ddt}.  Finally, one could attempt to exploit the existence of almost contact metric 3-structures on manifolds with a G$_2$-structure to try to generalize the example of \cite[Sect.~5.1]{delaOssa:2021cgd}. 

Finally, in \cite{Clarke:2020erl} T-dual solutions of the heterotic G$_2$ system are constructed. Applying a T-duality to our families of solutions---for example on the circle inside the 3-Sasakian fibre---to obtain new T-dual solutions is a tantalizing possibility. One could also attempt to perform a full non-abelian T-duality on the SU(2) fibre \cite{delaOssa:1992vci}.

\section*{Appendices}
\addcontentsline{toc}{section}{Appendices}

\setcounter{section}{0}
\renewcommand\theHsection{\thechapter.\Alph{section}}
\renewcommand\thesection{\thechapter.\Alph{section}}

\section{Sp(2) structure equations}
\label{sec:Sp2struceq}

A local coframe for Sp(2) is obtained from a basis of the Lie algebra $\mathfrak{sp}(2)$ as described in \cref{sec:Homogeneouscase}. We choose the following basis

{\singlespacing
\begin{align*}
    e^1&=\begin{pmatrix}
i &0 \\
0 &0
\end{pmatrix}, \qquad 
&e^2=\begin{pmatrix}
j &0 \\
0 &0
\end{pmatrix}, \qquad 
e^3&=\begin{pmatrix}
k &0 \\
0 &0
\end{pmatrix}, \\
e^4&=\begin{pmatrix}
0 &-1 \\
1 &0
\end{pmatrix}, \qquad 
&e^5=\begin{pmatrix}
0 &i \\
i &0
\end{pmatrix}, \qquad 
e^6&=\begin{pmatrix}
0 &j \\
j &0
\end{pmatrix}, \qquad 
e^7=\begin{pmatrix}
0 &-k \\
-k &0
\end{pmatrix}, \\
e^8&=\begin{pmatrix}
0 &0 \\
0 &i
\end{pmatrix}, \qquad 
&e^9=\begin{pmatrix}
0 &0 \\
0 &j
\end{pmatrix}, \qquad 
e^{10}&=\begin{pmatrix}
0 &0 \\
0 &k
\end{pmatrix}.
\end{align*}}

The nonzero structure constants on this basis are 
\begin{align}
\label{eq:structureconstantssp2}
\begin{split}
2&=f_{23}^1=-f_{45}^1=-f_{67}^1=f_{31}^2=-f_{46}^2=f_{57}^2=f_{12}^3=f_{47}^3=f_{56}^3 \, ,\\
1&=f_{15}^4=f_{26}^4=-f_{37}^4=-f_{14}^5=-f_{27}^5=-f_{36}^5=f_{17}^6=-f_{24}^6=f_{35}^6=\\
&=-f_{16}^7=f_{25}^7=f_{34}^7 \, ,\\
1&=f_{58}^4=f_{69}^4=-f_{7 \ 10}^4=-f_{48}^5=f_{6 \ 10}^5=f_{79}^5=-f_{49}^6=-f_{5 \ 10}^6=-f_{78}^6=\\
&=f_{4 \ 10}^7=-f_{59}^7=f_{68}^7 \, ,\\
2&=f_{9 \ 10}^8=f_{45}^8=-f_{6 7}^8=f_{10 \ 8}^9=f_{4  6}^{9}=f_{5 7}^{9}=f_{8 9}^{10}=-f_{4  7}^{10}=f_{56}^{10} \, .
\end{split}
\end{align}
Substituting \eqref{eq:structureconstantssp2} in \eqref{eq:generalstructureequation} we obtain the structure equations of the Sp(2) coframe\footnote{Note that the normalization of our basis, convenient for the description of the 3-Sasakian structure, makes our choice of metric not proportional to the Killing form of Sp(2).}
\begin{align*}
\dd e^1&=-\,2 \, e^2\wedge e^3+2 \, e^4\wedge e^5+2 \, e^6\wedge e^7, \\
\dd e^2&=-\,2 \, e^3\wedge e^1+2 \, e^4\wedge e^6-2 \, e^5\wedge e^7, \\
\dd e^3&=-\,2 \, e^1\wedge e^2-2 \, e^4\wedge e^7-2 \, e^5\wedge e^6, \\
\dd e^4&=-\,e^1\wedge e^5-e^2\wedge e^6+e^3\wedge e^7-e^5\wedge e^8-e^6\wedge e^9+e^7\wedge e^{10}, \\
\dd e^5&=+\,e^1\wedge e^4+e^2\wedge e^7+e^3\wedge e^6+e^4\wedge e^8-e^7\wedge e^9-e^6\wedge e^{10}, \\
\dd e^6&=-\,e^1\wedge e^7+e^2\wedge e^4-e^3\wedge e^5+e^7\wedge e^8+e^4\wedge e^9+e^5\wedge e^{10}, \\
\dd e^7&=+\,e^1\wedge e^6-e^2\wedge e^5-e^3\wedge e^4-e^6\wedge e^8+e^5\wedge e^9-e^4\wedge e^{10},
\end{align*}
\begin{align*}
\dd e^8&=-\,2 \, e^9\wedge e^{10}-2 \, e^4\wedge e^5+2 \, e^6\wedge e^7, \\
\dd e^9&=-\,2 \, e^{10}\wedge e^8-2 \, e^4\wedge e^6-2 \, e^5\wedge e^7, \\
\dd e^{10}&=-\,2 \, e^8\wedge e^9+2 \, e^4\wedge e^7-2 \, e^5\wedge e^6.
\end{align*}

\section{SU(3) structure equations}
\label{SU3struceq}

We construct a coframe for SU(3) using a convenient basis of the Lie algebra $\mathfrak{su}(3)$, given by an appropriate normalization of the Gell-Mann matrices

{\singlespacing
\begin{align*}
e^1&=\begin{pmatrix}
i &0 &0 \\
0 &-i &0 \\
0 &0 &0
\end{pmatrix}, &
e^2&=\begin{pmatrix}
0 &1 &0 \\
-1 &0 &0 \\
0 &0 &0
\end{pmatrix},  &
e^3&=\begin{pmatrix}
0 &i &0 \\
i &0 &0 \\
0 &0 &0
\end{pmatrix},\\
e^4&=\sqrt{2}\begin{pmatrix}
0 &0 &i \\
0 &0 &0 \\
i &0 &0
\end{pmatrix}, &
e^5&=\sqrt{2}\begin{pmatrix}
0 &0 &1 \\
0 &0 &0 \\
-1 &0 &0
\end{pmatrix}, &
e^6&=\sqrt{2}\begin{pmatrix}
0 &0 &0 \\
0 &0 &i \\
0 &i &0
\end{pmatrix},\\
e^7&=\sqrt{2}\begin{pmatrix}
0 &0 &0 \\
0 &0 &-1 \\
0 &1 &0
\end{pmatrix},  &
e^8&=-\frac{i}{3}\begin{pmatrix}
1 &0 &0 \\
0 &1 &0 \\
0 &0 &-2
\end{pmatrix}. & &
\end{align*}}

The nonzero structure constants satisfy $f_{\mu\nu}^\rho=-f_{\nu\mu}^\rho$ and take the values:
\begin{align}
\label{eq:structureconstantssu3}
\begin{split}
2&=f_{23}^1=-f_{45}^1=-f_{67}^1=f_{31}^2=-f_{46}^2=f_{57}^2=f_{12}^3=f_{47}^3=f_{56}^3 \, ,\\
1&=f_{15}^4=f_{26}^4=-f_{37}^4=-f_{14}^5=-f_{27}^5=-f_{36}^5=f_{17}^6=-f_{24}^6=f_{35}^6=\\
&=-f_{16}^7=f_{25}^7=f_{34}^7 \, ,\\
1&=f_{58}^4=-f_{48}^5=-f_{78}^6=f_{68}^7 \, ,\\
6&=f_{45}^8=-f_{6 7}^8.
\end{split}
\end{align}
Now from \eqref{eq:generalstructureequation} we deduce that the corresponding basis of left-invariant one-forms satisfy the following structure equations:
\begin{align*}
\dd e^1&=-\,2 \, e^2\wedge e^3+2 \, e^4\wedge e^5+2 \, e^6\wedge e^7 \, ,\\
\dd e^2&=-\,2 \, e^3\wedge e^1+2 \, e^4\wedge e^6-2 \, e^5\wedge e^7 \, , \\
\dd e^3&=-\,2 \, e^1\wedge e^2-2 \, e^4\wedge e^7-2 \, e^5\wedge e^6 \, , \\
\dd e^4&=-\,e^1\wedge e^5-e^2\wedge e^6+e^3\wedge e^7- e^5\wedge e^8 \, , \\
\dd e^5&=+\,e^1\wedge e^4+e^2\wedge e^7+e^3\wedge e^6+ e^4\wedge e^8 \, , \\
\dd e^6&=-\,e^1\wedge e^7+e^2\wedge e^4-e^3\wedge e^5+ e^7\wedge e^8 \, , \\
\dd e^7&=+\,e^1\wedge e^6-e^2\wedge e^5-e^3\wedge e^4- e^6\wedge e^8 \, , \\
\dd e^8&=-\,6 \,  e^4\wedge e^5+6 \,  e^6\wedge e^7 \, .
\end{align*}

\section{Explicit representation matrices of bundle adjoint action}

\subsection{Canonical connection}
\label{sec:ExpRepMatSp1AdjAct}

From the structure constants of the Lie algebra of Sp(2) \eqref{eq:structureconstantssp2} we can read off the adjoint action of SU(2) on the tangent bundle of the squashed 7-sphere, given by the following matrices

\vspace*{-0.5cm}

{\singlespacing
\begin{equation}
\setlength\arraycolsep{3pt}
I_8={\footnotesize\begin{pmatrix}
0 &0 &0 &0 &0 &0 &0 \\
0 &0 &0 &0 &0 &0 &0 \\
0 &0 &0 &0 &0 &0 &0 \\
0 &0 &0 &0 &-1 &0 &0 \\
0 &0 &0 &1 &0 &0 &0 \\
0 &0 &0 &0 &0 &0 &1 \\
0 &0 &0 &0 &0 &-1 &0 \\
\end{pmatrix}}, \quad
I_9={\footnotesize\begin{pmatrix}
0 &0 &0 &0 &0 &0 &0 \\
0 &0 &0 &0 &0 &0 &0 \\
0 &0 &0 &0 &0 &0 &0 \\
0 &0 &0 &0 &0 &-1 &0 \\
0 &0 &0 &0 &0 &0 &-1 \\
0 &0 &0 &1 &0 &0 &0 \\
0 &0 &0 &0 &1 &0 &0 \\
\end{pmatrix}}, \quad
I_{10}={\footnotesize\begin{pmatrix}
0 &0 &0 &0 &0 &0 &0 \\
0 &0 &0 &0 &0 &0 &0 \\
0 &0 &0 &0 &0 &0 &0 \\
0 &0 &0 &0 &0 &0 &1 \\
0 &0 &0 &0 &0 &-1 &0 \\
0 &0 &0 &0 &1 &0 &0 \\
0 &0 &0 &-1 &0 &0 &0 \\
\end{pmatrix}}.
\end{equation}}

\vspace*{-0.5cm}

It can be checked that these matrices satisfy the normalization $[I_a, I_b]=2\,\epsilon\indices{_a_b^c}\,I_c\,$, and we also find
\begin{equation}
\tr(I_a I_b)=-\,4\,\delta_{ab}\,.
\end{equation}
For the squashed Aloff--Wallach space case, we use the structure constants \eqref{eq:structureconstantssu3} and find the action described by

\vspace*{-0.5cm}

{\singlespacing
\begin{equation}
I_8={\footnotesize\begin{pmatrix}
0 &0 &0 &0 &0 &0 &0 \\
0 &0 &0 &0 &0 &0 &0 \\
0 &0 &0 &0 &0 &0 &0 \\
0 &0 &0 &0 &-1 &0 &0 \\
0 &0 &0 &1 &0 &0 &0 \\
0 &0 &0 &0 &0 &0 &1 \\
0 &0 &0 &0 &0 &-1 &0 \\
\end{pmatrix}},
\end{equation}}

\vspace*{-0.5cm}

and we have
\begin{equation}
\tr(I_8 I_8)=-\,4\,.
\end{equation}

\subsection{Clarke--Oliveira connection}
\label{sec:ExpRepMatCOconn}

From the structure constants of the Lie algebra of Sp(2) \eqref{eq:structureconstantssp2} and SU(3) \eqref{eq:structureconstantssu3} we can read off the adjoint action on the tangent bundle of the SU(2) associated to the 3-Sasakian triple of Killing fields. These matrices are valid for both the squashed 7-sphere and the squashed Aloff--Wallach space

\vspace*{-0.5cm}

{\singlespacing
\begin{equation}
\setlength\arraycolsep{3pt}
I_1={\footnotesize\begin{pmatrix}
0 &0 &0 &0 &0 &0 &0 \\
0 &0 &-2 &0 &0 &0 &0 \\
0 &2 &0 &0 &0 &0 &0 \\
0 &0 &0 &0 &1 &0 &0 \\
0 &0 &0 &-1 &0 &0 &0 \\
0 &0 &0 &0 &0 &0 &1 \\
0 &0 &0 &0 &0 &-1 &0 \\
\end{pmatrix}}, ~ ~ ~
I_2={\footnotesize\begin{pmatrix}
0 &0 &2 &0 &0 &0 &0 \\
0 &0 &0 &0 &0 &0 &0 \\
-2 &0 &0 &0 &0 &0 &0 \\
0 &0 &0 &0 &0 &1 &0 \\
0 &0 &0 &0 &0 &0 &-1 \\
0 &0 &0 &-1 &0 &0 &0 \\
0 &0 &0 &0 &1 &0 &0 \\
\end{pmatrix}}, ~ ~ ~
I_3={\footnotesize\begin{pmatrix}
0 &-2 &0 &0 &0 &0 &0 \\
2 &0 &0 &0 &0 &0 &0 \\
0 &0 &0 &0 &0 &0 &0 \\
0 &0 &0 &0 &0 &0 &-1 \\
0 &0 &0 &0 &0 &-1 &0 \\
0 &0 &0 &0 &1 &0 &0 \\
0 &0 &0 &1 &0 &0 &0 \\
\end{pmatrix}}.
\end{equation}}

\vspace*{-0.5cm}

It can be checked that these matrices satisfy the normalization $[I_i, I_j]=2\,\epsilon\indices{_i_j^k}\,I_k\,$, and we also find
\begin{equation}
\tr(I_i I_j)=-\,12\,\delta_{ij}\,.
\end{equation}

\section{Most general G${}_2$-compatible metric connection}
\label{sec:ConnectionCurvatureg2compatible}

We introduced in \cref{sec:tangentbundleinstantons} a one-parameter family of instantons on the tangent bundle given by metric connections compatible with the G${}_2$-structure. We list here their explicit expression, obtained from substituting the Levi-Civita connection and contorsion in \eqref{eq:oneformfinalformula}. Let
\begin{equation}
    \kappa(a,s)=(1+10\,a)\,s+(1-2\,a)\,\frac{1}{s}\,,
\end{equation}
the nonzero connection one-forms (up to antisymmetry $\omega\indices{^\mu_\nu}=-\,\omega\indices{^\nu_\mu}$) for the squashed 7-sphere are given by
\begin{align*}
\omega\indices{^1_2}&=-\,\kappa(a,s)\,\eta^3\,,  & \omega\indices{^2_3}&=-\,\kappa(a,s)\,\eta^1\,, \\
\omega\indices{^3_1}&=-\,\kappa(a,s)\,\eta^2\,,  & & \\
\omega\indices{^4_5}&=\frac{1}{2}\,\kappa(a,s)\,\eta^1 -\eta^8\,, &  \omega\indices{^6_7}&=\frac{1}{2}\,\kappa(a,s)\,\eta^1 +\eta^8\,, \\
\omega\indices{^4_6}&=\frac{1}{2}\,\kappa(a,s) \, \eta^2 -\eta^9\,, & \omega\indices{^5_7}&=-\,\frac{1}{2}\,\kappa(a,s)\,\eta^2 -\eta^9\,, \\
\omega\indices{^4_7}&=-\,\frac{1}{2}\,\kappa(a,s) \, \eta^3 +\eta^{10} \, , & \omega\indices{^5_6}&=-\,\frac{1}{2}\,\kappa(a,s) \, \eta^3 -\eta^{10} \, ,
\end{align*}
and for the squashed Aloff--Wallach space they are given by
\begin{align*}
\omega\indices{^1_2}&=-\,\kappa(a,s) \, \eta^3 \, ,  & \omega\indices{^2_3}&=-\,\kappa(a,s) \, \eta^1 \, , \\
\omega\indices{^3_1}&=-\,\kappa(a,s) \, \eta^2 \, ,  & & \\
\omega\indices{^4_5}&=\frac{1}{2}\,\kappa(a,s) \, \eta^1 -\eta^8 \, , &  \omega\indices{^6_7}&=\frac{1}{2}\,\kappa(a,s) \, \eta^1 +\eta^8 \, , \\
\omega\indices{^4_6}&=\frac{1}{2}\,\kappa(a,s) \, \eta^2 \, , & \omega\indices{^5_7}&=-\,\frac{1}{2}\,\kappa(a,s) \, \eta^2 \, , \\
\omega\indices{^4_7}&=-\,\frac{1}{2}\,\kappa(a,s) \, \eta^3 \, , & \omega\indices{^5_6}&=-\,\frac{1}{2}\,\kappa(a,s) \, \eta^3 \, ,
\end{align*}
From $F=\dd \omega+\omega\wedge\omega$ we compute the curvature. For the squashed 7-sphere (up to antisymmetry $F_{\mu\nu}=-\,F_{\nu\mu}$) the nonzero terms are
\begin{align*}
F_{12}&=-\,\kappa(a,s)\left[\left(\kappa(a,s)-\frac{2}{s}\right)\eta^1\wedge\eta^2+2 \, s \, \omega^3\right] \, , \\[1em]
F_{23}&=-\,\kappa(a,s)\left[\left(\kappa(a,s)-\frac{2}{s}\right)\eta^2\wedge\eta^3+2 \, s \, \omega^1\right] \, , \\[1em]
F_{31}&=-\,\kappa(a,s)\left[\left(\kappa(a,s)-\frac{2}{s}\right)\eta^3\wedge\eta^1+2 \, s \, \omega^2\right] \, , \\[1em]
F_{45}&=\frac{1}{2}\,\kappa(a,s)\left(\kappa(a,s)-\frac{2}{s}\right)\eta^2\wedge\eta^3 +s\left(\kappa(a,s)+\frac{2}{s}\right)\eta^4\wedge\eta^5 +\\
&\quad+s\left(\kappa(a,s)-\frac{2}{s}\right)\eta^6\wedge\eta^7 \, , \\[1em]
F_{67}&=\frac{1}{2}\,\kappa(a,s)\left(\kappa(a,s)-\frac{2}{s}\right)\eta^2\wedge\eta^3 +s\left(\kappa(a,s)-\frac{2}{s}\right)\eta^4\wedge\eta^5 +\\
&\quad+s\left(\kappa(a,s)+\frac{2}{s}\right)\eta^6\wedge\eta^7 \, , \\[1em]
F_{46}&=\frac{1}{2}\,\kappa(a,s)\left(\kappa(a,s)-\frac{2}{s}\right)\eta^3\wedge\eta^1 +s\left(\kappa(a,s)+\frac{2}{s}\right)\eta^4\wedge\eta^6 -\\
&\quad-s\left(\kappa(a,s)-\frac{2}{s}\right)\eta^5\wedge\eta^7 \, , \\[1em]
F_{57}&=-\,\frac{1}{2}\,\kappa(a,s)\left(\kappa(a,s)-\frac{2}{s}\right)\eta^3\wedge\eta^1 -s\left(\kappa(a,s)-\frac{2}{s}\right)\eta^4\wedge\eta^6 +\\
&\quad+s\left(\kappa(a,s)+\frac{2}{s}\right)\eta^5\wedge\eta^7 \, , \\[1em]
F_{47}&=-\,\frac{1}{2}\,\kappa(a,s)\left(\kappa(a,s)-\frac{2}{s}\right)\eta^1\wedge\eta^2 +s\left(\kappa(a,s)+\frac{2}{s}\right)\eta^4\wedge\eta^7 +\\
&\quad+s\left(\kappa(a,s)-\frac{2}{s}\right)\eta^5\wedge\eta^6 \, , \\[1em]
F_{56}&=-\,\frac{1}{2}\,\kappa(a,s)\left(\kappa(a,s)-\frac{2}{s}\right)\eta^1\wedge\eta^2 +s\left(\kappa(a,s)-\frac{2}{s}\right)\eta^4\wedge\eta^7 +\\
&\quad+s\left(\kappa(a,s)+\frac{2}{s}\right)\eta^5\wedge\eta^6 \, ,
\end{align*}
whereas for the squashed Aloff--Wallach space we find
\begin{align*}
F_{12}&=-\,\kappa(a,s)\left[\left(\kappa(a,s)-\frac{2}{s}\right)\eta^1\wedge\eta^2+2 \, s \, \omega^3\right] \, , \\[1em]
F_{23}&=-\,\kappa(a,s)\left[\left(\kappa(a,s)-\frac{2}{s}\right)\eta^2\wedge\eta^3+2 \, s \, \omega^1\right] \, , \\[1em]
F_{31}&=-\,\kappa(a,s)\left[\left(\kappa(a,s)-\frac{2}{s}\right)\eta^3\wedge\eta^1+2 \, s \, \omega^2\right] \, , \\[1em]
F_{45}&=\frac{1}{2}\,\kappa(a,s)\left(\kappa(a,s)-\frac{2}{s}\right)\eta^2\wedge\eta^3 +s\left(\kappa(a,s)+\frac{6}{s}\right)\eta^4\wedge\eta^5 +\\
&\quad+s\left(\kappa(a,s)-\frac{6}{s}\right)\eta^6\wedge\eta^7 \, , \\[1em]
F_{67}&=\frac{1}{2}\,\kappa(a,s)\left(\kappa(a,s)-\frac{2}{s}\right)\eta^2\wedge\eta^3 +s\left(\kappa(a,s)-\frac{6}{s}\right)\eta^4\wedge\eta^5 +\\
&\quad+s\left(\kappa(a,s)+\frac{6}{s}\right)\eta^6\wedge\eta^7 \, , \\[1em]
F_{46}&=\frac{1}{2}\,\kappa(a,s)\left[\left(\kappa(a,s)-\frac{2}{s}\right)\eta^3\wedge\eta^1+2 \, s \, \omega^2\right] \, , \\[1em]
F_{57}&=-\,\frac{1}{2}\,\kappa(a,s)\left[\left(\kappa(a,s)-\frac{2}{s}\right)\eta^3\wedge\eta^1+2 \, s \, \omega^2\right] \, , \\[1em]
F_{47}&=-\,\frac{1}{2}\,\kappa(a,s)\left[\left(\kappa(a,s)-\frac{2}{s}\right)\eta^1\wedge\eta^2+2 \, s \, \omega^3\right] \, , \\[1em]
F_{56}&=-\,\frac{1}{2}\,\kappa(a,s)\left[\left(\kappa(a,s)-\frac{2}{s}\right)\eta^1\wedge\eta^2+2 \, s \, \omega^3\right] \, .
\end{align*}

\newpage

\section{Coefficients of cubic equations}
\label{sec:cubiceqs}

Here we list the coefficients of the cubic equations that need to be satisfied in order to solve the heterotic Bianchi identity in \cref{subsec:canfam,subsec:COfam,subsec:famCO}.

The coefficients of the positive powers of $a$ are the same for all the cubics, whereas the independent term varies case by case. All of them are listed below.\footnote{In \cref{subsec:famCO} the parameter of the connection is denoted by $b$ instead of $a$.}

\begin{table}[h]
{
\centering
\renewcommand{\arraystretch}{2.2}
\begin{tabular}{|c|c|}
\hline 
Power of $a$ &
Coefficient \\
\hline
\hline
$a^3$ & $\dfrac{12 \left(5 \,  s^2-1\right)^3}{s^2 \left(2 \,  s^2-1\right)}$ \\ \hline
$a^2$ & $\dfrac{6 \left(5 s^2-1\right)^2 \left(7
   s^2-1\right)}{s^2 \left(2 \,  s^2-1\right)}$ \\ \hline
$a$ & $\dfrac{3 \left(5 \,  s^2-1\right)\left(11 \,  s^2-5\right)\left( s^2+1\right)}{s^2 \left(2 \,  s^2-1\right)}$  \\
\hline
\end{tabular}
\renewcommand{\arraystretch}{1}
\caption{List of cubic coefficients for positive powers of $a$.
}
\label{tab:cubicpositivecoeffs}
}
\end{table}

\begin{table}[h]
{
\centering
\renewcommand{\arraystretch}{2.2}
\begin{tabular}{|c|c|c|}
\hline 
Section &
Independent term &
Coefficient \\
\hline
\hline
\multirow{2}{*}{\ref{subsec:canfam}} 
  & 7-sphere & $\dfrac{2 \,  c \left( 2\, s^2-1\right) +15 \, s^6+21 \,  s^4-19 \,  s^2-1}{2 \,  s^2\left( 2 \,  s^2-1\right)}$ \\ \cline{2-3}
  & Aloff--Wallach space & $\dfrac{3 \left(2\,q \left(2 \,  s^2-1\right)+5 \,  s^6+7 \,  s^4-17 \,  s^2+5\right)}{2 \,  s^2\left( 2 \,  s^2-1\right)}$ \\
\hline
\multirow{2}{*}{\ref{subsec:COfam}} 
  & 7-sphere & $\dfrac{-\,2 \,  c \left( 2\, s^2-1\right) +15 \, s^6+21 \,  s^4-19 \,  s^2-1}{2 \,  s^2\left( 2 \,  s^2-1\right)}$ \\ \cline{2-3}
  & Aloff--Wallach space & $\dfrac{-\,2\,q \left(2 \,  s^2-1\right)+3 \left(5 \,  s^6+7 \,  s^4-17 \,  s^2+5\right)}{2 \,  s^2\left( 2 \,  s^2-1\right)}$ \\[1ex]
\hline
\multirow{2}{*}{\ref{subsec:famCO}} 
  & 7-sphere & $\dfrac{15 \,  s^6+117 \,  s^4-115 \,  s^2+23}{2 \,  s^2\left( 2 \,  s^2-1\right)}$ \\ \cline{2-3}
  & Aloff--Wallach space & $\dfrac{3 \left(5 \,  s^6+39 \,  s^4-49 \,  s^2+13\right)}{2 \,  s^2\left( 2 \,  s^2-1\right)}$ \\
\hline
\end{tabular}
\renewcommand{\arraystretch}{1}
\caption{List of independent terms of the cubics.
}
\label{tab:cubiccoeffs}
}
\end{table}

\renewcommand\theHsection{\thechapter.\arabic{section}}
\renewcommand\thesection{\thechapter.\arabic{section}}

\chapter{Conclusion}
\label{chap:conclusions}

Throughout this thesis we examined in detail string compactifications on manifolds equipped with a $G$-structure. In particular, we considered  compactifications on both 7-dimensional manifolds with a G$_2$-structure and 8-dimensional manifolds with a Spin(7)-structure. We adopted two different points of view that complement each other: a worldsheet perspective in the first part of the thesis, and a supergravity perspective in the second part. The rich interplay between geometry and physics played a major role in both approaches.

In the first part of the thesis we assumed a sigma model perspective and considered worldsheet aspects of string compactifications. Our focus was on Extra Twisted Connected Sum (ETCS) G$_2$-holonomy manifolds and Generalized Connected Sum (GCS) Spin(7)-holonomy manifolds. These manifolds are constructed by gluing together two open manifolds of reduced holonomy along a common ``neck'' region. We presented a worldsheet realization of this construction: the first step was associating superconformal algebras to the open manifolds, the neck region and the ETCS or GCS manifold. We then showed how these algebras can be arranged in a diamond of inclusions that perfectly matches the geometry.

The diamond was then used to explore several aspects of the geometry of ETCS and GCS manifolds. Explicit examples of ETCS manifolds have been constructed only for certain values of a parameter called \emph{gluing angle}. However, we found no obstructions to the gluing angle from the diamond perspective. We therefore expect that examples with arbitrary gluing angles exist. As for the GCS manifolds, there is no rigorous mathematical proof ensuring that the construction works in general. The existence of the diamond can thus be regarded as a proof of this fact from worldsheet string theory. We have also studied whether the ETCS and GCS constructions render manifolds of generic special holonomy. We have done so by looking at a certain intersection of algebras in the diamond and comparing it with the algebra of the ETCS or GCS manifold.

Finally, mirror maps for GCS manifolds were interpreted in terms of automorphisms of the diamond of superconformal algebras. We performed a systematic search for these automorphisms which resulted in a proposal for new mirror constructions of GCS manifolds. These mirror maps were shown explicitly in the case of GCS manifolds admitting a Joyce orbifold description.

In the second part of the thesis we adopted a perturbative approach and studied supergravity backgrounds. In particular, we considered heterotic supergravity with first order $\alpha'$ corrections compactified on a 7-dimensional manifold preserving minimal supersymmetry. These backgrounds are described by the heterotic G$_2$ system and must satisfy non-trivial conditions that we discussed thoroughly. For example, the compact manifold must have an integrable G$_2$-structure. The field content of the vacuum solution includes two G$_2$-instantons whose curvatures are intertwined with the geometry of the compact manifold via an anomaly cancellation condition.

We obtained new explicit solutions to the heterotic G$_2$ system. The compact manifolds we used are homogeneous 3-Sasakian manifolds---given by SU(2)-bundles over 4-dimensional orbifolds---with squashed metrics. There exists a one-parameter family of these metrics, obtained by rescaling the 3-Sasakian metric along the SU(2) fibres. We presented different G$_2$-instantons over these manifolds, including a one-parameter family on the tangent bundle. We then solved the anomaly cancellation condition by considering different combinations of these G$_2$-instantons.

These families of backgrounds possess several remarkable features. Unlike most of the solutions in the literature, the emerging spacetime is AdS$_3$ in our case. We obtained backgrounds with arbitrary small values of the string parameter $\alpha'$, in particular for the limiting cases of large and small squashing. For some solutions, we were able to vary the squashing parameter while keeping $\alpha'$ fixed, which yields a one-parameter family of finite deformations of that solution. This describes explicitly an unobstructed direction in the moduli space of heterotic G$_2$ systems.

\bigskip

Each of the main chapters in the thesis features an individual ``Conclusion'' section. We provided in those sections several future research directions following either the worldsheet or the supergravity approach. Furthermore, we stressed throughout the thesis how these two perspectives are complementary and benefit from each other. The thesis concludes with a brief outlook on how the results and techniques we have discussed could be combined in future work.

Worldsheet algebras for string compactifications have been mostly studied for torsion-free $G$-structures, which motivates the exploration of how non-zero torsion classes modify these algebras. The case of a G$_2$-structure with torsion class $\tau_0\neq 0$ was described in \cite{Fiset:2021azq} by looking at explicit type II backgrounds. An exciting possibility would be to perform a worldsheet study of the solutions of \cref{chap:heterotic} in the spirit of \cite{Fiset:2021azq}. Our backgrounds have $\tau_0\neq 0$ and $\tau_3\neq 0$ so we expect the resulting worldsheet algebra to be the one introduced in \cite{Fiset:2021azq}, or a generalization of it.\footnote{A natural question is whether the heterotic nature of our solution could affect the associated superconformal algebra. For example, one might worry that the perturbative gauge fields could have an impact on the worldsheet algebra. However, it was shown in \cite{delaOssa:2018azc} that the presence of a bundle does not alter the currents associated to the Howe-Papadopoulos symmetries of the sigma model---and their generalizations. Since these currents are classical versions of the chiral algebra operators, we do not expect differences in W-algebras coming from heterotic or type II theories.}

An intriguing idea would be to push the worldsheet analysis of the solutions even further to obtain a network of algebra inclusions analogous to the diamond from \cref{chap:scas}. To do so, we would have to choose local patches in our manifold and translate the local expressions of the G$_2$-forms, as well as the transition functions, to the language of W-algebras. The case of the 7-sphere is particularly appealing as only two patches are required to describe the whole manifold: this would lead again to a diamond structure in the worldsheet.

Finally, the study of T-duality for the backgrounds we have constructed in \cref{chap:heterotic} is a tantalizing possibility. Understanding the superconformal algebra associated to these solutions would allow a study from the worldsheet perspective. In that case T-dualities would be described as automorphisms of the chiral algebra. Furthermore, a detailed analysis of the automorphisms of this W-algebra could provide some insight into the possible existence of mirror symmetry for manifolds with torsionful $G$-structures. In the torsion-free case, mirror maps correspond to algebra automorphisms in the worldsheet, which raises the question of whether a similar statement holds when some of the torsion classes are non-zero.


\phantomsection
\addcontentsline{toc}{chapter}{Bibliography}
\providecommand{\href}[2]{#2}\begingroup\raggedright\endgroup

\end{document}